\documentclass[iop]{emulateapj}
\slugcomment{{\sc }}
\usepackage[section]{placeins}

\newcommand{\lsim}{\mathrel{\hbox{\rlap{\lower.55ex\hbox{$\sim$}} \kern-.3em \raise.4ex \hbox{$<$}}}}
\def\deg{\hbox{$^{\:\circ}$}}

\slugcomment{}

\shorttitle{0.85 -- 2.5 $\mu$m spectroscopy of nearby galaxies}
\shortauthors{Mason et al.}

\begin{document}

\title{The Nuclear Near-Infrared Spectral Properties of Nearby Galaxies}

\author{R. E. Mason$^1$, A. Rodr\'{i}guez-Ardila$^2$, L. Martins$^3$, R. Riffel$^4$, O. Gonz\'{a}lez Mart\'{i}n$^{5,6}$, C. Ramos Almeida$^{5,6}$,  D. Ruschel Dutra$^{4,5}$, L. C. Ho$^{7,8}$, K. Thanjavur$^{9}$, H. Flohic$^{10}$, A. Alonso-Herrero$^{11}$,  P. Lira$^{12}$, R. McDermid$^{1,13, 14}$, R. A. Riffel$^{15}$, R. P. Schiavon$^{16}$, C. Winge$^{17}$, M. D. Hoenig$^1$, E. Perlman$^{18}$ \vspace*{5mm}}

\affil{$^1$Gemini Observatory, Northern Operations Center, 670 N. A'ohoku Place, Hilo, HI 96720, USA}
\email{rmason@gemini.edu}
\affil{$^2$ Laborat\'orio Nacional de Astrof\'isica/MCT, Rua dos Estados Unidos 154, Itajub\'a, MG, Brazil}
\affil{$^3$ NAT -- Universidade Cruzeiro do Sul, Rua Galv\~ao Bueno, 868, S\~ao Paulo, SP, Brazil}
\affil{$^4$ Universidade Federal do Rio Grande do Sul, Instituto de F\'\i sica, CP 15051, Porto Alegre 91501-970, RS, Brazil}
\affil{$^5$ Instituto de Astrof\'isica de Canarias, Calle V\'ia L\'actea, s/n, E-38205, La Laguna, Tenerife, Spain}
\affil{$^6$ Departamento de Astrof\'isica, Universidad de La Laguna, E-38205, La Laguna, Tenerife, Spain}
\affil{$^7$ Kavli Institute for Astronomy and Astrophysics, Peking University, Beijing, China}
\affil{$^8$ Department of Astronomy, Peking University, Beijing, China}
\affil{$^9$ Department of Physics \& Astronomy, University of Victoria, Victoria, BC, V8W 2Y2, Canada}
\affil{$^{10}$ University of the Pacific, Department of Physics, 3601 Pacific Avenue, Stockton, CA 95211, USA}
\affil{$^{11}$ Instituto de F\'{i}sica de Cantabria, CSIC-UC, 39005 Santander, Spain}
\affil{$^{12}$ Departamento de Astronom\'ia, Universidad de Chile, Casilla 36-D, Santiago, Chile}
\affil{$^{13}$ Department of Physics and Astronomy, Macquarie University, Sydney NSW 2109, Australia}
\affil{$^{14}$ Australian Gemini Office, Australian Astronomical Observatory, PO Box 915, Sydney NSW 1670, Australia}
\affil{$^{15}$ Universidade Federal de Santa Maria, Departamento de F\'\i sica/CCNE, 97105-900, Santa Maria, RS, Brazil}
\affil{$^{16}$  Astrophysics Research Institute, Liverpool John Moores University, IC2, Liverpool Science Park 146 Brownlow Hill, Liverpool
L3 5RF, United Kingdom}
\affil{$^{17}$ Gemini Observatory, Southern Operations Center, Casilla 603, La Serena, Chile}
\affil{$^{18}$ Department of Physics \& Space Sciences, Florida Institute of Technology, 150 W.
University Blvd., Melbourne, FL 32901, USA}

\begin{abstract}

We present spectra of the nuclear regions of 50 nearby (D = 1 -- 92 Mpc, median = 20 Mpc) galaxies of morphological types E to Sm. The spectra, obtained with the Gemini Near-IR Spectrograph on the Gemini North telescope,  cover a wavelength range of approximately 0.85--2.5 $\mu$m at R$\sim$1300--1800. There is evidence that most of the galaxies host an active galactic nucleus (AGN), but the range of AGN luminosities (log (L$_{2-10 \; \rm keV}$ [erg s$^{-1}$])  =  37.0-43.2) in the sample means that the spectra display a wide variety of features. Some nuclei, especially the Seyferts, exhibit a rich emission-line spectrum. Other objects, in particular the type 2 Low Ionisation Nuclear Emission Region galaxies, show just a few, weak emission lines, allowing a detailed view of the underlying stellar population. These spectra display numerous absorption features sensitive to the stellar initial mass function, as well as molecular bands arising in cool stars, and many other atomic absorption lines. We compare the spectra of subsets of galaxies known to be characterised by intermediate-age and old stellar populations, and find clear differences in their absorption lines and continuum shapes. We also examine the effect of atmospheric water vapor on the signal-to-noise ratio achieved in regions between the conventional NIR atmospheric windows, of potential interest to those planning observations of redshifted emission lines or other features affected by telluric H$_2$O.  Further exploitation of this data set is in progress, and the reduced spectra and data reduction tools are made available to the community.

\end{abstract}

\keywords{infrared: galaxies, galaxies: active, galaxies: stellar content, galaxies: nuclei, galaxies:Seyfert}

\section{Introduction}
\label{intro}

Near-infrared (NIR; $\sim$0.8 - 2.5 $\mu$m) observations can provide a wealth of information addressing numerous outstanding questions about the physics and lifecycles of galaxies. Many diagnostic lines are present in this wavelength range: H$_2$ and coronal lines can be used to trace molecular gas and super-excited, often outflowing, material; obscured broad line regions (BLRs) can be revealed; [FeII] emission may probe shock-excited gas associated with supernova activity or jets; and stellar absorption lines may measure stages of stellar evolution that have less obvious optical signatures, to name just a few examples. NIR observations of inactive galaxies, starbursts, and active galactic nuclei (AGN) have already given rise to advances in areas from stellar population studies \citep[e.g.][]{Rieke80,McLeod93,Origlia93,Riffel09,Zibetti13,Martins13a} to black hole (BH) mass measurements \citep{Houghton06,Landt13,RARiffel13}.

One topic of much recent interest is the contribution of thermally-pulsing asymptotic  giant branch (TP-AGB) stars to the NIR luminosity of galaxies.  The complexity of this stage of stellar evolution, which occurs in stars of  M$\lesssim 6 M_\odot$ at ages of $\sim$0.3 - 2 Gyr, makes it very difficult to model accurately \citep[e.g.][]{Marigo08}, and the available observed spectra of single stars \citep[][and references therein]{Rayner09} span only a limited range of age and metallicity. Whereas some models predict only weak spectral features and a minor contribution from TP-AGB stars to the NIR luminosity of a galaxy \citep{Bruzual03}, others predict that these stars will cause strong molecular absorptions and contribute as much as 80\% of the K band light \citep{Maraston05}. This can have a large effect on the stellar masses and stellar population ages derived from observations \citep{Kannappan07,Melbourne12}.  Distinguishing between these models is not straightforward: recent studies have come to very different conclusions regarding whether a large contribution from TP-AGB stars is required \citep{Maraston06,Miner11,Riffel11,Lyubenova12} or ruled out \citep{Kriek10,Zibetti13} by observations of globular clusters, post-starburst galaxies, and high-redshift galaxies. High-quality NIR spectra of galaxies can be used to search for the molecular absorption bands, such as CN, C$_2$, and H$_2$O, expected from these stars \citep[e.g.][]{Riffel07}. Detailed investigations of these bands in nearby galaxies with a range of properties will aid the development of the models necessary to understand and predict the spectral energy distributions (SEDs) of galaxies at z$\sim$3, at the age when their initial stellar population was producing large numbers of TP-AGB stars.

Towards the opposite end of the stellar luminosity range,  it has been reported that the centers of elliptical galaxies contain an enhanced fraction of low-mass stars, indicating that the stellar initial mass function (IMF) may not be universal \citep[e.g.][]{vanDokkum10,Cappellari12,LaBarbera13,Spiniello14}. The relative numbers of high- and low-mass stars produced during an episode of star formation will affect the influence of the star-forming region on its surroundings, as well as determine the chemical evolution of a galaxy, making the stellar IMF an important area of study.  Much of the evidence for a non-universal IMF has come from studies of stellar absorption lines around 1 $\mu$m, such as the 0.99 $\mu$m Wing-Ford FeH band which is stronger in dwarf stars than in giants. IMF-sensitive features can also be found throughout the NIR \citep[][]{Ivanov04,Rayner09,Conroy12}. Modeling of the NIR features, especially in conjunction with optical spectra to break the degeneracy between the effects of the IMF and metallicity, may be used to infer the presence of a ``bottom-heavy'' IMF, and perhaps also to constrain the overall shape of the IMF of extragalactic stellar populations \citep{Conroy12}.

Both stellar- and AGN-related activity may play a role in producing the emission lines observed in objects like low-ionization nuclear emission region galaxies \citep[LINERs; e.g.][]{Heckman80,Ferland83,Terlevich85,Binette94,Dopita96,Ho08}. Detailed SED studies suggest that the weak AGN that exists in many LINERs cannot supply sufficient ionizing photons to power the optical emission lines \citep{Eracleous10b}. Other power sources must therefore contribute, such as jet or supernova shocks, or stellar photoionisation. \citet{Alonso-Herrero00} were able to explain the NIR [Fe II] 1.26 $\mu$m/Pa$\beta$ ratio in several LINERs with an aging starburst model, and correlations between line emission and optical stellar surface brightness/luminosity and soft X-ray emission suggest that post-AGB stars are important sources of ionizing photons  in many objects \citep{Macchetto96,Flohic06,Sarzi10}. In fact, \citet{Stasinska08} find
that post-AGB stars and white dwarfs could account for the placement of LINERs on the ``AGN branch'' of optical line diagnostic diagrams, and this could also explain their extended line emission \citep{Singh13}. If the emission lines in LINERs are indeed not directly indicative of the presence of a weak AGN, this would considerably alter our view of AGN demographics. By providing tracers of young stars (e.g. Pa$\alpha$, Pa$\beta$, and Br$\gamma$) and shocked gas (e.g. [Fe II], H$_2$), NIR spectra can be used to investigate the significance of these sources of ionization \citep{Larkin98,Rodriguez-Ardila04,Rodriguez-Ardila05,Riffel13}. In higher-luminosity active nuclei, high-excitation NIR coronal lines can be used to probe the AGN ionizing continuum and the density of the surrounding medium, as well as outflows and gas-jet interactions \citep[e.g.][]{Oliva94,Marconi96,Rodriguez-Ardila06,Rodriguez-Ardila11,Geballe09}. 

As well as stars and line-emitting gas, hot dust and the accretion disk also leave detectable IR signatures in a significant fraction of AGN \citep{Ivanov00,Rodriguez-Ardila06b,Riffel09,RamosAlmeida09,Mor12}. The IR spectra of many Seyfert 1 nuclei and quasars show an inversion around 1 $\mu$m, interpreted as the minimum between the AGN-heated dust whose emission peaks in the mid-infrared (MIR), and the accretion disk that emits strongly in the optical/UV \citep[e.g.][]{Landt11}. At low accretion rates, however, this situation is expected to change. The inner region of the accretion disk is predicted to be replaced by an advection-dominated accretion flow \citep{Esin97,Narayan08}, implying that the disk emission should shift redwards, perhaps into the NIR \citep{Ho08,Taam12}. Furthermore, various models suggest that the obscuring torus of the AGN unified model should disappear when the accretion rate becomes sufficiently low \citep{Elitzur06,Vollmer08}. The dust continuum in such nuclei should therefore be weak. IR studies of the accretion disk and dust have so far been carried out on only a few low-luminosity AGN, using high-resolution imaging in the NIR and MIR \citep{Asmus11,Mason12,Mason13}. Using NIR spectroscopy to reveal the disk and dust emission over a range of AGN luminosities may unearth systematic trends that will allow a better understanding of accretion and obscuration in AGN at low accretion rates.

Several sizeable compendia of NIR spectra of local galaxies exist in the literature. In addition to a large number of compilations covering individual NIR atmospheric windows \citep[e.g.][]{Kawara87,Goldader97,Vanzi97,Ivanov00,Reunanen02,Kotilainen12,vanderLaan13}, a handful of groups have presented spectra covering the whole $\sim$0.8--2.5 $\mu$m range.
\citet{Mannucci01} compiled large-aperture spectra of 28 normal galaxies, creating template spectra of various morphological types and calculating k-corrections for these objects. \citet{Riffel06} presented an atlas of 47 AGN nuclear spectra in which a rich set of emission lines was revealed, while \citet{Landt11,Landt13} used NIR spectroscopy of 23 broad-line AGN to estimate the black hole masses of those objects. \citet{Mould12} showed spectra of the central $\sim$1\arcsec\ of 136 local, radio-detected elliptical galaxies, with a view to future adaptive optics-assisted integral field spectroscopy. Most recently, a NIR atlas of 23 star-forming galaxies was published by \citet{Martins13a}, and their stellar populations analyzed in \citet{Martins13b}. 

In this paper, we present Gemini NIR spectroscopy of the nuclei of 50 galaxies, almost all from the Palomar survey \citep{Ho95,Ho97}. The Palomar galaxy survey provides a spectral atlas of a near-complete, flux-limited (B$<$12.5 mag) sample of nearby, northern galaxies. The 486 Palomar galaxies were observed in the optical with a 2\arcsec\ slit at a spectral resolution of R$\sim$2000, and the resulting spectra revealed 420 emission-line objects - Seyferts, LINERs, transition objects (TO; objects with [OI] strength intermediate between HII region galaxies and classical LINERs)  and HII nuclei - tracing nuclear activity at low luminosities. The uniform, statistically-significant data set has been an invaluable aid to research on a wide range of issues. 

The galaxies selected for this work were chosen from among the Palomar galaxies showing emission lines in their optical spectra, with Seyfert, LINER, or, occasionally, TO classification. The data complement published NIR galaxy spectra in several ways. Some of the more luminous, nearby AGN in the subsample provide an opportunity to study numerous, strong AGN emission lines. In most of the galaxies, however, the emission lines are weak and the continuum emission is dominated by starlight. This, together with the good signal-to-noise ratio (S/N) of the spectra, allows detailed study of undiluted atomic and molecular stellar absorption features. The wide, simultaneous wavelength coverage, in some cases with good enough telluric  line removal to reveal the regions between the conventional $J$-, $H$- and $K$-band atmospheric windows, offers an extensive overview of the properties of nearby galaxy nuclei in the NIR. And, as the Palomar galaxies are bright and nearby, it has been possible to obtain high-quality NIR spectra with a relatively modest investment of telescope time.

In \S\ref{sample}, we describe the sample of galaxies selected for this study. \S\ref{obsdr} gives details about the observations and data reduction, as well as information about accessing the spectra and the code used to produce them. In \S\ref{results} we show some illustrative spectra, compare the absorption-line and continuum spectra of intermediate-age and old galaxies, and discuss the emission lines observed in the data. Additional material, including notes about individual objects, and the optical and NIR spectrum of each one, is given in Appendix \ref{A1}. In Appendix \ref{A2}, we examine how the quality of the data between the NIR atmospheric windows depends on the atmospheric water vapor column at the time of the observations. The intent of this paper is to present the spectra, point out some noteworthy features of the data, and act as a useful reference for the general NIR properties of local galaxy nuclei. Further analysis of various aspects of the data set will be presented in forthcoming papers\footnote{These include: (1) a comparison of velocity dispersions derived from the calcium triplet and CO band heads \citep{Riffel14}; (2) modeling of the stellar absorption features in the highest-quality spectra (Riffel et al.); (3) analysis of the emission-line spectra (Gonz\'{a}lez-Mart\'{i}n, Martins et al.); (4) presentation of the extended emission-line spectrum of NGC~4388 (Rodr\'{i}guez-Ardila et al.); (4) the dust and accretion disk emission in the low-luminosity AGN (Biddle, Mason et al.); and (5) full stellar population synthesis of the whole data set (Riffel et al.).}. 

\vspace*{5mm}

\section{The Sample}
\label{sample}

The galaxies selected for this work were drawn from the Palomar nearby galaxy survey \citep{Ho95,Ho97}. With the exception of NGC~205, an inactive galaxy chosen as an example of an intermediate-age stellar population (\S\ref{abs}), the targets are all classified by \citet{Ho97} as Seyferts, LINERs and TOs. As well as these 48 galaxies, we also observed two non-Palomar objects (1H1934-063 and NGC~7469)  in common with \citet{Riffel06}, to compare the results with previously published spectra. 

\begin{figure*}
\includegraphics[scale=0.8]{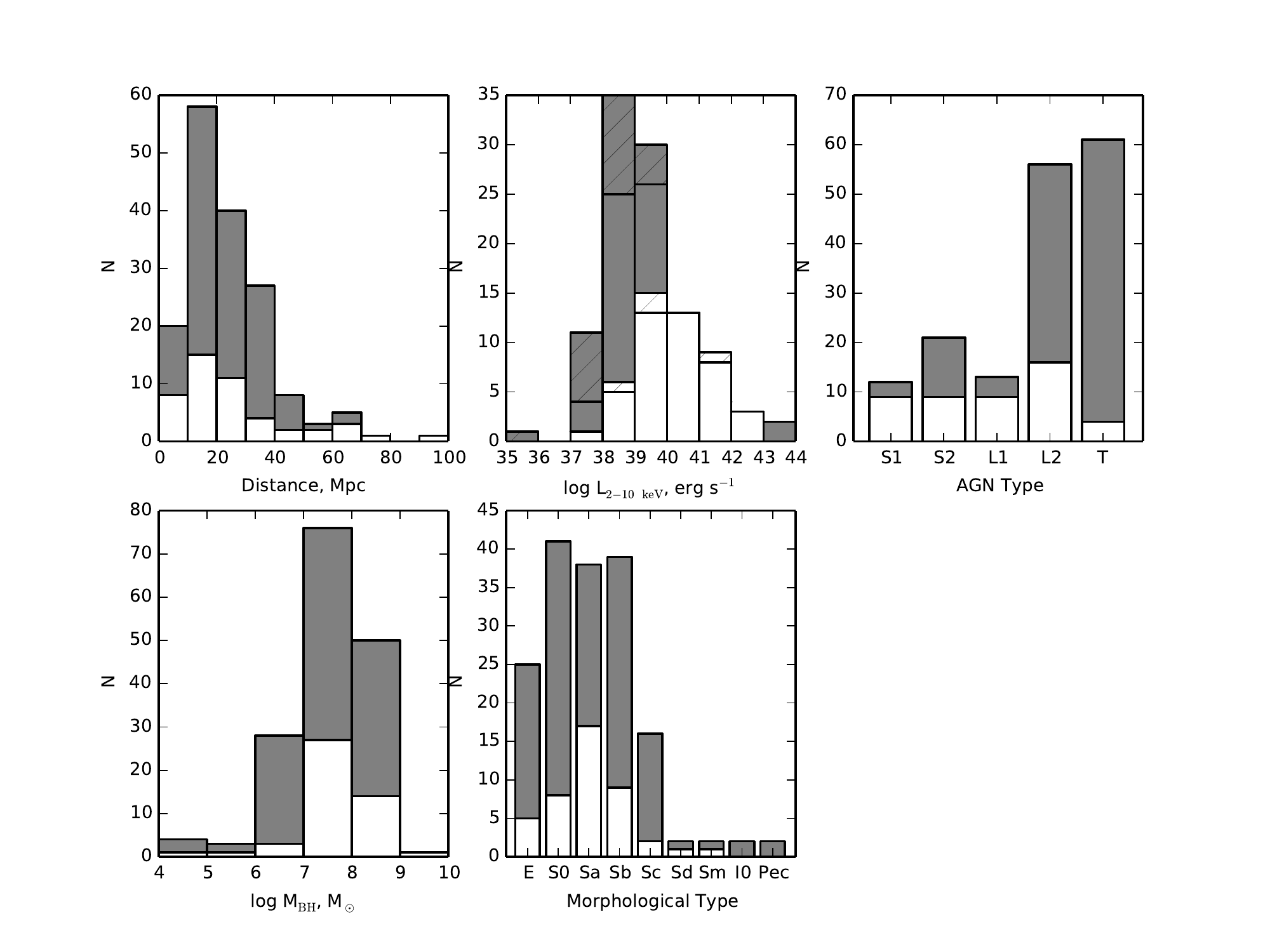}
\caption{ {\small Characteristics of the subsample of Palomar emission-line galaxies presented in this paper (white), and the remainder of the Palomar Seyferts, LINERs and TOs (gray). Distances, AGN types and morphological types are from \citet{Ho97}. ``S1'' encompasses Seyfert 1 - 1.9. Black hole masses are derived from the stellar velocity dispersion compilation of \citet{Ho09a}, using equation 3 of \citet{Kormendy13}. While estimated hard (2-10 keV) X-ray luminosities are available for all of the Palomar objects in our sample in \citet{Ho09}, the 67 galaxies in the parent sample with no hard X-ray luminosity in that paper are omitted from the log L$\rm_X$ panel. Those objects probably lie towards the low end of the L$_X$ distribution. Hatched areas in the log L$\rm_X$ panel indicate upper limits.}}

\label{histo}
\end{figure*}

The characteristics of the Palomar emission-line galaxies in this sample, and those of the remainder of the Seyferts, LINERs and TOs in the original Palomar study, are shown in Figure \ref{histo}. The objects studied here were chosen primarily to extend the range of AGN with NIR spectra to lower luminosities than currently available. The hard X-ray luminosities of the AGN are therefore generally low relative to well-known Seyferts such as NGC~1068 and NGC~4151 \citep[log (L$_{2-10 \; \rm keV}$ {[erg s$^{-1}$]}) = 42.8 and 42.5, respectively; ][]{Panessa06}. They do, however, lie towards the high end of the X-ray luminosity distribution of the Palomar Seyferts/LINERs/TOs as a whole. This is because (i) in future work we are interested in decomposing the spectra into stellar and AGN (accretion disk and dust) components, and recovering AGN emission from seeing-limited spectra of very low-luminosity AGN may not be possible; and (ii) brighter nuclei were prioritized in order to enable observations in sub-optimal conditions (\S\ref{obsdr}). The brighter nuclei also tend to be well-studied objects with plentiful ancillary data, potentially allowing fuller exploitation of the data.

The sample contains comparable numbers of Seyfert and LINER nuclei of both types 1 and 2, and their host galaxies span a range of morphological types from E to Sm. The distances to the galaxies range from 0.7 - 92 Mpc, with a median of 20 Mpc. This, and other basic information about each object, is given in Table \ref{table1}.

\begin{deluxetable}{ccccccccc}
\tabletypesize{\scriptsize}
\vspace*{-10mm}
\tablecaption{Basic information about the galaxies \label{table1} \vspace*{-2mm}}
\tablehead{
\colhead{Galaxy} & \colhead{Distance (Mpc)} & \colhead{Morph. Type} & \colhead{Inclination (${\deg}$)} & \colhead{Activity} & \colhead{log L$\rm_{2-10 \; keV}$\tablenotemark{a}} & log M$\rm_{BH}$\tablenotemark{b} & \colhead{Ref.\tablenotemark{c}}
}
\startdata
\cutinhead{Seyfert 1.2--1.9}
NGC 4235  & 35.1 & SA(s)a & 84 & S1.2 & 42.25 & 8.0 &  13,14\\
NGC 3031 & 1.4 & SA(s)ab & 60 & S1.5 & 39.38 & 8.1 &  3,6,7\\
NGC 5033 & 18.7 & SA(s)c & 64 & S1.5 & 40.70 & 8.0 &  14,21\\
NGC 4395 & 3.6 & SA(s)m: & 34 & S1.8 & 39.58 & 4.9 &   -- \\
NGC 2639 & 42.6 & (R)SA(r)a? & 54 & S1.9 & 40.85 & 8.3  & 3,13\\
NGC 4258 & 6.8 & SAB(s)bc & 70 & S1.9 & 40.89 & 7.9 &  6,8,12,14,23 \\
NGC 4388 & 16.8 & SA(s)b:  & 83 & S1.9 & 42.14 & 7.0 &  9,14,15,16,17,18,24 \\
NGC 4565 & 9.7 & SA(s)b?  & -- & S1.9 & 39.56 & 7.8 & 32 \\
NGC 4579 & 16.8 & SAB(rs)b & 38 & S1.9/L1.9 & 41.15 & 8.1 &  3,19,23\\
\cutinhead{Seyfert 2}
NGC 1167 & 65.3 &  SA0- & 32 & S2 & 40.32 & 8.7 &  1,23\\
NGC 1358 & 53.6 & SAB(r)0/a & 38 & S2 & 42.68 & 8.7 &  23 \\
NGC 2273 & 28.4 & SB(r)a: & 41 & S2 & 40.02 & 7.9 &  1,2,14,24\\
NGC 2655 & 24.4 & SAB(s)0/a & 34 & S2 & 41.87 & 8.1 &   -- \\
NGC 3079 & 20.4 & SB(s)c  & -- & S2 & 38.60 & 8.3 &  8,33 \\
NGC 3147 & 40.9 & SA(rs)bc & 27 & S2 & 41.87 & 8.7 &  10\\
NGC 4725 & 12.4 & SAB(r)ab pec & 46 & S2: & 39.11 & 7.8 &   21 \\
NGC 5194 & 7.7 & SA(s)bc pec & 53 & S2 & 41.03 & 7.1 &  12,23 \\
NGC 7743 & 24.4 & (R)SB(s)0+ & 32 & S2 & 39.71 & 7.0 &  1,3,12,23 \\
\cutinhead{LINER 1.9}
NGC 266 & 62.4 & SB(rs)ab & 12 & L1.9 & 40.88 & 8.8 &   -- \\
NGC 315 & 65.8 & E+: & 52 & L1.9 & 41.63 & 9.3 &  1\\
NGC 1052 & 17.8 & E4 & -- & L1.9 & 41.53 & 8.6 &  1,2,5\\
NGC 3718 & 17.0 & SB(s)a pec & 62 & L1.9 & 40.44 & 8.1 &  --  \\
NGC 3998 & 21.6 & SA(r)0? & 34 & L1.9 & 41.34 & 9.3 &  2,3,11,12\\
NGC 4203 & 9.7 &  SAB0-: & 21 & L1.9 & 39.69 & 8.2 &  1,21\\
NGC 4450 & 16.8 & SA(s)ab & 43 & L1.9 & 40.02 & 7.8 &   -- \\
NGC 4750 & 26.1 & (R)SA(rs)ab & 24 & L1.9 & 39.94 & 7.8 &   -- \\
NGC 5005 & 21.3 & SAB(rs)bc & 63 & L1.9 & 39.94 & 8.2 &  21\\
\cutinhead{LINER 2}
NGC 404 & 2.4 & SA(s)0-: & -- & L2 & 37.02 & 5.4 &  2,4,12\\
NGC 474 & 32.5 & (R')SA(s)0 & 27 & L2:: & $<$38.36 & 8.1 &  1 \\
NGC 1961 & 53.1 & SAB(rs)c & 51 & L2 & 40.31 & 8.9 &  -- \\
NGC 2768 & 23.7 & E6: & -- & L2 & 40.59 & 8.3 &   1 \\
NGC 2832 & 91.6 & E+2: & -- & L2:: & $<$41.49 & 9.5 &   -- \\
NGC 3169 & 19.7 & SA(s)a pec & 52 & L2 & 41.05 & 8.4 &  -- \\
NGC 3190 & 22.4 & SA(s)a pec  & 72 & L2 & 39.54 & 8.4 &   -- \\
NGC 3607 & 19.9 & SA(s)0: & 62 & L2 & 38.63 & 8.8 &  1,6 \\
NGC 4346 & 17.0 & SA0 & 70 & L2:: & $<$39.82 & 7.9 &  -- \\
NGC 4548 & 16.8 & SB(rs)b & 38 & L2 & 39.74 & 7.4 &  -- \\
NGC 4594 & 20.0 & SA(s)a  & 68 & L2 & 40.69 & 8.9 &  20\\
NGC 4736 & 4.3 & (R)SA(r)ab & 36 & L2 & 38.48 & 7.4 & 7,12 \\
NGC 5371 & 37.8 & SAB(rs)bc & 38 & L2 & 40.84 & 8.3 &  21 \\
NGC 5850 & 28.5 & SB(r)b & 30 & L2 & $<$39.54 & 7.8 &  21 \\
NGC 6500 & 39.6 & SAab: & 43 & L2 & 39.73 & 8.6 &  2 \\
NGC 7217 & 16.0 & (R)SA(r)ab & 34 & L2 & 39.87 & 7.8 &  12 \\
\cutinhead{Transition Object, Inactive}
NGC 410 & 70.6 & E+: & -- & T2: & 40.09 & 9.3 &  1,5 \\
NGC 660 & 11.8 & SB(s)a pec & 70 & T2/H: & 39.55 & 7.6 & -- \\
NGC 4569 & 16.8 & SAB(rs)ab & 65 & T2 & 39.41 & 7.8 &  2,3,19 \\
NGC 7331 & 14.3 & SA(s)b & 72 & T2 & 38.66 & 7.8 &  22 \\
NGC 205 & 0.7 & dE5 pec & -- & -- & $<$36.41 & 4.4 & -- \\
\cutinhead{Non-Palomar Galaxies}
1H1934-063\tablenotemark{d} & 42.7 & E & 35 & NLS1 & 42.8 & 8.3 &  25,26,28\\
NGC 7469\tablenotemark{d} & 57.0 & SBa & 30 & S1 & 43.2 & 7.7 & 8,27,28,29,30,31
\enddata
\tablecomments{All data from \citet{Ho97} and references therein, except where stated. S=Seyfert, L=LINER, T=Transition Object, H=HII region galaxy. ``:'' indicates an uncertain classification, ``::" a very uncertain one, as in \citet{Ho97}.}
\tablenotetext{a}{From \citet{Ho09}, who used spectral fitting where available to convert published luminosities to the 2--10 keV bandpass, and otherwise assumed a standard spectral slope and no extinction correction.}
\tablenotetext{b}{Calculated using the stellar velocity dispersions ($\sigma_*$) of \citet{Ho09a}, and equation 3 of \citet{Kormendy13}.}
\tablenotetext{c}{Previous spectroscopy in the 1-2.5 $\mu$m range (including only papers in which the spectra are presented). 
1: \citet{Mould12}. 2: \citet{Rhee05}. 3: \citet{Alonso-Herrero00}. 4: \citet{Seth10}. 5: \citet{Mobasher96}. 6: \citet{Kang13}. 7: \citet{Walker88}. 8: \citet{Sosa-Brito01}. 9: \citet{Veilleux97}. 10: \citet{Tran11}. 11: \citet{Walsh12}. 12: \citet{Larkin98}. 13: \citet{Imanishi04b}. 14: \citet{Ivanov00}. 15: \citet{Imanishi04}. 16: \citet{Knop01}. 17: \citet{Goodrich94}. 18: \citet{Blanco90}. 19: \citet{Mazzalay13}. 20: \citet{Cesetti09}. 21: \citet{Bendo04}. 22: \citet{Matsuoka12}. 23: \citet{Oi10}. 24: \citet{vanderLaan13}. 25: \citet{Rodriguez-Ardila00}. 26: \citet{Rodriguez-Ardila02}. 27: \citet{Riffel06}. 28: \citet{Landt08}. 29:   \citet{Davies05}. 30: \citet{Hawarden99}. 31: \citet{Goldader97b}. 32: \citet{Doyon89}. 33: \citet{Hawarden95}}
\tablenotetext{d}{Distances from NED, morphologies and inclinations from Hyperleda \citep{Paturel13}. NGC~7469: AGN type from \citet{Osterbrock93}, X-ray luminosity from \citet{Pereira-Santaella11}. 1H-1934-063: AGN type from \citet{Rodriguez-Ardila00}, X-ray luminosity from \citet{Panessa11}. Black hole masses calculated as above, using $\sigma_*$ from \citet{Nelson04} and \citet{RARiffel13}.}
\end{deluxetable}

\section{Observations and Data Reduction}
\label{obsdr}

The spectra were obtained using the cross-dispersed (XD) mode of the Gemini Near-IR Spectrograph \citep[GNIRS; ][]{Elias06} on the 8.1 m Gemini North telescope. With the ``short blue" camera, 32 l/mm grating and 0.3\arcsec\ slit, this mode gives simultaneous spectral coverage from $\sim 0.83 - 2.5 \; \mu$m at R$\sim$1300/1800\footnote{Work on the GNIRS grating turret in mid-2012 fixed a problem with the mechanism at the expense of degrading the resolution achieved with the 32 l/mm grating and 0\farcs3 slit. The spectral resolution of data obtained in that specific configuration before November 2012 was R$\sim$1700, but has been R$\sim$1300 since then.} with a pixel scale of 0.15\arcsec/pix. To minimize the effects of differential atmospheric refraction, which can be important over this wide wavelength range especially at low elevation, the slit was usually orientated close to the mean parallactic angle during the observations of both the science target and standard star. Exceptions are noted in Table \ref{table2}. Because the galaxies are extended objects, and the slit used in this XD mode is only 7.5\arcsec\ in length, the telescope was nodded to blank sky (typically 50\arcsec\ distant) in an ABBA-type pattern. Individual and total exposure times varied depending on the object's brightness and the likely observing conditions (see below), and are given in Table \ref{table2}. 

The data were acquired in queue mode between May 2011 and August 2014. Some observations were taken from standard queue programs, while others came from dedicated poor weather programs, designed to be used in cloudy or poor seeing conditions. The requested (and actual) observing conditions for the objects in this study therefore cover a large range. Some data sets from these programs have been omitted from this paper. Most of these are spectra that were obtained under very poor conditions, resulting in data with very low S/N and not useful for the planned scientific analysis. Less commonly, telluric OH lines did not subtract out well, leaving large residuals in the data, particularly in the H band. 

The data were reduced using version 1.9 of the ``XDGNIRS'' pipeline. The code, created for this project and based mainly on the Gemini IRAF package, produces a 1D, roughly flux-calibrated spectrum from a set of raw science and calibration files. First, the data are cleaned of any electronic striping. ``Radiation events'', cosmic ray-like features caused by $\alpha$ particles from the coatings of the camera lenses, are identified by comparing each 2D image to a relatively clean minimum image formed from all data taken at that nod position, and interpolated over.  The files are then divided by a flat field constructed from observations of quartz halogen and infrared lamps taken immediately after the galaxy spectroscopy. Following sky subtraction, the 2D data are rectified using daytime pinhole flats. GNIRS' detector is slightly tilted such that sky and arc lines do not fall exactly along detector rows, and this tilt is removed using observations of an argon arc lamp taken along with the science data at night. The arc spectra are also used to wavelength calibrate the data at this stage, and a spectrum of each spectral order is then extracted. For this work, an aperture of 1.8\arcsec\ along the slit was used, corresponding to about twice the typical seeing in the $K$-band (Table \ref{table2}). Additional spectra were extracted in steps along the slit, but we do not discuss them in this paper.

To cancel out telluric absorption lines in the data, a standard star, usually of spectral type A0 - A5, was observed immediately before or after each galaxy. Prior to using a star to correct the galaxy spectrum, its intrinsic absorption lines are removed either by using a model spectrum of Vega\footnote{http://kurucz.harvard.edu/stars.html} or by automatically fitting Lorentz profiles to the lines. XDGNIRS allows the user to interactively improve the results from the chosen automatic algorithm, and this option was used in most cases. This process can leave residuals, and we found this particularly common around the wavelengths of Pa$\beta$ (1.282 $\mu$m) and Br10 (1.737 $\mu$m). The detection and interpretation of weak H recombination lines in the spectra should be treated with some caution.

IRAF's ``telluric'' task is then used to shift and scale the standard star spectrum, thereby optimizing the telluric line removal. NIR telluric absorption is caused by various species; for example, the strong absorption around 1.9 $\mu$m is largely due to H$_2$O, whereas the bands between $\sim$2.0 - 2.1 $\mu$m are produced by CO$_2$. The best cancellation of one set of bands does not necessarily result in good cancellation of others, and a subjective judgment was made as to the most useful result in these cases.

For the purposes of flux calibration the standard stars are assumed to have the continuum spectrum of a blackbody of the appropriate temperature \citep{Pecaut13}. An absolute flux scale is applied by simply scaling the blackbody spectrum to the 2MASS Ks-band magnitude of the standard star \citep{Skrutskie06}. To estimate the accuracy of the flux calibration we identified several nights on which more than one standard star was observed for our program, and used one star on each night to flux calibrate the other(s). On clear nights the resulting magnitudes differed by 5--40\% from 2MASS, with errors likely dominated by seeing variations in the 0\farcs3 slit.

The individual spectral orders generally match well in flux, with the exception of fairly frequent small (few per cent) offsets between orders 7 and 8. XDGNIRS allows interactive scaling of the orders to remove any inter-order offsets, and also editing of the range of pixels from each order that is used to create the final, combined spectrum. These operations were carried out in a few cases. On rare occasions an obvious, spurious artefact remained in the final spectrum, usually resulting from imperfect removal of a radiation event. These were interpolated over at the end of the reduction.

The raw and reduced data are available through the Canadian Advanced Network for Astronomical Research (CANFAR), at http://www.canfar.phys.uvic.ca/vosui/\#/karun/xdgnirs\_Dec2014. An account with the  Canadian Astronomy Data Center is needed to access the storage area, and can be obtained from http://www3.cadc-ccda.hia-iha.nrc-cnrc.gc.ca/en/auth/request.html. The data files are accompanied by ``readme'' files that contain the commands used to reduce the data. For readers who may be interested in, for example, re-extracting spectra in different apertures, the latest version of the XDGNIRS code and manual can be found on the Gemini data reduction forum: http://drforum.gemini.edu/forums/gemini-data-reduction/.

\clearpage

\begin{deluxetable}{cccccccc}
\tabletypesize{\scriptsize}
\tablecaption{Observing log \label{table2}}
\tablewidth{0pt}
\tablehead{
\colhead{Galaxy} & \colhead{Program ID} & \colhead{Date\tablenotemark{a}} & \colhead{Texp (s)  $\times$ Nexp} & Seeing\tablenotemark{b} & \colhead{Clear?\tablenotemark{c}} & \colhead{Slit angle} & \colhead{Aperture\tablenotemark{d}} \\
 & & & & \arcsec & & \deg E of N & pc}
\startdata
\cutinhead{Seyfert 1.2--1.9}
NGC 4235 & GN-2013A-Q-16 & 20130202 & 240 $\times$ 4 & 0.76 & Y & 293 & 51 $\times$ 306 \\
NGC 3031 & GN-2012A-Q-23 & 20120305 & 120 $\times$ 6 & 0.92 & Y & 194 & 2 $\times$ 12 \\
NGC 5033 & GN-2012A-Q-23 & 20120301 & 120 $\times$ 6 & 0.62 & N & 130 & 27 $\times$ 163 \\
NGC 4395 & GN-2012B-Q-80 & 20121216 & 240 $\times$ 4 & 0.84 & Y & 267 & 5 $\times$ 31 \\
NGC 2639 & GN-2012A-Q-23 & 20120130 & 120 $\times$ 6 & 1.00 & Y & 133\tablenotemark{e} & 62 $\times$ 372\\
NGC 4258 & GN-2012A-Q-23 & 20120531 & 120 $\times$ 6 & 0.77 & N & 130 & 10 $\times$ 59 \\
NGC 4388 & GN-2013A-Q-16 & 20130204 & 240 $\times$ 4 & 0.58 & Y & 64 & 24 $\times$ 147\\
NGC 4565 & GN-2013A-Q-16 & 20130205 & 240 $\times$ 4 & 0.71 & Y & 277 & 14 $\times$ 85 \\
NGC 4579 & GN-2012A-Q-23 & 20120604 & 120 $\times$ 3 & 0.85 & N & 140\tablenotemark{e} & 24 $\times$ 147\\
\cutinhead{Seyfert 2}
NGC 1167 & GN-2011B-Q-111 & 20111204 & 120 $\times$ 6 & 0.46 & N & 97 & 95 $\times$ 570\\ 
NGC 1358 & GN-2011B-Q-111 & 20111205 & 120 $\times$ 6 & 0.69 & N & 90 & 78 $\times$ 468\\
NGC 2273 & GN-2012B-Q-112 & 20121204 & 240 $\times$ 4 & 0.66 & Y & 169 & 41 $\times$ 248\\
NGC 2655 & GN-2013A-Q-16 & 20130212 & 240 $\times$ 4 & 0.86 & Y & 177 & 36 $\times$ 213\\
NGC 3079 & GN-2012A-Q-23 & 20120123 & 120 $\times$ 6 & 0.67 & Y & 180 & 30 $\times$ 178\\
NGC 3147 & GN-2012B-Q-80 & 20121215 & 240 $\times$ 4 & 1.01 & Y & 207 & 60 $\times$ 357\\
NGC 4725 & GN-2012A-Q-120 & 20120530 & 300 $\times$ 6 & 1.15 & N & 100 & 18 $\times$ 108\\
NGC 5194 & GN-2012A-Q-23 & 20120531 & 120 $\times$ 6 & 0.55 & N & 190 & 11 $\times$ 67\\
NGC 7743 & GN-2012A-Q-120 & 20120607 & 300 $\times$ 4 & 0.82 & N & 290 & 36 $\times$ 213\\
\cutinhead{LINER 1.9}
NGC 266 & GN-2012B-Q-112 & 20121113 & 240 $\times$ 4 & 0.65 & Y & 210 & 91 $\times$ 545\\
NGC 315 & GN-2012B-Q-112 & 20121115 & 240 $\times$ 4 & 0.69 & N & 0 & 96 $\times$ 574\\
NGC 1052 & GN-2011B-Q-111 & 20111207 & 120 $\times$ 6 & 0.61 & N & 90\tablenotemark{e} & 26 $\times$ 155\\
NGC 3718 & GN-2013A-Q-16 & 20130202 & 240 $\times$ 4 & 0.83 & Y & 241 & 25 $\times$ 148\\
NGC 3998 & GN-2012A-Q-23 & 20130101 & 120 $\times$ 6 & 0.78 & N & 0 & 31 $\times$ 189\\
NGC 4203 & GN-2012A-Q-23 & 20130102 & 120 $\times$ 6 & 0.51 & N & 17 & 14 $\times$ 85\\
NGC 4450 & GN-2012A-Q-23 & 20120501 & 120 $\times$ 6 & 0.96 & Y & 79 & 24 $\times$ 147\\
NGC 4750 & GN-2013A-Q-16 & 20130215 & 240 $\times$ 4 & 0.81 & Y & 227\tablenotemark{e} & 38 $\times$ 228\\
NGC 5005 & GN-2013A-Q-16 & 20130215 & 240 $\times$ 4 & 0.75 & Y & 260 & 31 $\times$ 186\\
\cutinhead{LINER 2}
NGC 404 & GN-2011B-Q-111 & 20111031 & 120 $\times$ 6 & 0.55 & N & 125 & 4 $\times$ 21\\
NGC 474 & GN-2012B-Q-112 & 20121109 & 300 $\times$ 2 & 0.88 & N & 60 & 47 $\times$ 284\\
NGC 1961 & GN-2012B-Q-112 & 20121204 & 240 $\times$ 4 & 0.66 & Y & 166 & 77 $\times$ 463 \\
NGC 2768 & GN-2012B-Q-80 & 20121115 & 240 $\times$ 4 & 0.77 & Y & 240 & 35 $\times$ 207\\
NGC 2832 & GN-2013A-Q-16 & 20130212 & 240 $\times$ 4 & 0.90 & Y & 56 & 133 $\times$ 799\\
NGC 3169 & GN-2012A-Q-120 & 20120418 & 300 $\times$ 6 & 1.22 & N & 63 & 29 $\times$ 172\\
NGC 3190 & GN-2011B-Q-111 & 20111127 & 120 $\times$ 6 & 0.89 & N & 275 & 37 $\times$ 196\\
NGC 3607 & GN-2012A-Q-120 & 20120417 & 300 $\times$ 6 & 1.51 & Y & 81 & 29 $\times$ 174 \\
NGC 4346 & GN-2013A-Q-16 & 20130202 & 240 $\times$ 4 & 0.66 & Y & 226 & 25 $\times$ 148\\
NGC 4548 & GN-2013A-Q-16 & 20130204 & 240 $\times$ 4 & 0.59 & Y & 74 & 24 $\times$ 147\\
NGC 4594 & GN-2013A-Q-16 & 20130202 & 240 $\times$ 4 & 0.62 & Y & 31 & 29 $\times$ 175\\
NGC 4736 & GN-2011A-Q-126 & 20110516 & 300 $\times$ 2 & 0.48 & N & 357\tablenotemark{e} & 6 $\times$ 38 \\
NGC 5371 & GN-2013A-Q-16 & 20130215 & 240 $\times$ 4 & 0.68 & Y & 247 & 55 $\times$ 330 \\
NGC 5850 & GN-2013A-Q-16 & 20130209 & 240 $\times$ 4 & 0.93 & Y & 150& 42 $\times$ 249\\
NGC 6500 & GN-2012A-Q-120 & 20120530 & 300 $\times$ 6 & 0.94 & N & 280 & 58 $\times$ 346\\
NGC 7217\tablenotemark{f} & GN-2013A-Q-16 & 20130505/22 & 240 $\times$ 4 & 0.96, 0.70 & N, Y & 266, 254& 23 $\times$ 140\\
\cutinhead{Transition Object, Inactive}
NGC 410 & GN-2011B-Q-111 & 20111031 & 120 $\times$ 6 & 0.40 & N & 110 & 103 $\times$ 616\\
NGC 660 & GN-2011B-Q-111 & 20111101 & 126 $\times$ 6 & 0.40 & N & 75 & 17 $\times$ 103\\
NGC 4569 & GN-2012B-Q-80 & 20121217 & 240 $\times$ 4 & 0.65 & N & 285 & 24 $\times$ 147\\
NGC 7331 & GN-2011B-Q-111 & 20111123 & 120 $\times$ 6 & 0.46 & N & 137 & 21 $\times$ 125\\
NGC 205 & GN-2012B-Q-80 & 20121202 & 240 $\times$ 4 & 0.66 & N & 227 & 1 $\times$ 6\\
\cutinhead{Non-Palomar Galaxies}
1H1934-063 & GN-2013A-Q-120 & 20130711 & 120 $\times$ 4 & 0.81 & Y & 310 & 76 $\times$ 454 \\
NGC 7469 & GN-2013A-Q-120 & 20130807 & 120 $\times$ 4 & 0.76 & N & 299 & 62 $\times$ 373 
\enddata
\tablenotetext{a}{Data obtained before 20121101 have R$\sim$1700, otherwise R$\sim$1300; see \S\ref{obsdr} for details.}
\tablenotetext{b}{Full width at half maximum (FWHM) of standard star in the $K$-band. For NGC\,4395 and NGC\,7743, FWHM$\rm_{star} >$ FWHM$\rm_{galaxy}$, so the galaxy FWHM is given instead.}
\tablenotetext{c}{Whether or not sky was clear at the time of observation, as judged by the observing log and the Mauna Kea All Sky Infrared and Visible Sky Monitor (http://www.cfht.hawaii.edu/{\raise.17ex\hbox{$\scriptstyle\sim$}}asiva/)}
\tablenotetext{d}{Slit width $\times$ extraction aperture}
\tablenotetext{e}{Galaxy and/or standard star observed with slit $>$15\deg\  from parallactic angle and at airmass $>$1.2. See e.g. http://www.gemini.edu/sciops/instruments/gnirs/spectroscopy/observing-strategies\#refraction for a table giving the magnitude of differential atmospheric refraction effects as a function of wavelength and airmass.}
\tablenotetext{f}{Spectrum is mean of 2 epochs}
\end{deluxetable}

\pagebreak
\clearpage

\section{Results}
\label{results}

In this section we show some illustrative spectra and discuss the overall features of the data set. The individual galaxies and their spectra are discussed in more depth in Appendix \ref{A1}. 

\subsection{General Characteristics}
\label{cont}

\begin{figure*}
\hspace*{-15mm}
\includegraphics[scale=0.95]{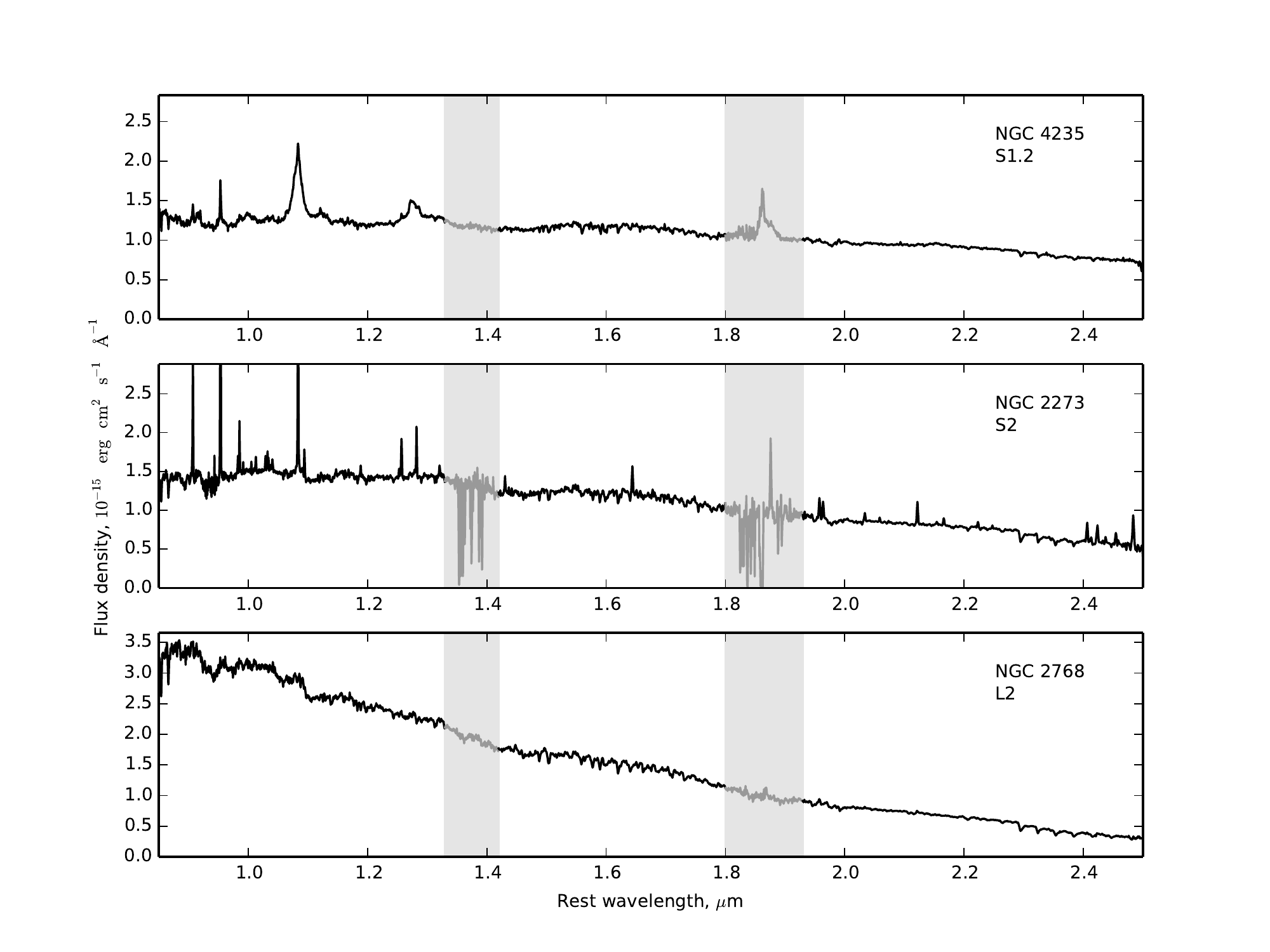}
\caption{ {\small Example spectra. NGC~4235, a type 1.2 Seyfert, shows several broad emission lines while NGC~2273 (type 2 Seyfert) exhibits numerous narrow lines. In contrast, NGC~2768, a LINER 2, has no detectable emission lines and a much bluer continuum. Many absorption features arising in the galaxies' stellar populations are visible. Regions of poor atmospheric transmission ($\lesssim$20\% for 3 mm H$_2$O at zenith) are indicated in gray. See Figure \ref{emlines} and Appendix \ref{A1} for emission line identifications and Fig. \ref{abslines_fig} for absorption lines.}}
\label{examples}
\end{figure*}

Three example spectra are presented in Figure \ref{examples}. NGC~2768 is typical of the LINER 2 and transition object classes: with few exceptions, these objects exhibit a blue continuum, many stellar absorption lines, and few or no detectable emission lines. The line emission in the LINER 1.9 and Seyfert 2 classes is more prominent, but on the whole they also tend to contain few, weak emission lines. With strong lines from both low- and high-excitation species, NGC~2273 (S2) is a notable exception. The Seyfert 1.2 - 1.9 class is the most varied in this sample. It contains objects with few, weak emission lines (NGC~4258, NGC~5033); with numerous strong, relatively narrow lines (NGC~4395, NGC~4388); and one object (NGC~4235) with very broad, asymmetric features.

With some exceptions, the continuum emission of the galaxies generally slopes down towards longer wavelengths (in F$_{\lambda}$). This overall spectral shape is characteristic of stellar emission from the host galaxy of the weak AGN. The red tail of the optical/UV accretion disk emission, and hot dust emission in the K band, often visible in higher-luminosity AGN \citep[e.g.][]{Riffel09,Landt11} are not directly detected in most of these objects. This is likely an effect of both AGN luminosity and angular resolution; for instance, previous work has shown that hot dust is detected in adaptive optics-assisted spectroscopy of Mrk~1066, but not in seeing-limited data \citep{RARiffel10}. 
A minority of the spectra exhibit a flattening or turnover at short wavelengths. This is most pronounced in NGC~1961, NGC~3079, NGC~3718, and NGC~4388. These are all spiral galaxies with dust lanes crossing the line of sight (see Notes in Appendix \ref{A1}), suggesting that this spectral shape is due to extinction. 

A broad, weak ``bump''  is frequently visible in the $H$ band superimposed on the general stellar continuum and extending from about 1.45 -- 1.75 $\mu$m. This likely results from a minimum in the opacity of  H$^-$ in cool stars in the galaxies \citep{Rayner09}, whose absorption features are discussed below.  As can be seen in Figure \ref{examples}, the quality of the data in the regions around 1.35 $\mu$m and 1.9 $\mu$m, between the $J$/$H$ and $H$/$K$ bands, varies widely. This depends primarily on the telluric water vapor column at the time of the observations and is investigated in Appendix \ref{A2}.

\subsection{Absorption Lines}
\label{abs}

\begin{figure*}
\hspace*{-15mm}
\includegraphics[scale=0.95]{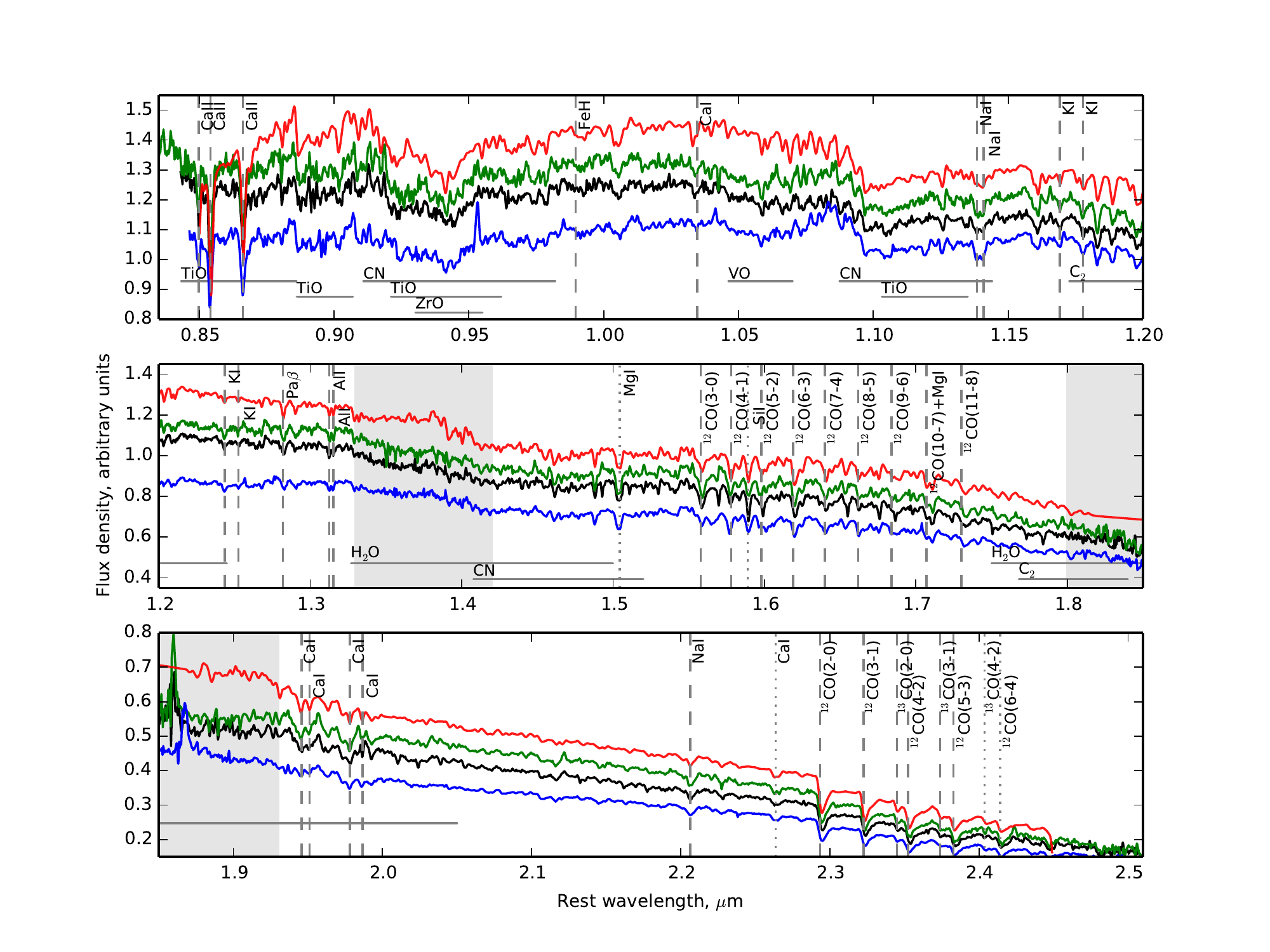}
\caption{ {\small From bottom to top in each panel: Spectra of NGC~4565 (blue), NGC~5371 (black) and NGC~5850 (green), and a fit of empirical stellar spectra to NGC~5850 (red), illustrating various stellar absorption lines. The wavelengths of IMF-sensitive features (see text) are indicated by dashed vertical lines, and those of molecular bands (other than FeH and CO) by solid horizontal lines. Other lines of potential interest are shown by dotted vertical lines. The spectra have been scaled by separate multiplicative factors in each panel for clarity. Regions of poor atmospheric transmission, as defined in Figure \ref{examples}, are indicated in gray.}}
\label{abslines_fig}
\end{figure*}

Numerous stellar absorption lines are present in the spectra. In particular, the optically-classified type 2 LINERs often exhibit only few, very weak  emission lines in the NIR, clearly revealing the absorption line spectrum of the nuclear stellar population. Figure \ref{abslines_fig} compares three spiral galaxies (NGC~4565, NGC~5371 and NGC~5850) with high S/N, good telluric line removal, and little detectable line emission, illustrating the absorption features in the spectra. The detailed structure of all three spectra is remarkably similar, revealing weak bands common to all the galaxies. Systematic effects are unlikely to be responsible for the presence of these features at the same rest-frame wavelength in all the spectra. The redshifts of the nearest and most distant of the three objects, z=0.0041 (NGC~4565) and z=0.0085 (NGC~5371), mean that the shifts necessary to move the spectra to the rest frame differ by about 0.005 $\mu$m. This is approximately the width of the Na I 1.14 $\mu$m absorption, and should be sufficient to ensure that artifacts from telluric line removal, or pixel-dependent effects (e.g. flatfielding errors), do not affect all three spectra in the same way. In support of this interpretation, Figure \ref{abslines_fig} also shows a fit of empirical stellar spectra to the spectrum of NGC~5850. The fit was carried out with the STARLIGHT spectral synthesis software \citep{CidFernandes04, CidFernandes05a,CidFernandes05b,Asari07},
using the empirical IRTF NIR stellar spectral library \citep{Rayner09} as a base set\footnote{The base set also included hotter stars than found in the IRTF library (Coelho 2014, private communication), a $F_{\nu} \propto \nu^{-0.5}$ power-law to represent a possible contribution from the AGN featureless continuum, and a set of blackbody spectra to represent hot dust. The aim of the modeling is not to derive information about the stellar population, but simply to investigate which of the spectral features may be due to starlight.}. The spectrum is very well reproduced by this model fit, implying that the numerous, weak features are real and arise in the stellar populations of the galaxies.

These absorption features have a number of practical applications. For example, in a recent study, \citet{Conroy12} identified several bands in the NIR spectral region that are sensitive to stellar surface gravity and can therefore aid in determining the IMF of a galaxy.  The locations of the bands identified in their Figure 10 are marked in Figure \ref{abslines_fig}. Most of these lines -- from CaII, CaI, NaI, KI, Pa$\beta$, AlI, and CO -- are not difficult to detect in large-telescope observations of nearby galaxy nuclei; with the exception of the weak, broad 0.99 $\mu$m FeH line, they are all clearly visible in the spectra in Figure \ref{abslines_fig}. While features at $\lambda \sim$0.8 -- 1 $\mu$m have been widely used to investigate the IMF in various galaxy samples \citep[e.g.][]{Cenarro03,vanDokkum10,Smith12}, to the best of our knowledge the longer-wavelength features have not yet been used in this way.  Stellar population synthesis modeling of NIR galaxy spectra like these, together with optical spectroscopy to break the degeneracy between the effects of IMF variations and elemental abundances, may be a powerful way of measuring the IMF of local objects\footnote{The galaxies in this sample host weak active nuclei. In most cases the AGN accretion disk and dust are not bright enough relative to the stellar emission to leave a noticeable imprint on the spectrum (\S\ref{cont}). However, given the small magnitude of the effect of IMF variations on the features, it is possible that weak AGN emission could cause systematic errors in the measurement of their strengths. Fitting the spectra with stellar libraries should reveal whether AGN emission must be present in the nuclear aperture, and extracting extranuclear spectra would entirely avoid the possibility of AGN contamination. We note that the sample of \citet{vanDokkum10}, who analysed features at 0.8 -- 1.0 $\mu$m, contained some Virgo cluster LINERs. Their results did not depend on aperture size, suggesting that AGN contamination was negligible in their sample. }. 

Some other absorption features of potential interest are also indicated in Figure \ref{abslines_fig}. The numerous, relatively strong and evenly-spaced features in the $H$ and $K$ bands are CO ro-vibrational bands \citep[sometimes blended with bands of other species;][]{Origlia93}.  Velocity dispersions derived from the CO (and CaT) bands are presented in a companion paper \citep{Riffel15}. \citet{Origlia93} characterize the dependence of the equivalent widths of the CO(6,3) (1.62 $\mu$m) and CO(2,0) (2.29 $\mu$m) bands, as well as the Si (1.59 $\mu$m) band, on stellar spectral type, and also demonstrate that the $H$ band CO lines are much less prone to dilution by AGN-heated dust emission than the $K$ band lines. Based on that work, the Si (1.59 $\mu$m)/CO (1.62 $\mu$m) equivalent width ratio has been used to derive the average stellar spectral type in Seyfert and inactive spiral galaxies \citep{RamosAlmeida09b,Kotilainen12}. \citet{Ivanov04} also suggest the use of the Mg 1.50 $\mu$m band as a stellar effective temperature indicator.

As well as the features discussed so far, many broad, shallow molecular absorption features are also apparent in Figure \ref{abslines_fig}. The bands arise in the atmospheres of cool stars --  M dwarfs, giants, and the TP-AGB stars that are of concern to stellar population synthesis models (\S\ref{intro}). TP-AGB stars are variable, mass-losing, stars  \citep[oxygen-rich, late-M giants; s-process enhanced S-stars; and the ``N'' variety of carbon star;][]{Habing03}  in which a thermally-unstable He-burning shell causes repeated thermal pulses. The locations of molecular bands (other than CO) that are prominent in spectra of late-M giants, S stars, and C-N stars, as shown by \citet{Rayner09}, are indicated by horizontal lines\footnote{The short-wavelength ends of the bars are located at the band heads quoted by \citet{Rayner09}. Their lengths illustrate the approximate extent of the absorption features, subjectively judged from figures 34 and 39 of  \citet{Rayner09}.} in Figure \ref{abslines_fig}. These bands are generally weak, broad, and often overlapping in wavelength, and stars other than TP-AGB stars may contribute to them. For instance, the noticeable absorption around 0.94~$\mu$m in our spectra may be a blend of ZrO (M giants, S stars), CN (C-N stars, supergiants), and TiO (late-M giants and dwarfs). Similarly, the broad, shallow dip between $\sim$ 1.32 -- 1.5 $\mu$m is likely caused by both H$_2$O and CN. The relatively isolated VO feature at $\lambda \sim$1.05 $\mu$m is particularly strong in late-type M giants \citep{Rayner09}.

\begin{figure*}
\hspace*{-15mm}
\includegraphics[scale=0.95]{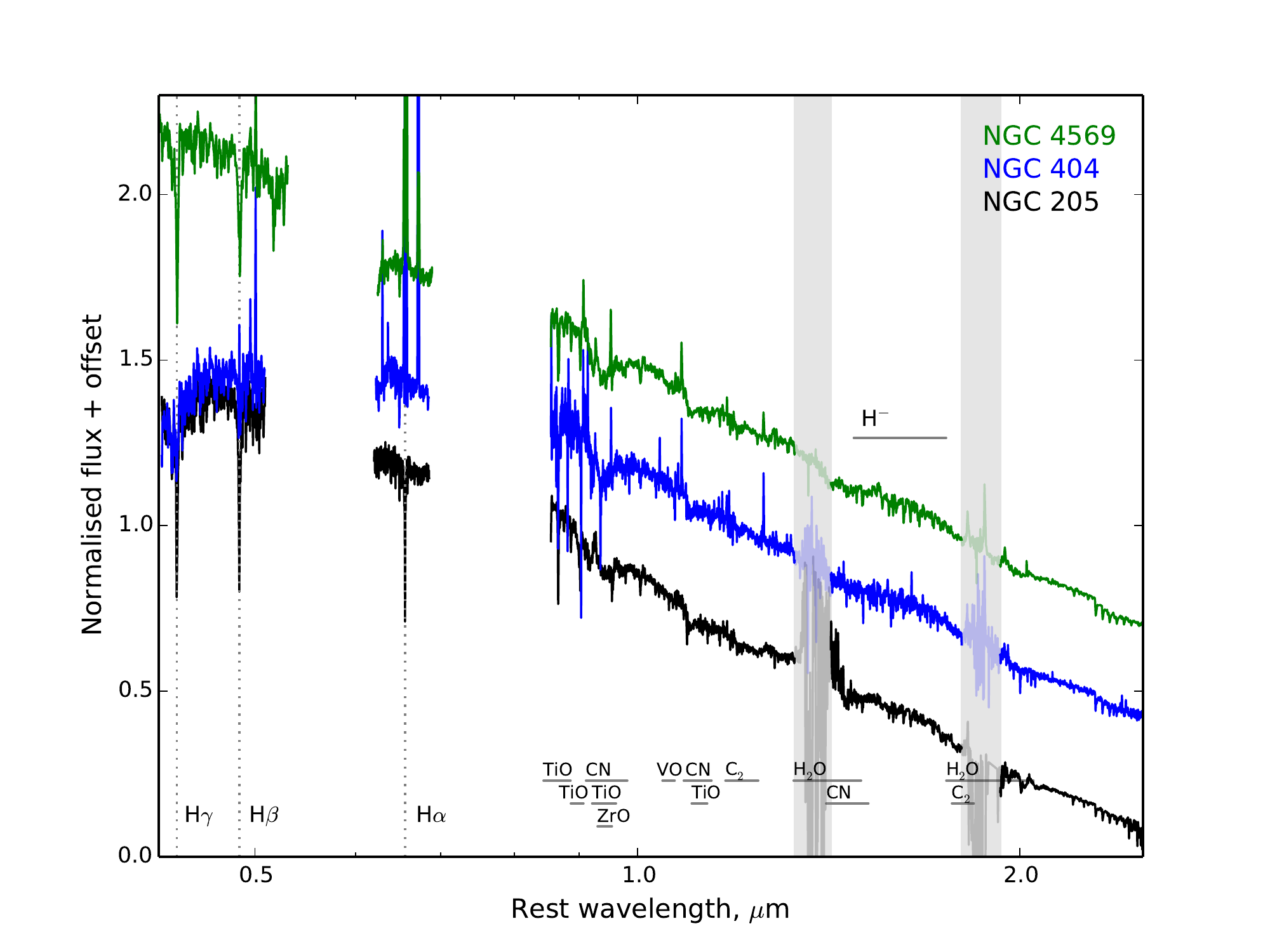} 
\caption{ {\small Optical -- NIR spectra of ``intermediate-age'' galaxies (see text). The same molecular bands as shown in Figure \ref{abslines_fig} are indicated, as well as the ``$H$-band bump'' (stellar H$^-$ opacity minimum). Optical Balmer absorption lines are also indicated. Note that the wavelength coverage of the optical data \citep{Ho95}  does not include the strong higher-order Balmer lines that are present in the \citet{CidFernandes04} spectra of these nuclei.  The NIR spectra were normalised around 0.9 $\mu$m and offset for clarity. As the optical spectra were taken through a wider slit, they have been scaled in flux to match the extrapolation of  a low-order polynomial fit to the IR spectra. Regions of poor atmospheric transmission, as defined in Figure \ref{examples}, are shown in gray. }}
\label{Igals}
\end{figure*}

\begin{figure*}
\hspace*{-15mm}
\includegraphics[scale=0.95]{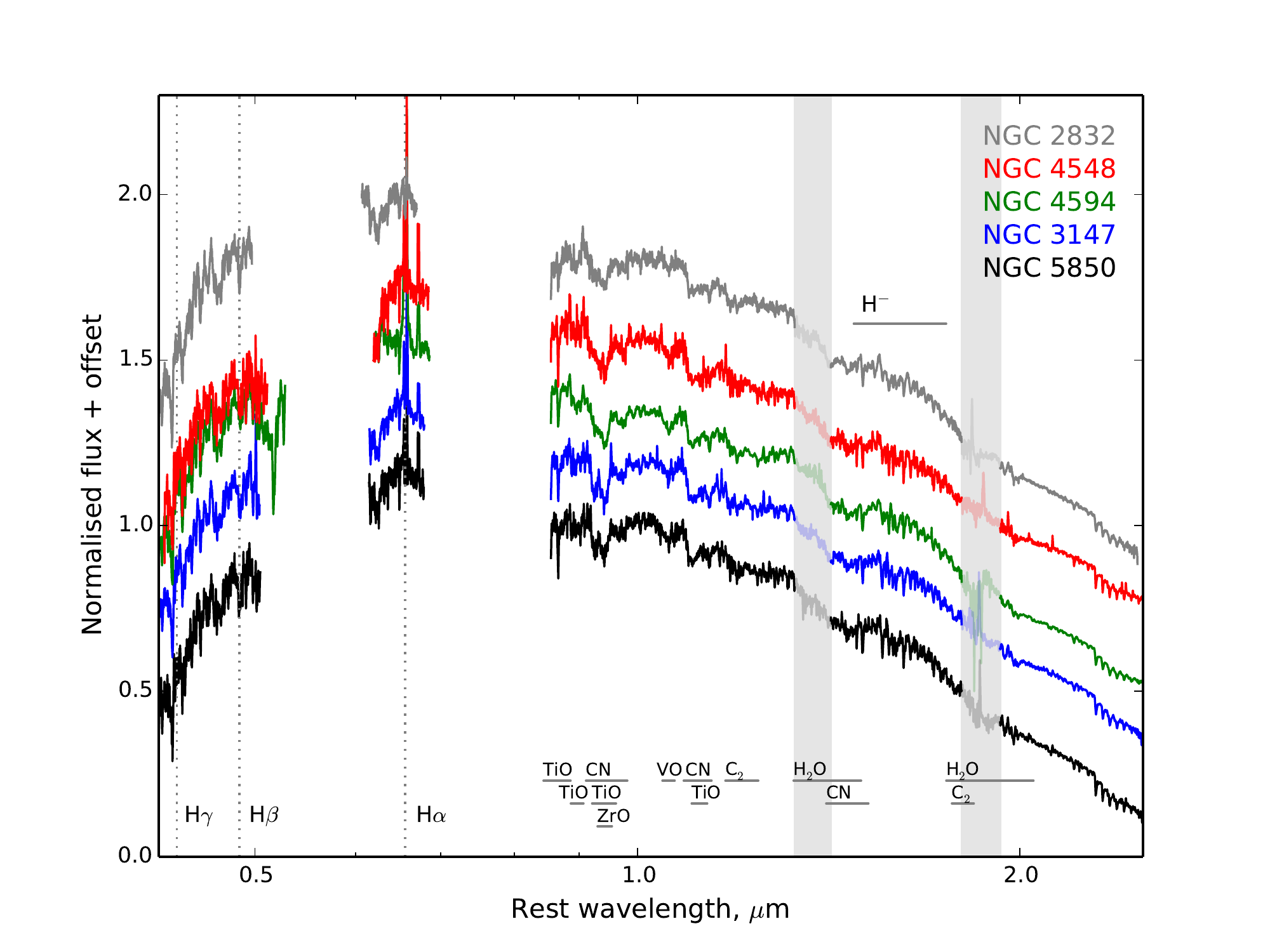} 
\caption{ {\small As for Fig. \ref{Igals}, for ``old'' galaxies. }}
\label{Ogals}
\end{figure*}

A full determination of the dependence of these molecular features on the stellar populations of these galaxies requires accurate stellar population synthesis modelling, and is beyond the scope of this work. 
However, as a preliminary exploration, in Figures \ref{Igals} and \ref{Ogals} we compare the IR and optical spectra of galaxies identified as ``intermediate-age'' and ``old''.

The intermediate-age category consists of three objects that are well-known for their strong Balmer absorption lines, indicating star formation occurring within the last 1.5 Gyr. These galaxies are NGC~205, NGC~404, and NGC~4569 \citep{Keel96,Ho03,Bouchard10,Seth10}. Roughly 70\% of the optical light of NGC~205 is emitted by stars with ages $\sim$0.1 -- 1 Gyr \citep{Bica90,GonzalezDelgado99}, and this galaxy is used by \citet{CidFernandes04} as the intermediate-age template object in their optical stellar population synthesis work on low-luminosity AGN.  They find that NGC~404 and NGC~4569 are the galaxies in their sample with the most significant intermediate-age populations, with upwards of 70\% of the 4020\AA\ emission accounted for by the NGC~205 template. 

The old galaxies are those with ages $\gtrsim$5 Gyr, based on published analysis of their optical spectra. The central optical spectra and/or line indices of these objects are best described by stellar populations with ages of $\sim$5 -- 11 Gyr (see references in Appendix \ref{A1}). Many of the galaxies in this sample fit this criterion, so for clarity of presentation we restrict that category in Figure \ref{Ogals} to those with the cleanest telluric line removal (defined as S/N$>$20 at 1.32 - 1.37 $\mu$m). These are NGC~2768, NGC~2832, NGC~3147, NGC~4548, NGC~4594, and NGC~5850. 

A qualitative comparison of Figures \ref{Igals} and \ref{Ogals} suggests some systematic differences between the NIR spectra of intermediate-age and old galaxies. First, the NIR continuum shape of the old galaxies differs from that of the younger galaxies, becoming noticeably flatter at short wavelengths. This is not due to extinction; dust is not detected in HST optical images of NGC~2832, for example \citep{Martel04,Laine03}, and a variable nuclear UV source is detected in NGC~4594 \citep[despite the galaxy being highly inclined; ][]{Maoz05}. Neither is AGN activity likely to be responsible for altering the shape of the spectra. NGC~205 and NGC~2832 have very different ages and continuum shapes, yet neither shows evidence for AGN activity (beyond the presence of a few, very weak emission lines in NGC~2832; Appendix A). The different spectral shape of the intermediate-age galaxies therefore appears to be due to their younger, bluer stellar populations relative to those of the old galaxies.

The 1.6~$\mu$m ``bump'', which is strong in the atmospheres of K- and M-type giants and supergiants \citep{Rayner09}, is also more pronounced in the old galaxies. Additionally, the broad, molecular absorption features are generally stronger in the old objects. This is noteworthy given that the stellar populations of the intermediate-age galaxies should contain large numbers of the TP-AGB stars discussed above and in \S\ref{intro}. Depending on their age, metallicity, and pulsation cycles, these stars have deep features from molecules such as CN, C$_2$, H$_2$O, and VO \citep{Lancon02}. This is reflected by the evolutionary population synthesis models of \citet{Maraston05}, which predict strong molecular features in stellar populations with ages $\sim$1 Gyr. However, the 1.1 $\mu$m CN band in the galaxies in Figure \ref{Igals} is weak compared to that in the old galaxies, as are the 1.05 $\mu$m VO band and the feature at $\sim$0.93 $\mu$m (likely a blend of CN, TiO, and ZrO). Gauging the strength of the $\sim$1.4 $\mu$m H$_2$O band is difficult because of the likely blending with CN absorption, the adjoining H$^-$ bump, and telluric residuals in NGC~205 and NGC~404, but any minimum around this wavelength appears no stronger in the intermediate-age galaxies than in the old ones. 

This simple comparison, although based on just a few objects, suggests that models that predict a relatively featureless NIR spectrum will provide a better description of galaxies with significant intermediate-age populations. One caveat is that the intermediate-age galaxies all have relatively low stellar velocity dispersions (20 $< \sigma_* <  \; 140 \rm \; km \; s^{-1}$), while with the exception of NGC~4548 ($\sigma_*$ = 113 km s$^{-1}$),  the old galaxies are high-dispersion systems with 140 $< \sigma_* <  \; 340 \rm \; km \; s^{-1}$ \citep[]{Riffel15,Ho09a}. The well-known relations between $\sigma_*$ and metallicity \citep{Bender93}  and the abundances of carbon, nitrogen, and $\alpha$-elements \citep{Schiavon07} then suggest that the intermediate-age galaxies are metal-poor, while the old galaxies are metal-rich. The spectral differences between these sets of galaxies may therefore reflect differences in metallicity as well as in age.

This conclusion also relies on the optical spectra being a reasonably reliable indicator of the age of the stellar population that dominates in the NIR.  Differences in stellar populations derived from optical and NIR spectra have been observed in star-forming and Seyfert galaxies \citep{Riffel09,Martins13b}. These differences have been attributed to the inclusion of regions of star formation in the wider optical slit; the detection of dust-enshrouded stellar populations in the NIR; and the differing sensitivity of the optical and NIR regimes to stars of different ages \citep[e.g.][]{Riffel11c}. It has also been found that post-starburst spectra indicative of intermediate-age stars can arise in spatially isolated regions of a galaxy \citep{Sanmartim14}. We are therefore working to obtain aperture-matched optical data for in-depth population synthesis of the galaxies in this sample. Nonetheless, these observations provide an initial illustration of the NIR characteristics of intermediate-age and old stellar populations in galaxies.

\subsection{Emission Lines}
\label{em}

\begin{figure*}
\hspace*{-15mm}
\includegraphics[scale=0.95]{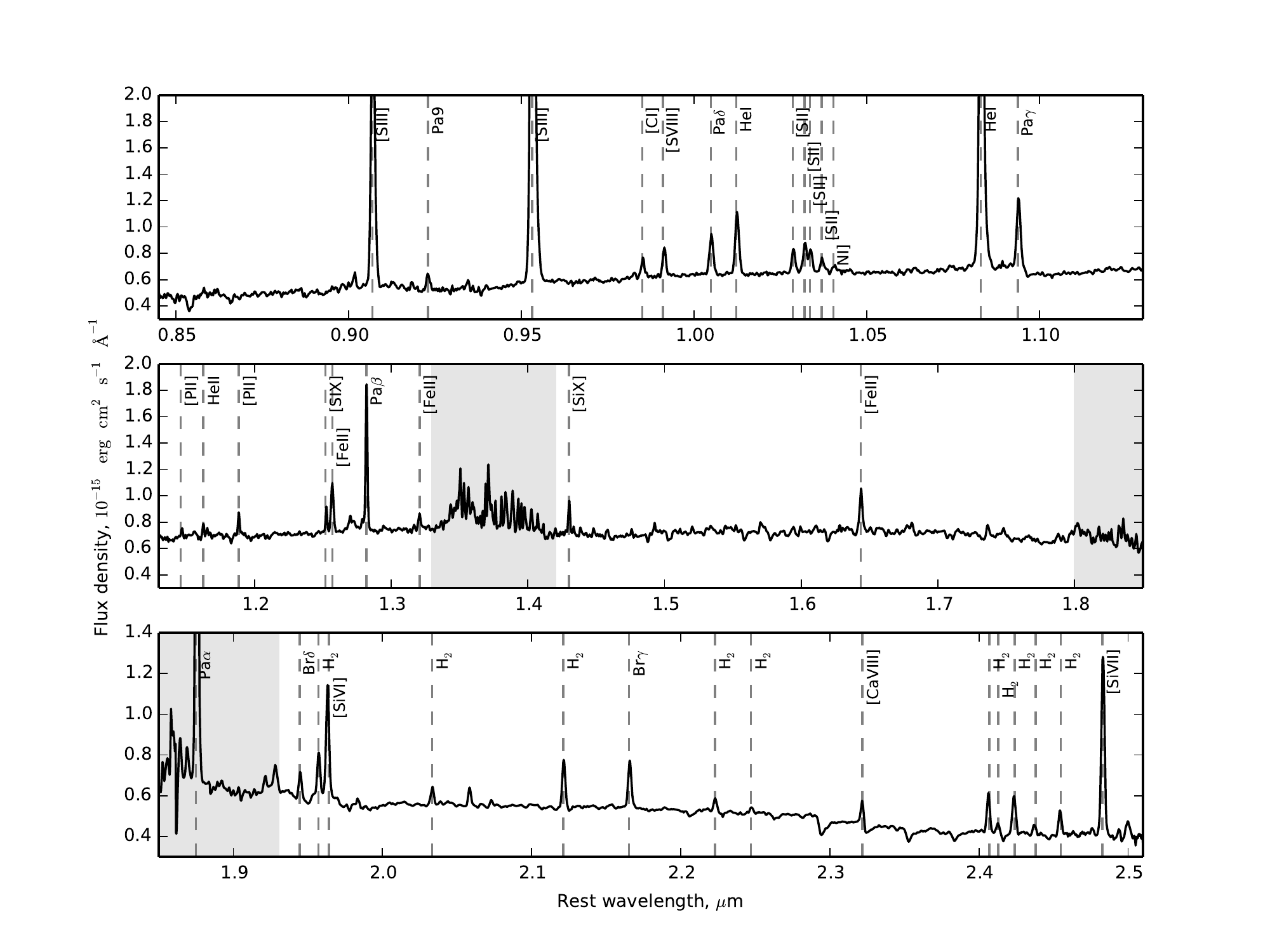}
\caption{ {\small Spectrum of NGC~4388 (Sy 1.9), indicating the emission lines in the data. Regions of poor atmospheric transmission, as defined in Figure \ref{examples}, are indicated in gray.}}
\label{emlines}
\end{figure*}

Emission lines are also present in many of the spectra, being particularly numerous and prominent 
in some of the Seyferts. In Figure \ref{emlines}, the Seyfert 1.9 NGC~4388 is used to illustrate and identify many of these lines. Detailed and quantitative analysis of the line emission is being presented in a separate paper (Gonz\'{a}lez-Mart\'{i}n et al., in prep), so here we make some general remarks about the detected lines and their potential utility.

Broadly speaking, the optical spectrum is a good predictor of the IR emission-line spectrum: objects with plentiful, strong lines in their optical spectrum have a rich IR line spectrum, and vice versa. The most commonly detected, and generally strongest, features are [S III]$\lambda\lambda$0.907,0.953; 
He I $\lambda$1.083; and [Fe II]$\lambda\lambda$1.257,1.644 (all wavelengths in $\mu$m). At least some of these lines are present in most of the Seyferts and type 1.9 LINERs alike, although they are not detected in several type 2 LINERs.  Lower-intensity lines are also present in some of the spectra. [C I]$\lambda$0.985, [S II]$\lambda$1.029,1.032  (the redder line being a blend of three [S II] transitions), and [N I]$\lambda$1.041 are fairly common, while [P II]$\lambda$1.147,1.188; He II$\lambda\lambda$1.012,1.163 , and [Fe II]$\lambda$ 1.321 are detected in a small fraction of the Seyferts. 

The most intense hydrogen recombination line in the near-infrared range, Pa$\alpha$, may be present in some spectra but its location in a region of very poor atmospheric transmission often makes its detection highly uncertain (see Appendix \ref{A2}). The next-most intense H recombination lines, Br$\gamma$ and Pa$\beta$, are also visible in a minority of the spectra, along with Pa$\gamma$ (and higher-order H recombination lines) in a few. When detected, these lines and the He I $\lambda$1.083  recombination line show broad pedestals in the spectra of most of the Seyfert 1 --1.9 (NGC~2639, 
NGC~3031, NGC~4235, NGC~4258, NGC~4395, NGC\,4579, NGC~5033, and NGC~7469) and LINER 1.9 (NGC\,315, NGC\,1052, NGC\,3998, NGC\,4203, NGC\,4450 and NGC\,4750) objects. In most other cases the emission lines are too faint to distinguish any broad components, if present.  On the other hand, almost none of the type 2 Seyferts or LINERs show broad components to the recombination lines. 

Some of these emission lines can be used as tracers of the dominant ionization mechanisms. For example,
strong [Fe II] emission is believed to be indicative of shock-excited gas, demonstrated by the fact that it occurs in the 
filaments of supernova remnants, in contrast to the weak [Fe II] emission characteristic of the photoionized
gas in H II regions \citep{Mouri90,Rodriguez-Ardila04}. Several mechanisms can contribute to the [Fe II] emission in galaxy nuclei: photoionization by extreme UV to soft X-ray radiation from the central source \citep{Simpson96}, shocks induced by interaction of the radio jets with the surrounding medium, and shocks produced by supernova 
remnants present in star-forming regions \citep{Forbes93,Alonso-Herrero00}. 

The [Fe II]$\lambda$1.257/Pa$\beta$ line ratio (or, equivalently [Fe II]$\lambda$1.644/Br$\gamma$) has been widely used 
to investigate the origin of the 
[Fe II] emission. These ratios increase from H II regions (photoionization by hot stars) 
to supernova remnants (shock excitation), with starbursts and active galaxies showing intermediate values
\citep{Alonso-Herrero97,Rodriguez-Ardila04}.
Galaxies exhibiting [Fe II]$\lambda$1.257/Pa$\beta <$  0.6 are usually classified as 
starbursts, those with [Fe II]$\lambda$1.257/Pa$\beta >$  2 as LINERs, and Seyfert galaxies usually have 0.6 $<$ [Fe II]$\lambda$1.257/Pa$\beta <$ 2 \citep[][although see Martins et al. 2013a for counterexamples]{Larkin98,Rodriguez-Ardila04,Rodriguez-Ardila05,Riffel13}. 

High-ionisation transitions (e.g. [S VIII]$\lambda$0.991, [S IX]$\lambda$1.252, [Si X]$\lambda$1.430, [Si VI]$\lambda$1.963, 
and [Ca VIII]$\lambda$2.322) are also detected in a few of the Seyferts in this sample (NGC~2273, NGC~4388, NGC~4395, NGC~7469 and 1H1934-063). These ``coronal'' lines (ionization potential $>$100 eV) can only exist very 
close to the ionization source, making them unique tracers of AGN activity and energetics. The simultaneous observation of very high- and 
low-ionization lines in AGN spectra implies a wide variety of physical conditions over the 
regions probed by the spectra, and the detection of strong coronal lines  
 is indicative of the presence of extreme UV and X-ray photons \citep{Prieto00}. The association of strong coronal lines with soft, thermal X-ray emission suggests that photoionization by the central source is the dominant excitation mechanism for coronal line emission in AGN \citep{Rodriguez-Ardila11}.
 
Finally, emission from the H$_{2}$ molecule is 
detected in around half of the galaxies. The strongest lines are from the
H$_{2}$ 1-0S(3) 1.96 $\mu$m and H$_{2}$ 1-0S(1) 2.12 $\mu$m transitions, although lower intensity lines such as 
H$_{2}$ 1-0S(2) 2.03 $\mu$m, H$_{2}$ 1-0Q(2) 2.41 $\mu$m and H$_{2}$ 1-0Q(3) 2.42 $\mu$m are also detected in some of the spectra. There are two 
main H$_{2}$ excitation mechanisms: ``thermal'' and ``non-thermal'' \citep[][and references therein]{Moorwood88,Rodriguez-Ardila04}. 
In the thermal case, the molecules are heated by shocks, UV photons
or X--rays, whereas in the non-thermal case the molecules fluoresce after being excited by absorption of a UV photon 
or collision with a fast electron from an X--ray ionized plasma.  
These two mechanisms produce different relative emission line intensities, which
can be used to identify the dominant mechanism. In particular,
the H$_{2}$ 1-0S(1)/2-1S(1) line ratio is generally higher for thermal excitation ($\sim$5-10) 
than for UV fluorescence ($\sim$1.82), as proposed by \citet{Mouri94}. Alternatively, 
rotational and vibrational temperatures can be determined 
using the expressions given by \citet{Reunanen02}, which involve the H$_{2}$ 1-0S(1)/2-1S(1) and 
H$_{2}$ 1-0S(2)/1-0S(0) line ratios. In the case of thermal excitation, both temperatures 
are similar, whereas in the case of fluorescent excitation, the vibrational temperature 
is larger. 

\section{Conclusions}
\label{conclusions}

We have presented moderate-resolution (R$\sim$1300-1800) NIR spectra of the nuclear regions of 50 nearby (D = 1 -- 92 Mpc) galaxies, with complete spectral coverage from $\sim$0.85 -- 2.5 $\mu$m. The galaxies span a variety of morphological types, and host comparable numbers of Seyfert and LINER nuclei (as well as a handful of transition objects and inactive nuclei). This, together with the wide wavelength coverage and good signal-to-noise ratio of the spectra, establishes the NIR spectral properties of galaxies hosting a range of nuclear activity and stellar populations.  The emission lines detected in many of the spectra may be used to examine the ionization mechanisms at work in the nuclei, while the underlying absorption line spectra are potentially of use to studies of the IMF and interesting stages of stellar evolution. We are using the data to investigate a range of issues related to low-luminosity AGN and their host galaxies, and we encourage interested readers to explore whether the spectra -- now available to the public -- may be of use to their own research as well.

\appendix
\section{Appendix A: Spectra and notes for individual galaxies}
\label{A1}

In this Appendix we present the optical and near-IR spectrum of each object, along with brief descriptions of the galaxies. The optical data are from \citet{Ho95}. For ease of comparison, the spectra in the top panels of Figures \ref{f1} -- \ref{f48} are all shown on a scale of 0.1 -- 2.5 $\times$ the median $H$-band flux. 

To aid the intercomparison of the spectra, a ``reference'' set of emission line wavelengths has been indicated on each one. These lines are given in Table \ref{table3}. They consist mainly of those lines that are most frequently detected, and include the five brightest H$_2$ lines in the $K$-band. The set also contains two lines, Pa$\beta$ and Br$\gamma$, which are not detected in most objects but which have been of interest for stellar population and kinematic studies of active galaxies \citep[e.g.][]{Davies07,Riffel11}. For the Seyfert galaxies with many emission lines (NGC~4395, NGC~4388, NGC2273, 1H 1934-063, NGC7469), some additional lines are also indicated.

\begin{deluxetable}{lc}
\tabletypesize{\scriptsize}
\tablecaption{The ``reference'' set of emission lines shown  in Figures \ref{f1} -- \ref{f49} \label{table3}}
\tablehead{\colhead{Line} & \colhead{Wavelength, $\mu$m}}
\startdata
{[SIII]} & 0.907 \\
{[SIII]} & 0.953 \\
{[CI]} & 0.985\\
{[SII]} & 1.029 \\
{[SII]} & 1.032 \\ 
{[NI]} & 1.041\\
HeI & 1.083\\
Pa$\beta$ & 1.282\\
Pa$\gamma$ & 1.093\\
{[FeII]} & 1.257\\
{[FeII]} & 1.644\\
H$_2$ 1-0 S(3) & 1.958\\
H$_2$  1-0 S(2) & 2.034\\
H$_2$ 1-0 S(0) & 2.122\\
Br$\gamma$ & 2.165\\
H$_2$  1-0 Q(1) & 2.407\\
H$_2$ 1-0 Q(3) & 2.424
\enddata
\end{deluxetable}

\subsection {Seyfert 1 - 1.9}

{\bf NGC~4235 (S1.2)}\\

The center of this nearly edge-on galaxy in the Virgo Cluster is bisected by a dust lane. The nucleus displays an emission
line spectrum characteristic of type 1 Seyfert galaxies, although optical
spectroscopy shows that the broad H$\alpha$ line has a peculiar
secondary hump redward of [N\,{\sc ii}]. The NLR appears
compact with a possible extension in [O\,{\sc iii}]  towards the north-east
with a PA$\sim 48\deg$ and a length of $\sim 4.4\arcsec$ or
$\sim$0.9\,kpc \citep{Pogge88}.

\begin{figure*}
\hspace*{-10mm}
\includegraphics[scale=0.9]{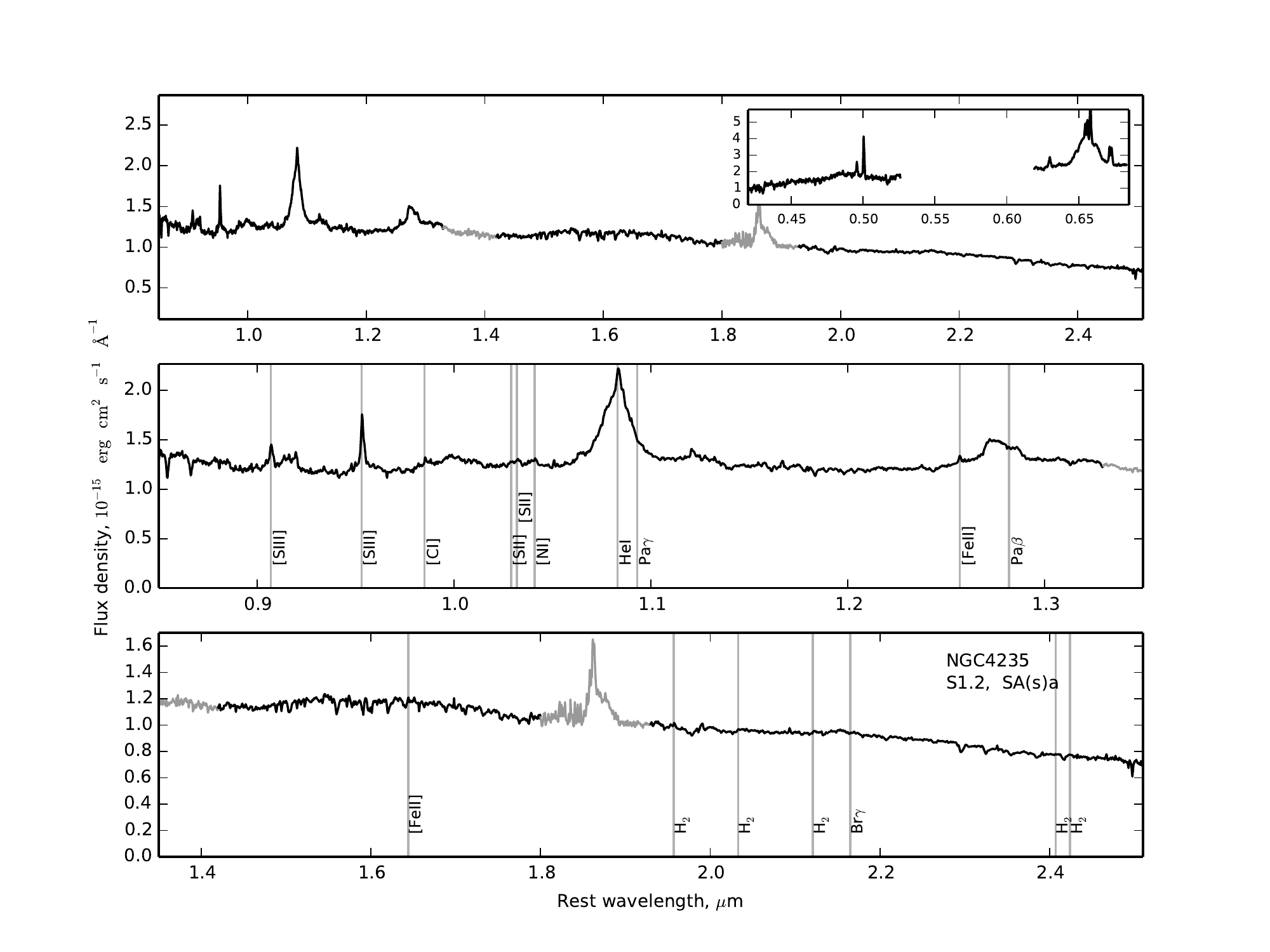}
\caption{ {\small NIR spectrum of NGC~4235, along with the optical spectrum of \citet[][inset]{Ho95}. Regions of very poor atmospheric transmission ($\lesssim$20\% for 3 mm H$_2$O at zenith) are plotted in gray. 
Vertical lines indicate the set of emission lines given in Table \ref{table3}}. For absorption line identifications, see Figure \ref{abslines_fig}.}
\label{f1}
\end{figure*}

The GNIRS nuclear spectrum (Figure \ref{f1}) shows that NGC\,4235 contains very broad permitted He\,{\sc i} lines, the He\,{\sc i} 1.083 $\mu$m line being the brightest line in the observed wavelength interval.
The broad Pa$\beta$ (and possibly also Pa$\delta$) line has a complex profile with a prominent blue
peak and a very broad, weak red wing centred about 7700\,km\,s$^{-1}$ from the systemic
velocity. This value is significantly higher than
the shift of $\sim$5400\,km\,s$^{-1}$ found in the red peak of the
H$\alpha$ line measured from the SDSS optical spectrum, possibly indicating that this AGN is highly variable. 
Forbidden lines of [S\,{\sc iii}]\,$\lambda \lambda$ 0.907,0.953\,$\mu$m and [Fe\,{\sc ii}]\,1.257\,$\mu$m are also identified in the spectrum. 

The continuum shows prominent stellar absorption features, from the very blue edge, where the CaT is strong by the standards of a Seyfert nucleus, to the red end,
where deep CO absorption bands are observed. The $H$-band displays a broad
bump which is interpreted as a minimum in H$^-$ opacity in cool stars in the
galaxy \citep{Rayner09}. \citet{Peletier07} find the central stellar population of NGC~4235 to be old, although they note that AGN emission makes the derived age rather uncertain. $K$-band observations of this source have been published by \citet{Ivanov00} and \citet{Imanishi04b}, but to the best of our knowledge this is the first spectroscopy of the other NIR bands. \\

{\bf NGC~3031 (S1.5)} \\

\begin{figure*}
\hspace*{-10mm}
\includegraphics[scale=0.9]{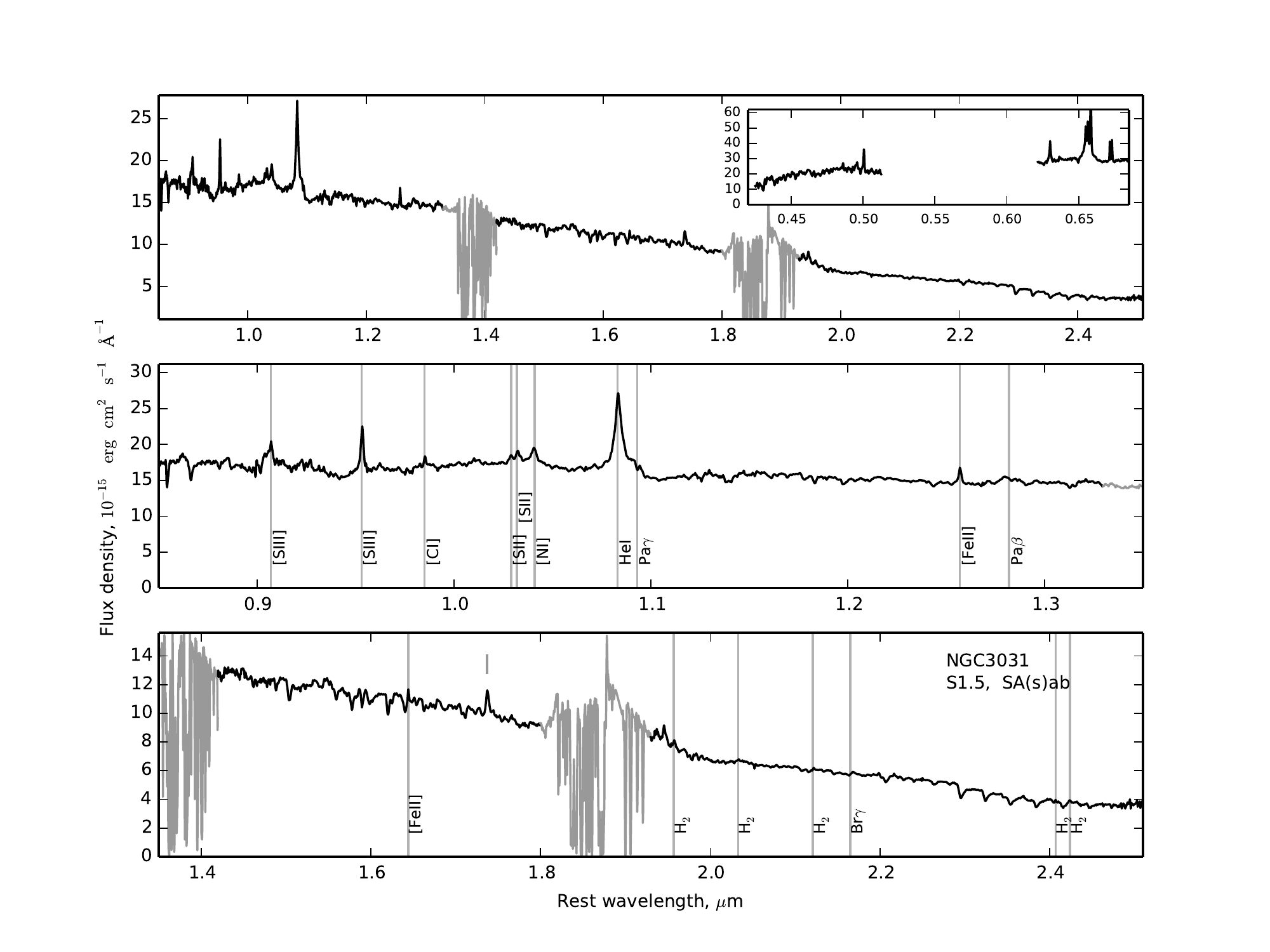}
\caption{ {\small As for Fig. \ref{f1}, for NGC~3031. The short vertical line near 1.73 $\mu$m indicates an artefact from division by the standard star that is common to several of the spectra.}}
\label{f2}
\end{figure*}

Also known as M\,81, this spiral galaxy is part of an interacting galaxy group that also includes M\,82 and NGC\,3077 \citep{Yun94}.
Although the active nucleus is classified as a low-luminosity Seyfert 1.5 by \citet{Ho97},  it is often referred to as a LINER, and \citet{Ho96} note that its ionization parameter is low relative to typical Seyferts. The nucleus appears unresolved
in \emph{HST} optical images \citep{Devereux97}. At radio wavelengths, NGC\,3031 shows a compact morphology and nuclear variability \citep[][and references therein]{Ho99b}, and at high angular resolution the infrared/optical SED is flatter than that of typical Seyfert templates \citep{Mason12}.  Based on multicolor photometry, \citet{Kong00} find the central stellar population of NGC~3031 to be old ($>$8 Gyr), and that the outer regions of the galaxy are significantly older. This is in agreement with optical spectroscopy analysed by \citet{Boisson00}.
The near-infrared spectrum presented here (Figure \ref{f2}) 
shows a prominent He I 1.083 $\mu$m line with a broad pedestal, and possibly broad Pa$\beta$ as well. The [Fe II] 1.257 $\mu$m line detected by \citet{Alonso-Herrero00} is also present, along with several other forbidden transitions including [S III], [S II], and [N I].  \\

{\bf NGC~5033 (S1.5)} \\

\begin{figure*}
\hspace*{-10mm}
\includegraphics[scale=0.9]{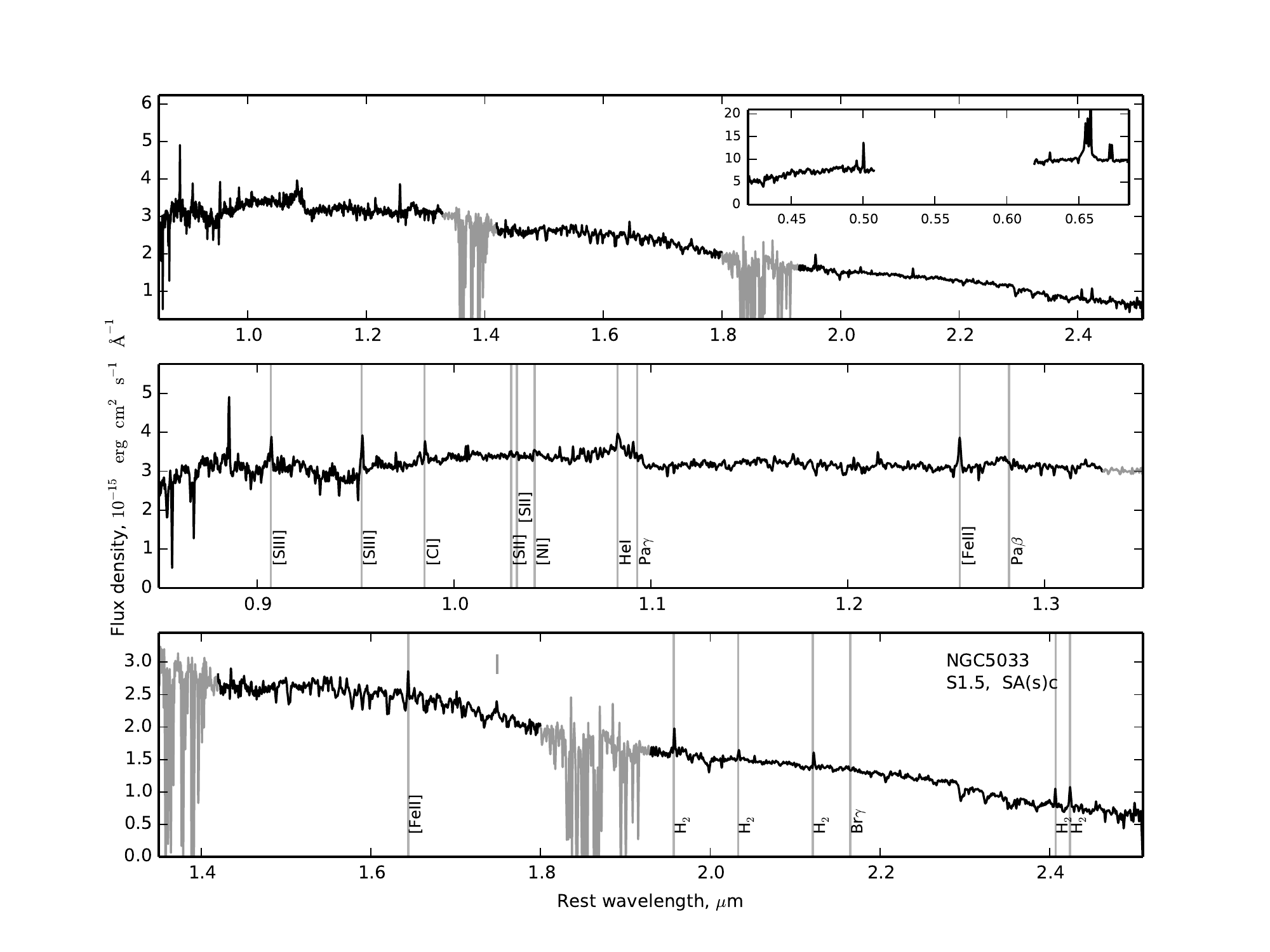}
\caption{ {\small As for Fig. \ref{f1}, for NGC~5033. The short vertical line near 1.73 $\mu$m indicates an artefact from division by the standard star that is common to several of the spectra.}}
\label{f3}
\end{figure*}

This highly inclined spiral galaxy forms a pair with NGC~5005 \citep{Helou82}. It hosts a Seyfert 1.5 nucleus, obscured by strong spiral dust lanes revealed
by HST/NICMOS and WFPC2 images \citep{Martini03,Erwin04}. The galaxy nucleus is conspicuously brighter in the 
near-infrared than in the optical, confirming the presence of nuclear dust lanes that are probably responsible for the flattening of the continuum at short wavelengths in the GNIRS spectrum (Fig. \ref{f3}; cf. NGC~3031, for example). High internal reddening was also noted by \citet{Boisson00}, who find that the central stellar population of NGC~5033 is old, with just a few per cent of the 5450\AA\ emission coming  from A-type stars.
The optical emission-line spectrum of this galaxy has characteristics of both LINERs and 
Seyfert galaxies, and the broad H$\alpha$ emission is variable \citep{Ho95}. The near-infrared spectrum is fairly typical of a Seyfert nucleus, with He I lines showing broad pedestals and lines of [S III], [Fe II] and [C I] detected, as well as several H$_2$ emission lines in the K-band (also noted by \citet{Ivanov00} and \citet{Bendo04}).  \\

{\bf NGC~4395 (S1.8)} \\

\begin{figure*}
\hspace*{-10mm}
\includegraphics[scale=0.9]{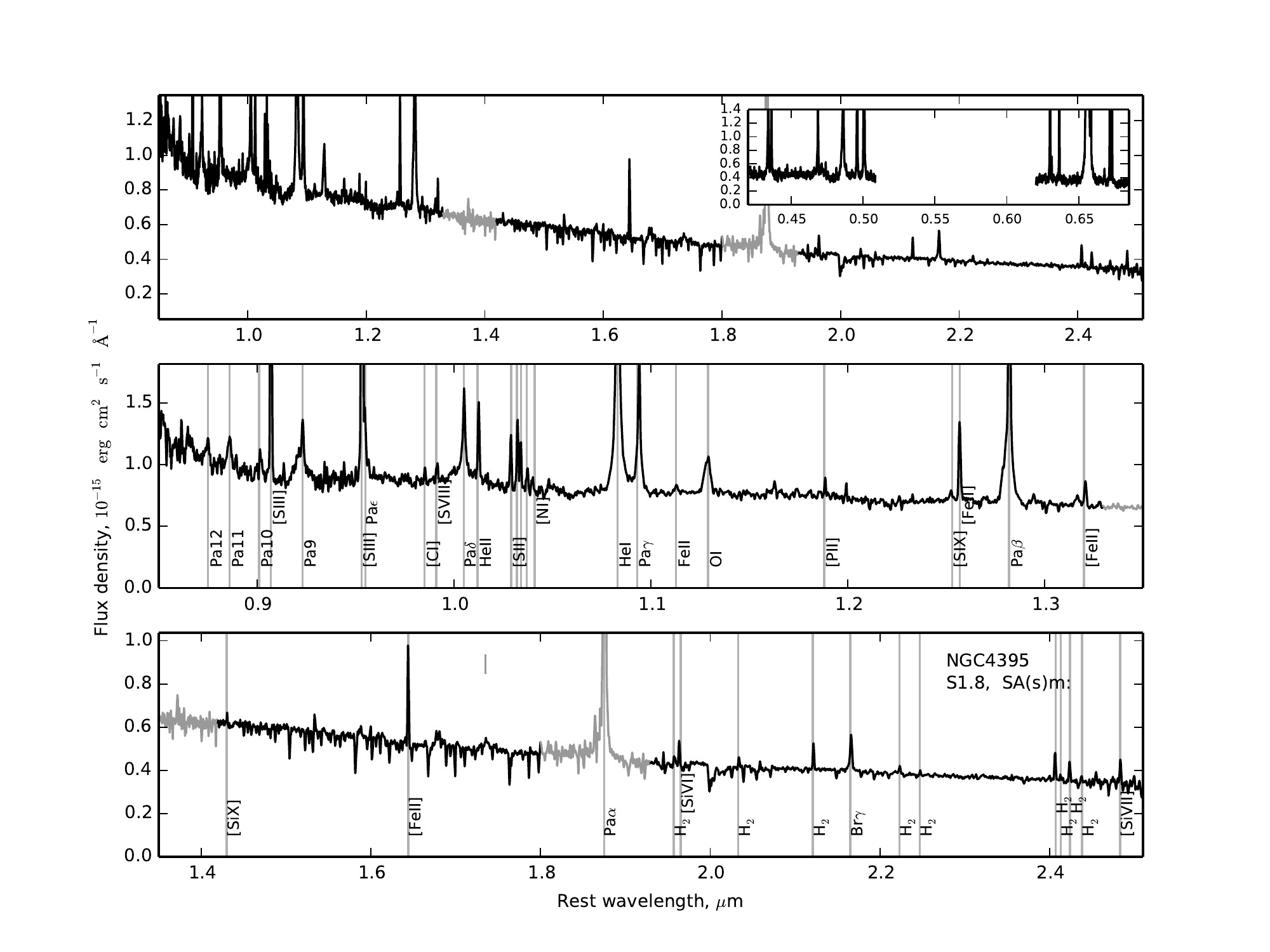}
\caption{ {\small As for Fig. \ref{f1}, for NGC~4395. Removal of atmospheric OH emission is unusually poor for this source, as shown by the strong residuals appearing as negative ``spikes'' between $\sim$1.5 - 1.8 $\mu$m. Lines other than those in the ``reference'' set are indicated for this object, and the short vertical line near 1.73 $\mu$m indicates an artefact from division by the standard star that is common to several of the spectra.}}
\label{f4}
\end{figure*}

This galaxy (Fig. \ref{f4}) holds several records: it is the nearest and faintest Seyfert 1 
nucleus \citep{Filippenko89}, and it is one of the intrinsically weakest nuclear X-ray 
sources observed to date \citep{Lira99,Moran99}. NGC 4395 is also highly 
unusual as an AGN host because of its late Hubble type: it is classified as a Magellanic 
spiral \citep[Sm; ][]{Ho01}. Last but not least, it harbors one of the smallest 
supermassive black holes with an accurately-determined mass \citep[3$\times 10^5$ M$_\odot$][]{ 
Peterson05}. With all these superlatives, it is not surprising that the radio, IR, 
optical and X-ray characteristics of this AGN have been extensively discussed by several 
authors \citep[e.g.][]{Lira99,Iwasawa00,Moran05,Peterson05,Skelton05,Minezaki06,Wrobel06}. Curiously, very few spectroscopic observations in the NIR are found in the literature. \citet{Kraemer99}, for instance,
reported optical/NIR spectroscopy from the atmospheric cutoff (3200~\AA) to almost 
1~$\mu$m, along with $L-$band spectroscopy of this object. To the best of our knowledge, the GNIRS spectrum 
shown here is the first 0.85 -- 2.5 $\mu$m spectrum in the literature. It is dominated by nebular lines, 
both permitted and forbidden, on top of a weak power-law
continuum with very little sign of the underlying stellar population. 
[S\,{\sc iii}]~0.907, 0.953\,$\mu$m, He\,{\sc i} 1.083~$\mu$m
and the Paschen series of H\,{\i} are, by far, the brightest emission lines in the spectrum. Broad permitted lines of H\,{\i}, He\,{\sc i}, Fe\,{\sc ii}
and O\,{\sc i}, typical of the Seyfert\,1 nature of this object, are present, and weak, 
high-ionization lines of [S\,{\sc viii}], [S\,{\sc ix}], [Si\,{\sc vi}], [Si\,{\sc vii}] and 
[Si\,{\sc x}] are also detected. Molecular hydrogen lines are bright in the 
$K-$band spectrum of this source.  \\

{\bf NGC~2639 (S1.9)} \\

\begin{figure*}
\hspace*{-10mm}
\includegraphics[scale=0.9]{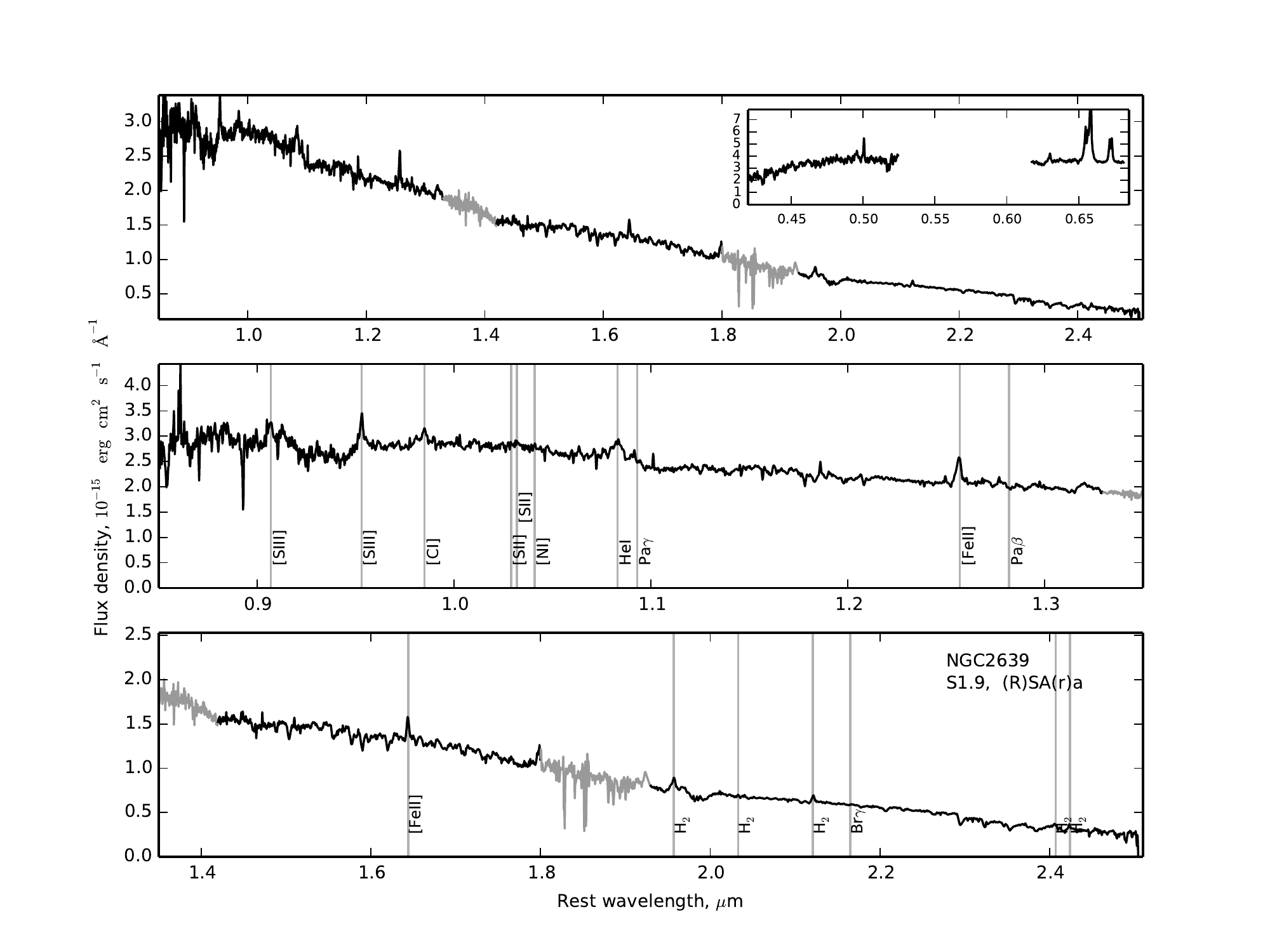}
\caption{ {\small As for Fig. \ref{f1}, for NGC~2639.}}
\label{f5}
\end{figure*}

Like NGC~4258, NGC~2639 hosts a rare nuclear H$_2$O megamaser, interpreted as arising in a dense accretion disk \citep{Braatz94,Wilson95}. The stellar velocity structure of the nucleus is also noteworthy, with measurements of the calcium triplet lines indicating a drop in velocity dispersion towards the nucleus that may be due to the presence of a young stellar population in a circumnuclear disk \citep{Marquez03}. The IR spectrum of NGC~2639 (Fig. \ref{f5}) is rather unremarkable, with only a handful of forbidden and permitted emission lines detected, along with some H$_2$ emission in the $K$-band. The [Fe II] emission in the J band was also noted by \citet{Alonso-Herrero00}. Stellar absorption features similar to those shown in Figure \ref{abslines_fig} are visible throughout the spectrum. \\

{\bf NGC~4258 (S1.9)} \\

\begin{figure*}
\hspace*{-10mm}
\includegraphics[scale=0.9]{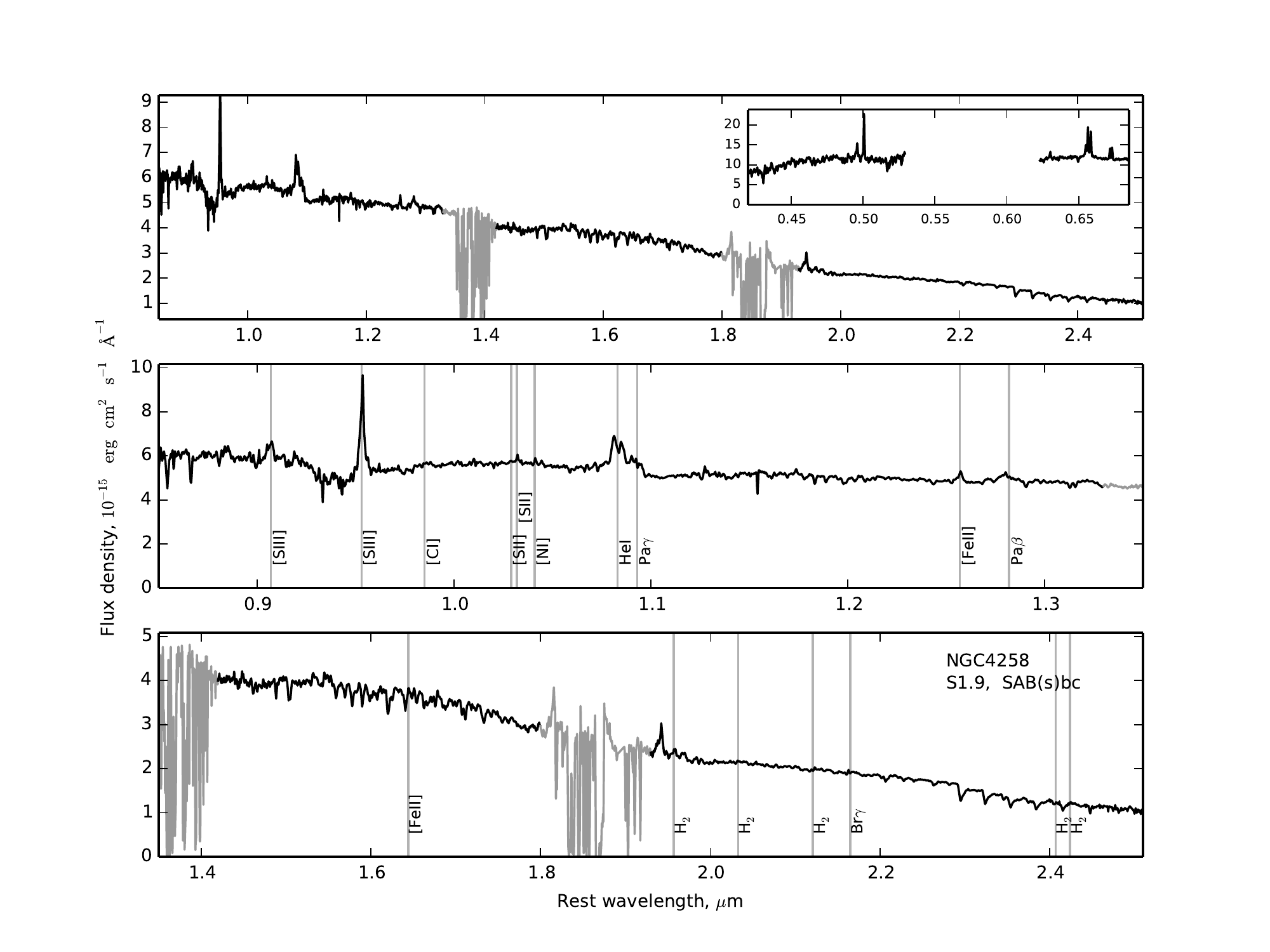}
\caption{ {\small As for Fig. \ref{f1}, for NGC~4258.}}
\label{f6}
\end{figure*}

NGC~4258 is well-known for its nuclear H$_2$O megamaser, which traces a dense, edge-on disk on sub-parcsec scales \citep{Watson94,Greenhill95,Miyoshi95}. As such, it has been the target of numerous distance measurements using several different techniques \citep{Macri06,Mager08,Humphreys13}. The nucleus of NGC~4258  also contains a variable, relativistic radio jet \citep{Doi13}, which may, on larger scales, be interacting with molecular clouds in the galaxy's disk \citep{Krause07}. Overall, the infrared spectrum (Fig. \ref{f6}) rather closely resembles that of NGC~2639 (Fig. \ref{f5}), another megamaser source, although the stellar absorption features appear somewhat deeper in NGC~4258. Differences in the emission lines are also apparent, such as stronger [S III] lines in NGC~4258 than in NGC~2639, and weaker or undetected H$_2$ and [Fe~II] emission. It appears that no optical stellar population synthesis has been published for this galaxy. \\

{\bf NGC~4388 (S1.9)} \\

\begin{figure*}
\hspace*{-10mm}
\includegraphics[scale=0.9]{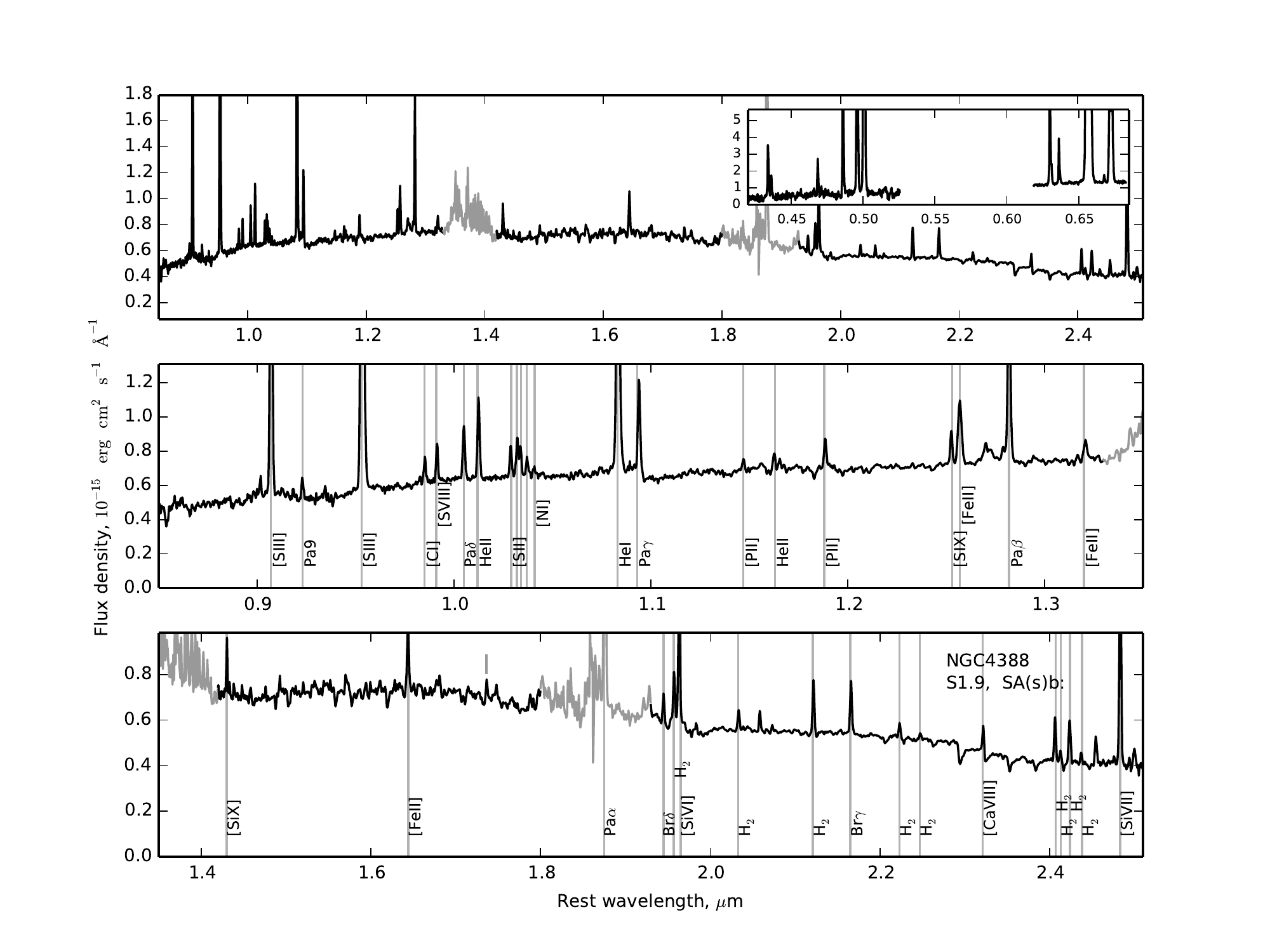}
\caption{ {\small As for Fig. \ref{f1}, for NGC~4388. The short vertical line near 1.73 $\mu$m indicates an artefact from division by the standard star that is common to several of the spectra. See Fig. \ref{emlines} for clearer labelling of the numerous emission lines in this object. }}
\label{f7}
\end{figure*}

Classified as a Seyfert\,2 galaxy by \citet{Phillips82} and Seyfert 1.9 by \citet{Ho97}, NGC~4388 (Fig. \ref{emlines}, \ref{f7}) is one of the first galaxies in which a conically shaped NLR was detected \citep{Pogge88}. This
AGN, which is host to an H$_2$O megamaser \citep{Kuo11}, lies in an SB(r)a galaxy close to edge-on. For this reason, it has an extremely 
chaotic appearance at small scales, in part due to a host galaxy dust lane
passing to the immediate north of the nucleus. Its continuum emission
reflects the dusty nature of this source, being heavily absorbed from 1.4~$\mu$m
bluewards. NGC\,4388 has been studied extensively in various wavelength regimes, including the 
NIR (see references in Table \ref{table2}). The GNIRS spectrum reveals
a wealth of emission lines, many of them never reported in the literature. In particular,
spectral features located in the region $1-1.15~\mu$m and redwards of 2.36~$\mu$m 
are new detections. Emission lines of [S\,{\sc iii}]~0.907, 0.953\,$\mu$m, 
He\,{\sc i}\,1.083~$\mu$m and the Paschen series of H\,{\i} are the brightest ones in
the $0.84-2.5~\mu$m interval. The
strong high-ionization spectrum of NGC\,4388 is also noteworthy, with [Si\,{\sc vii}]\,2.483\,$\mu$m,
for instance, being the second brightest line in the $K-$band after Pa$\alpha$. 
Moreover, [S\,{\sc viii}], [Si\,{\sc vii}] and [Si\,{\sc x}] were not previously 
reported in the literature, as well as the low-ionization forbidden emission of 
[C\,{\sc i}],  [S\,{\sc ii}], [N\,{\sc i}] and [P\,{\sc ii}], which are also
prominent in the spectrum. Most of the above emission lines extend across the slit, 
from NE to SW.   The CaT and CO absorption bands are visible at the blue
and red edges of the spectrum, respectively, as is CN absorption at 1.1~$\mu$m. The analysis of the nuclear and extended emission of this object is the subject of a 
separate publication (Rodr\'{\i}guez-Ardila et al., in prep.). \\

{\bf NGC~4565 (S1.9)} \\

\begin{figure*}
\hspace*{-10mm}
\includegraphics[scale=0.9]{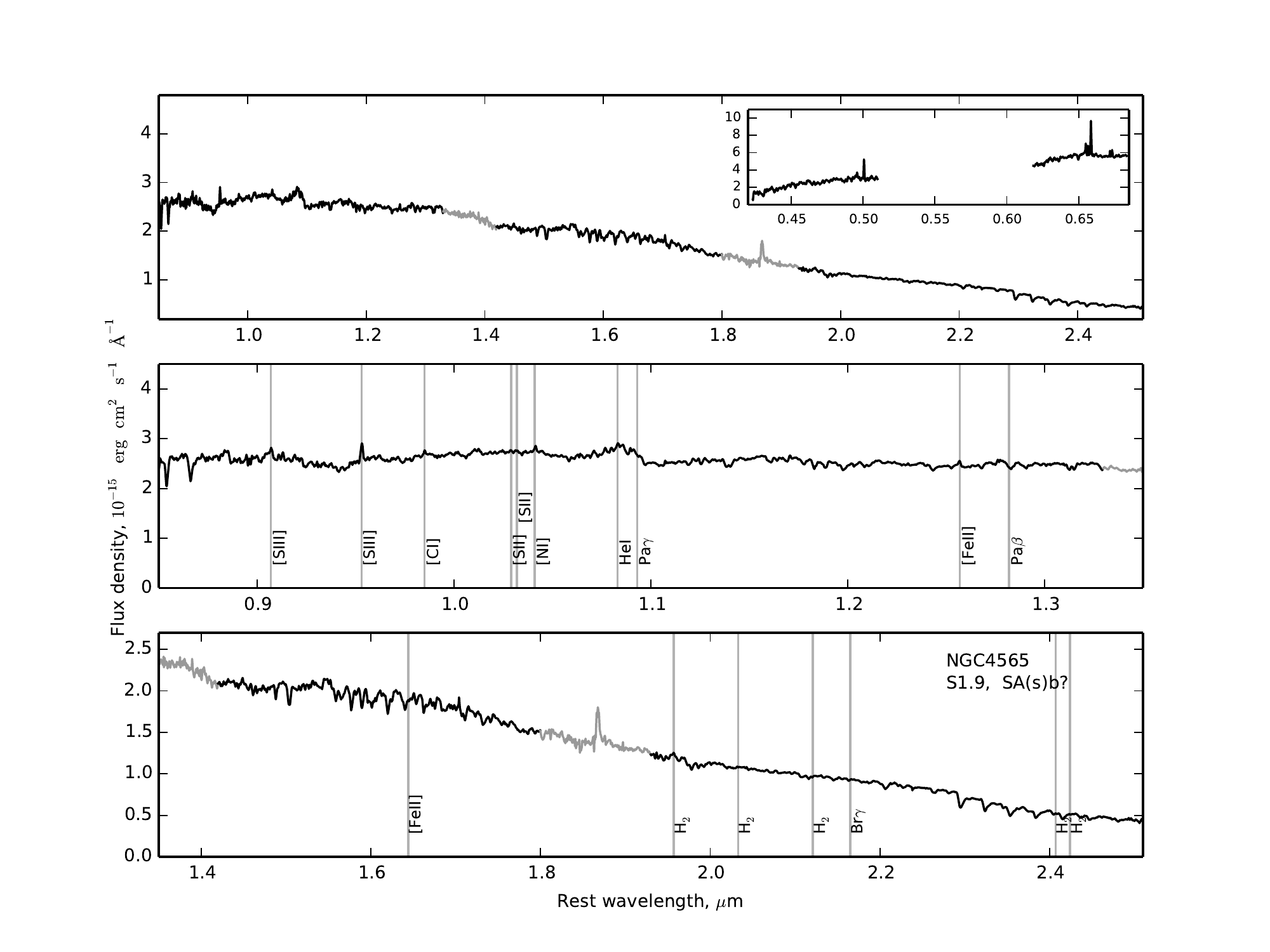}
\caption{ {\small As for Fig. \ref{f1}, for NGC~4565.}}
\label{f8}
\end{figure*}

NGC~4565 is an SAb galaxy that is highly inclined to the line of sight. 
 \citet{Proctor00} present evidence for a kinematically cold population at the centre of NGC~4565 that may be associated with ongoing star formation. As might be expected from the relatively weak emission lines in the optical spectrum of NGC~4565, the GNIRS spectrum (Fig. \ref{f8})  of this object contains just a few, weak lines from [SIII], [CI], HeI and [FeII]. H$_2$ emission has been detected at 17 and 28 $\mu$m \citep{Laine10}, but molecular hydrogen lines are not seen in Fig. \ref{f8}, nor in the $K$-band spectrum presented by \citet{Doyon89}. Absorption lines from the nuclear stellar population, including the CaT, CO band heads, and a slew of features in the $H$-band, are clearly visible in the spectrum. \\

{\bf NGC~4579 (S1.9/L1.9)} \\

\begin{figure*}
\hspace*{-10mm}
\includegraphics[scale=0.9]{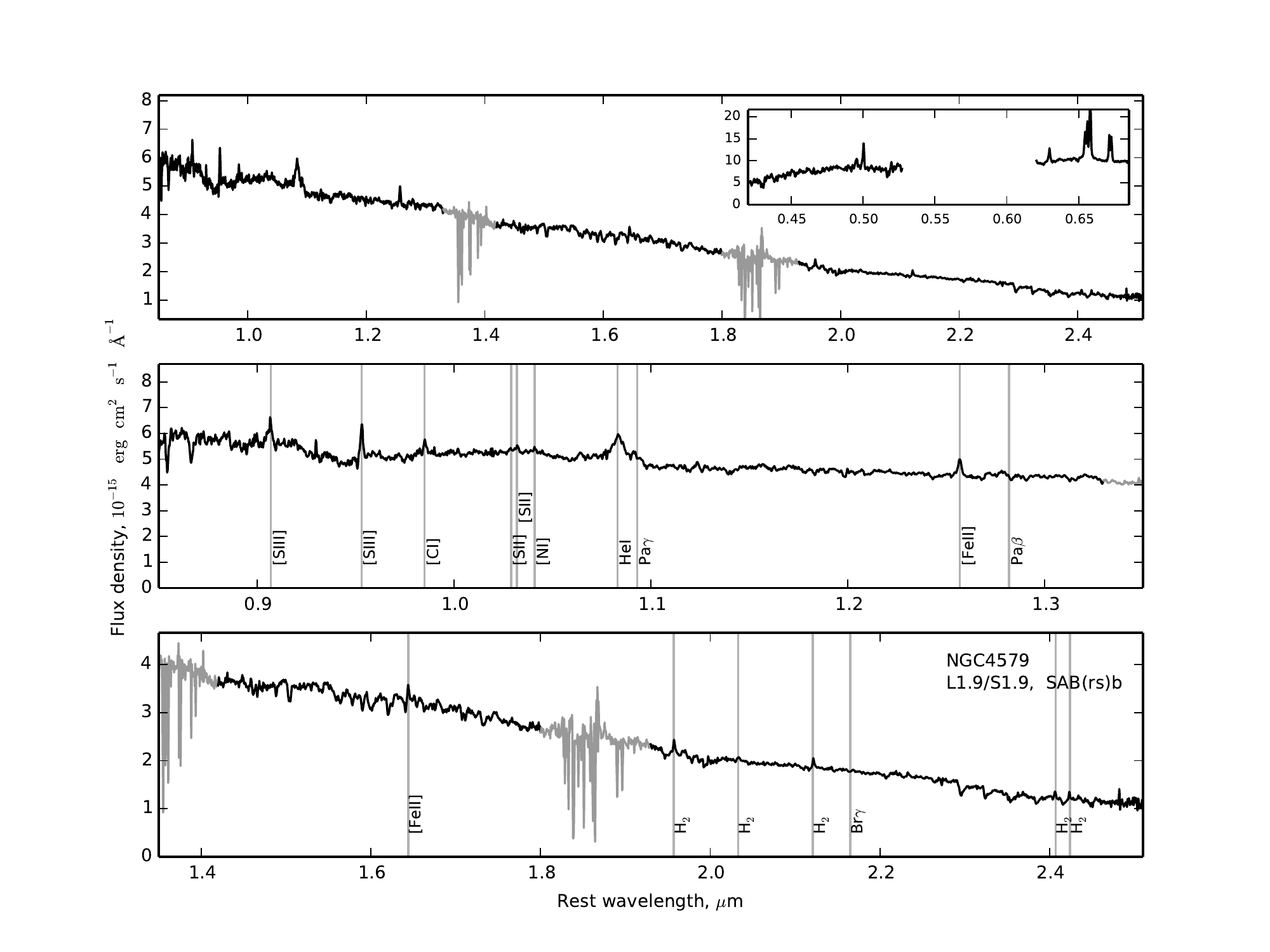}
\caption{ {\small As for Fig. \ref{f1}, for NGC~4579.}}
\label{f9}
\end{figure*}

NGC~4579 contains a low-luminosity AGN that may be accreting via an advection-dominated accretion flow \citep{Quataert99,Yu11}. Mapping of the gravitational torques in the galaxy suggest that material is being driven to within 50 pc of the AGN \citep{Garcia-Burillo09}. \citet{Palacios97} present evidence based on optical and NIR Mg indices that the stellar population in NGC~4579 may be fairly young, with a characteristic age of roughly 2.5 Gyr. Absorption lines from the nuclear stellar population in NGC~4579 are clearly visible in the GNIRS spectrum (Fig. \ref{f9}), as are emission lines of [SIII], [CI], HeI, [FeII] and H$_2$ that are common in this class of objects. The [FeII] 1.26 $\mu$m line was previously reported by \citet{Alonso-Herrero00}, and the morphology and kinematics of the $K$-band H$_2$ emission have been analyzed by \citet{Mazzalay13,Mazzalay14}. \\

\pagebreak


\subsection {Seyfert 2}

{\bf NGC 1167 (S2)} \\

\begin{figure*}
\hspace*{-10mm}
\includegraphics[scale=0.9]{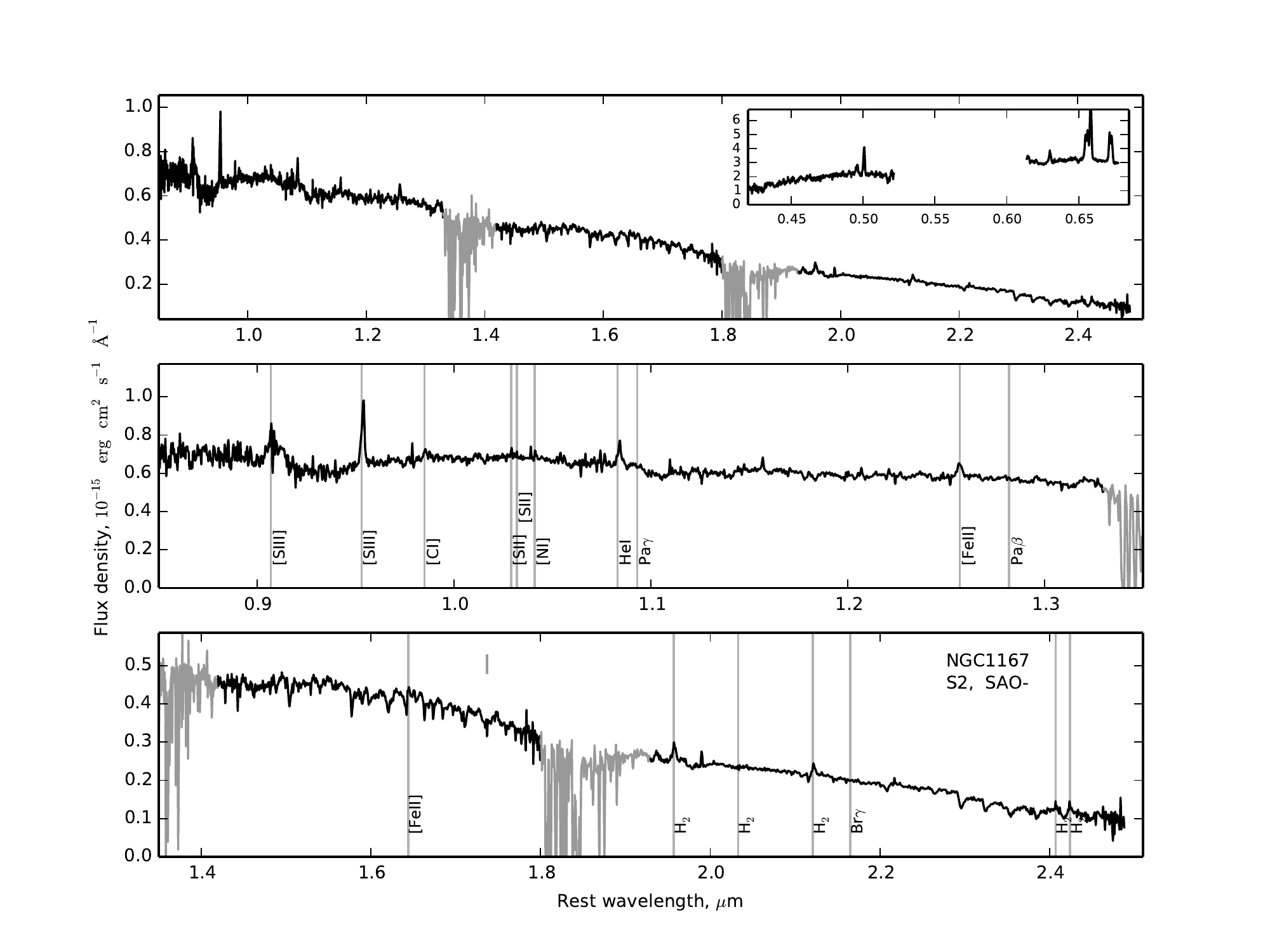}
\caption{ {\small As for Fig. \ref{f1}, for NGC~1167. The short vertical line near 1.73 $\mu$m indicates an artefact from division by the standard star that is common to several of the spectra.}}
\label{f10}
\end{figure*}

This lenticular galaxy resides in a rich environment, with several companions within a projected distance of 350 kpc \citep{Struve10}. The galaxy hosts an active nucleus with sub-parsec scale jets \citep{Nagar05,Giovannini01}, suspected to be obscured by a Compton-thick column of material \citep{Panessa06,Akylas09}. The stellar population within the half-light radius is fairly old, with a luminosity-weighted mean age of $\sim$6 Gyr \citep{GonzalezDelgado14}.
The GNIRS spectrum of this object (Fig. \ref{f10}) exhibits a fairly sparse set of emission lines, from [SIII], HeI, [FeII] and H$_2$.\\

{\bf NGC 1358 (S2)} \\

\begin{figure*}
\hspace*{-10mm}
\includegraphics[scale=0.9]{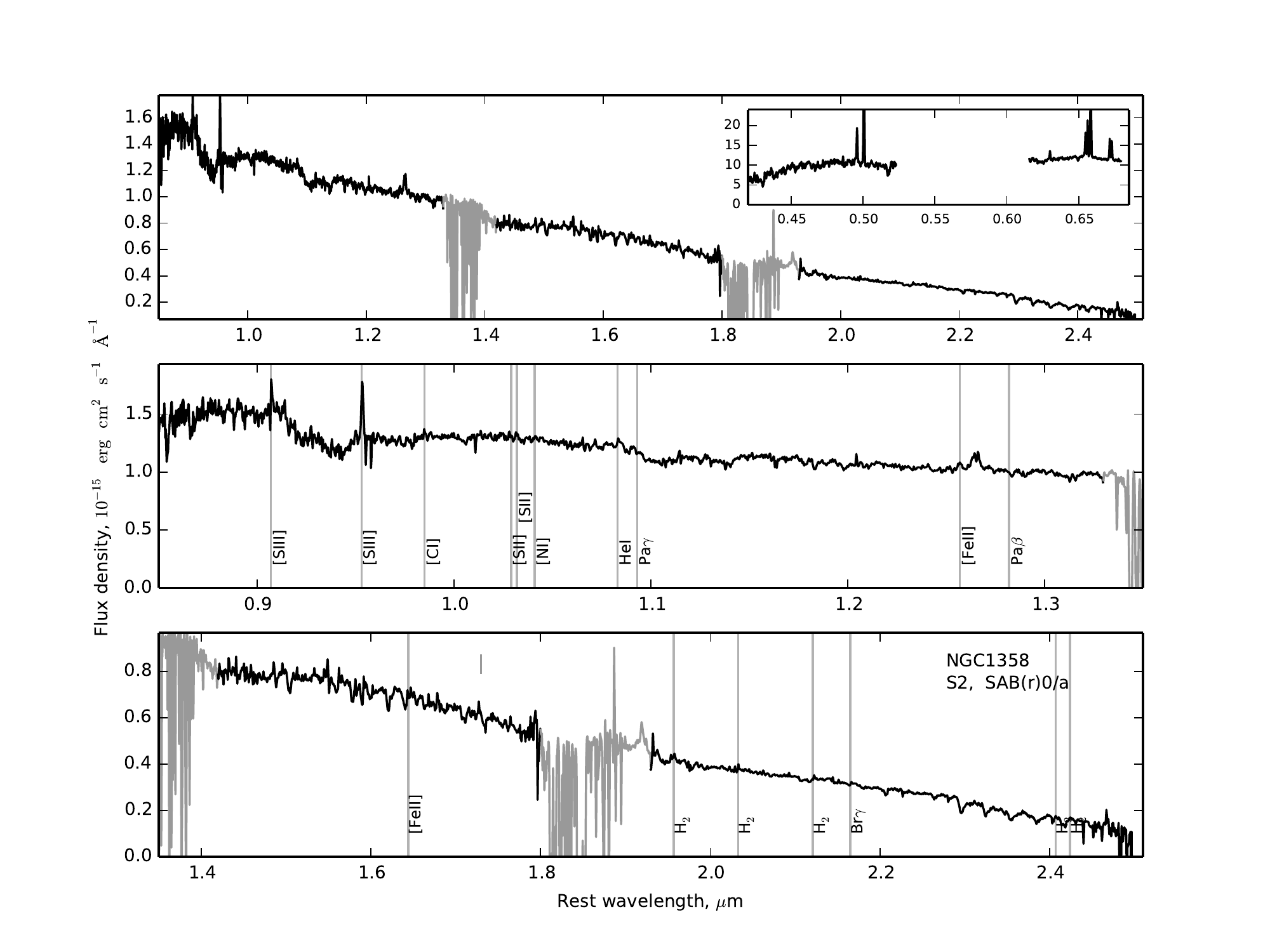}
\caption{ {\small As for Fig. \ref{f1}, for NGC~1358. The short vertical line near 1.73 $\mu$m indicates an artefact from division by the standard star that is common to several of the spectra.}}
\label{f11}
\end{figure*}

The nucleus of this barred galaxy shows irregular dusty structures \citep{Malkan98}.
\citet{Storchi-Bergmann98} find that 64\% of the nuclear optical emission of NGC~1358 can be accounted for by a 10 Gyr template, with almost all of the remainder from a 1 Gyr template. On the other hand, \citet{Perez11} estimate the age of the central stellar population to be in the region of 6 Gyr. The NIR spectrum of this object (Fig. \ref{f11}) bears a broad resemblance to that of NGC~1167, containing only a handful of weak emission lines. The $\sim$0.93 $\mu$m absorption, likely a blend of ZrO, CN, and TiO (\S\ref{abs}), appears to be unusually deep in this object. \\

{\bf NGC 2273 (S2)} \\

\begin{figure*}
\hspace*{-10mm}
\includegraphics[scale=0.9]{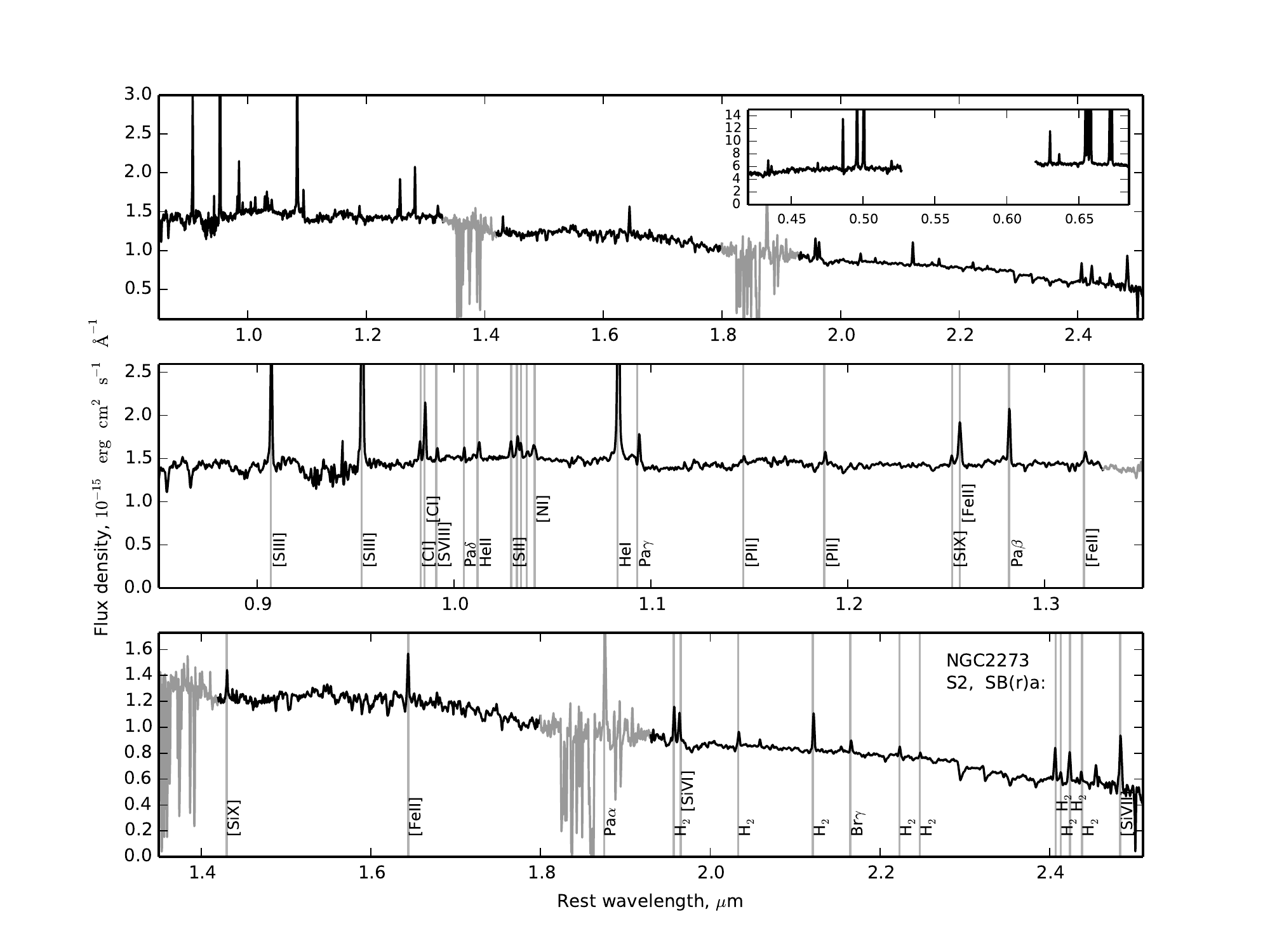}
\caption{ {\small As for Fig. \ref{f1}, for NGC~2273. Some lines other than those in the ``reference'' set are indicated for this object.}}
\label{f12}
\end{figure*}

This galaxy, classified as Seyfert\,2 by \citet{Ho97}, has been extensively studied in the 
literature, from radio to X-ray wavelengths \citep{Anderson05,Terashima02,Falcon-Barroso06}. It is a well-known barred galaxy \citep{Ferruit00}, with nuclear spiral arms inside a ringlike dust feature \citep{Laurikainen05}. VLBA imaging shows 
an elongated multiple-component structure aligned east-west on the sky \citep{Anderson05}.
The brightest [O\,{\sc iii}]~$\lambda$5007 emission is a linear, jetlike structure extending
2$\arcsec$ and aligned with the radio components. Fainter [O\,{\sc iii}] emission is also
present north and south of the nucleus. The GNIRS spectrum (Fig. \ref{f12}) was taken at a PA of 169$\deg$,
that is, nearly perpendicular to the radio and most prominent [O\,{\sc iii}]~$\lambda$5007 emission.
It is dominated by strong forbidden emission of [S\,{\sc iii}]~0.953\,$\mu$m and permitted He\,{\sc i}
at 1.083~$\mu$m. High ionization lines of [S\,{\sc viii}], [S\,{\sc ix}], [Si\,{\sc vi}], 
[Si\,{\sc vii}], and [Si\,{\sc x}] are also detected as well as 
low-ionization lines of [Fe\,{\sc ii}], [S\,{\sc ii}], [N\,{\sc i}] and [C\,{\sc i}]. In the
$K$-band, the nuclear spectrum includes numerous H$_2$ lines. The shape of the continuum emission indicates that it is mostly starlight, with prominent CaT and CO absorption bands at the blue and red edges of the 
spectrum, respectively. In the $H$-band, absorption lines of CO, Si, Mg and Fe are also
observed. \citet{Peletier07} find that the strength of the H$\beta$ line in the centre of this galaxy indicates a significant $\sim$1 Gyr stellar population. \\

{\bf NGC 2655 (S2)} \\

\begin{figure*}
\hspace*{-10mm}
\includegraphics[scale=0.9]{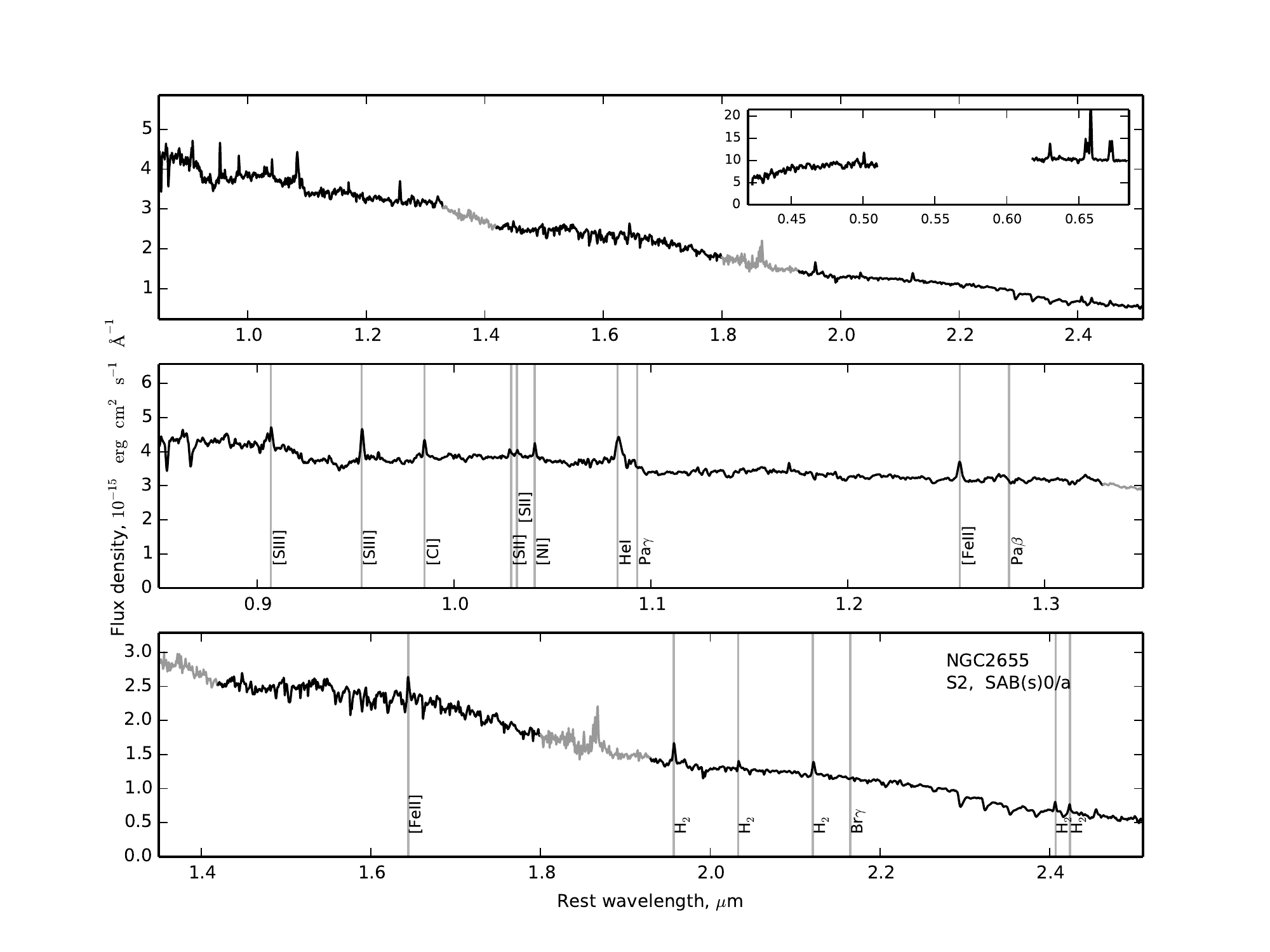}
\caption{ {\small As for Fig. \ref{f1}, for NGC~2655.}}
\label{f13}
\end{figure*}

NGC~2655 is a lenticular galaxy in a loose grouping of galaxies. Emission-line gas extends over several kpc \citep{Keel88}, and the galaxy's diffuse, extended optical structure and complex HI dynamics have been interpreted as indicating a succession of mergers \citep{Sparke08}. Optical line indices suggest recent star formation activity in this galaxy, and \citet{Silchenko06} find a luminosity-weighted mean age of 2 Gyr for the nucleus of NGC~2655. The NIR spectrum (Fig. \ref{f13}) shows several fairly weak, narrow emission lines, including several H$_2$ transitions in the $K$-band. \\

{\bf NGC 3079 (S2)} \\

\begin{figure*}
\hspace*{-10mm}
\includegraphics[scale=0.9]{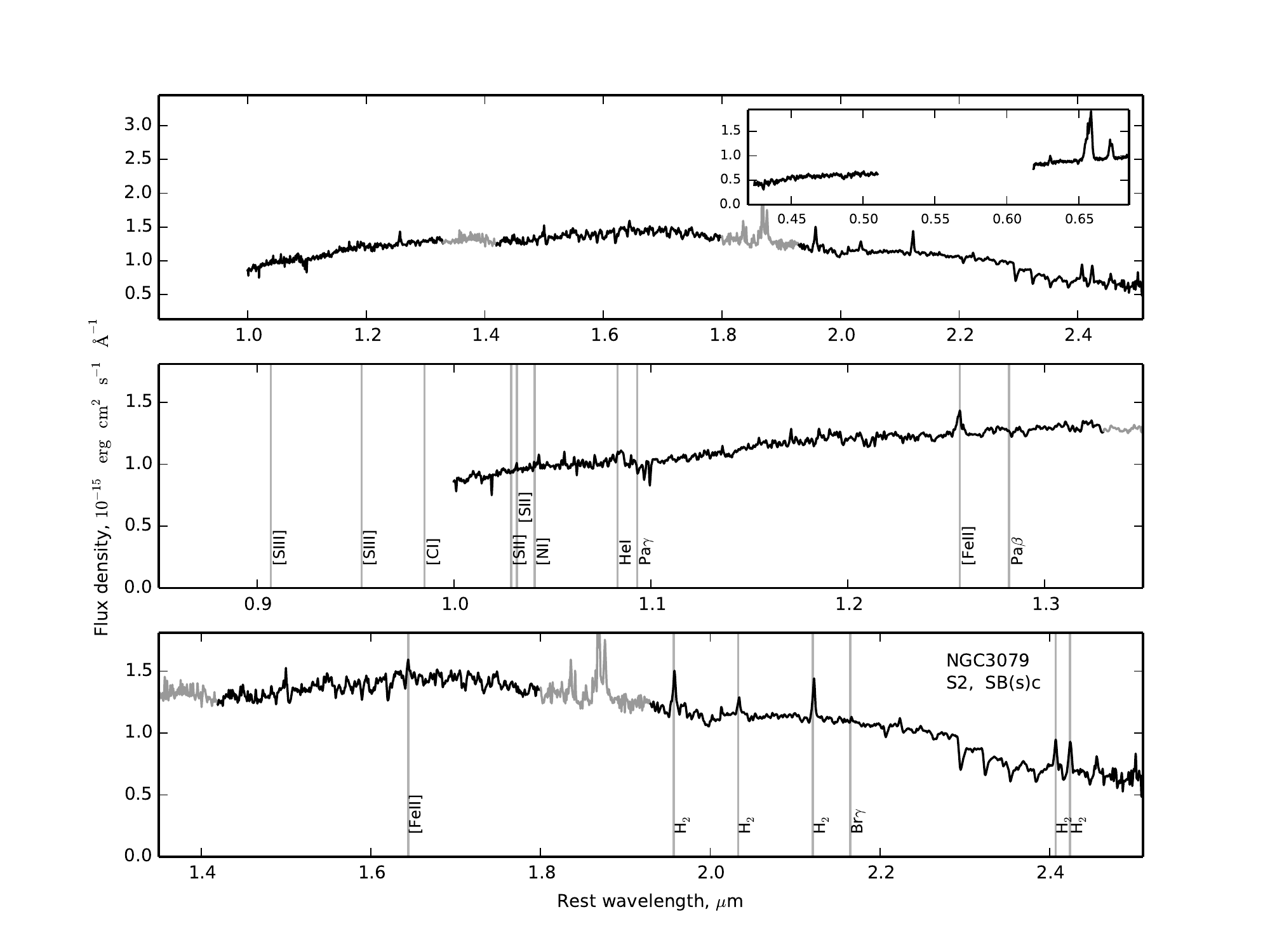}
\caption{ {\small As for Fig. \ref{f1}, for NGC~3079. The S/N at the extinguished, short-wavelength end of the spectrum is very low, and this region has been omitted from the figure.}}
\label{f14}
\end{figure*}

This edge-on spiral galaxy is known for its nuclear H$_2$O megamaser \citep{Henkel84}. It also contains a powerful outflow that appears to be influencing its satellite galaxy, NGC~3073 \citep{Irwin87,Filippenko92}. The GNIRS spectrum of NGC~3079 (Fig. \ref{f14}) is characterised by a curved continuum, likely due to the high extinction towards the nucleus \citep{Israel98}. H$_2$ emission lines are relatively strong in this object, and \citet{Hawarden95} argue that they arise in material excited by shocks in the nuclear outflow. The stellar absorption lines in the $H$- and $K$- bands broadly resemble those observed in many of the other galaxies examined here (although the CO bandheads are rather strong), but the 1.1 $\mu$m CN absorption appears somewhat weaker than usual. \\

{\bf NGC 3147 (S2)} \\

\begin{figure*}
\hspace*{-10mm}
\includegraphics[scale=0.9]{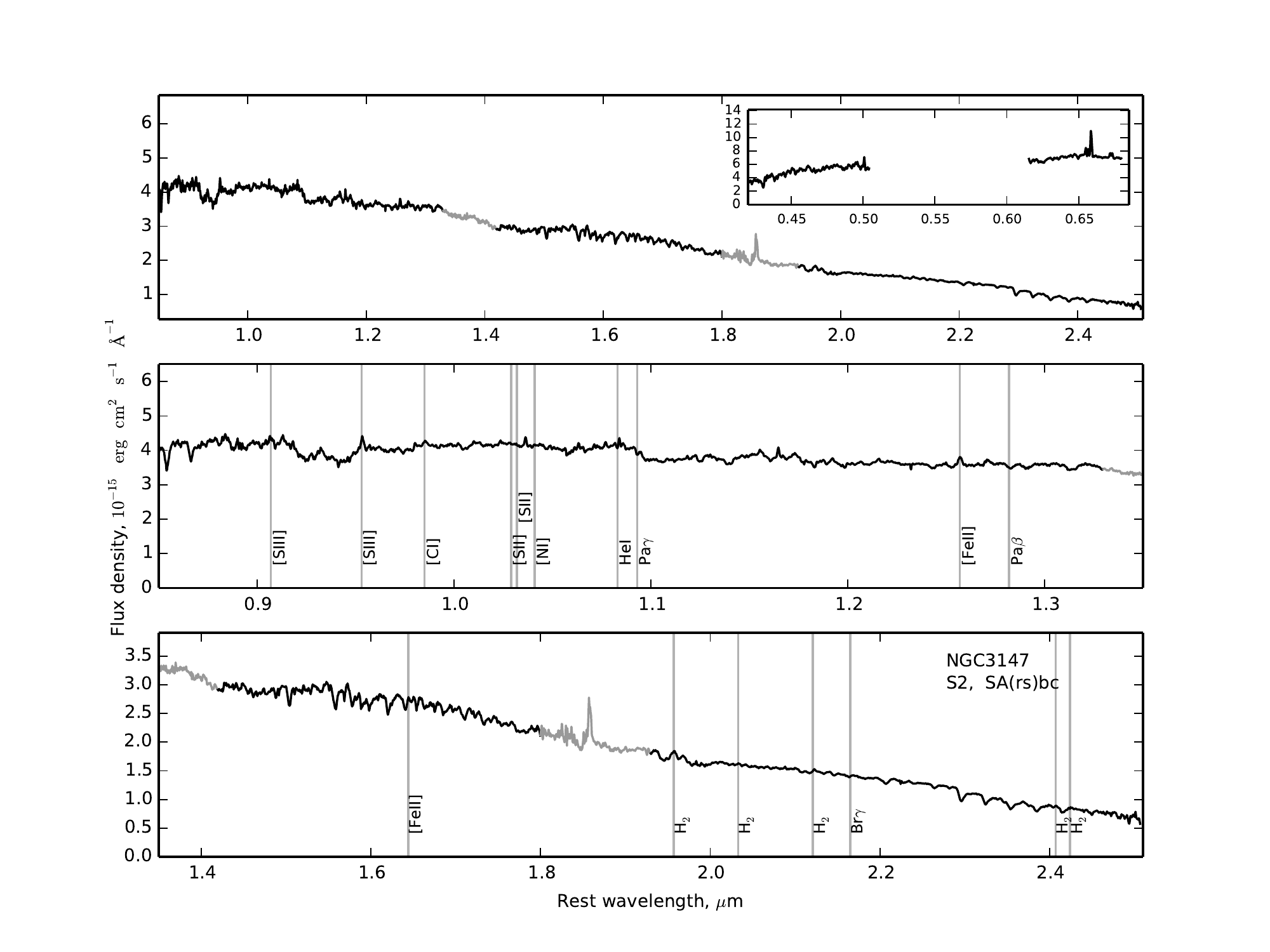}
\caption{ {\small As for Fig. \ref{f1}, for NGC~3147.}}
\label{f15}
\end{figure*}

NGC~3147 is a promising candidate for a ``true'' type 2 Seyfert nucleus, in which the broad line region is genuinely lacking \citep[rather than simply obscured; ][]{Bianchi08,Matt12}. The optical spectrum of the source is well fit by a stellar population model with age=8 Gyr \citep{Bruzual93}. The GNIRS spectrum (Fig. \ref{f15}) shows only a few, weak emission lines, along with many atomic and molecular stellar absorption lines.  The good telluric line removal in the regions between the NIR atmospheric windows provides an opportunity to study in detail the stellar population properties of this source. \\

{\bf NGC 4725 (S2:)} \\

\begin{figure*}
\hspace*{-10mm}
\includegraphics[scale=0.9]{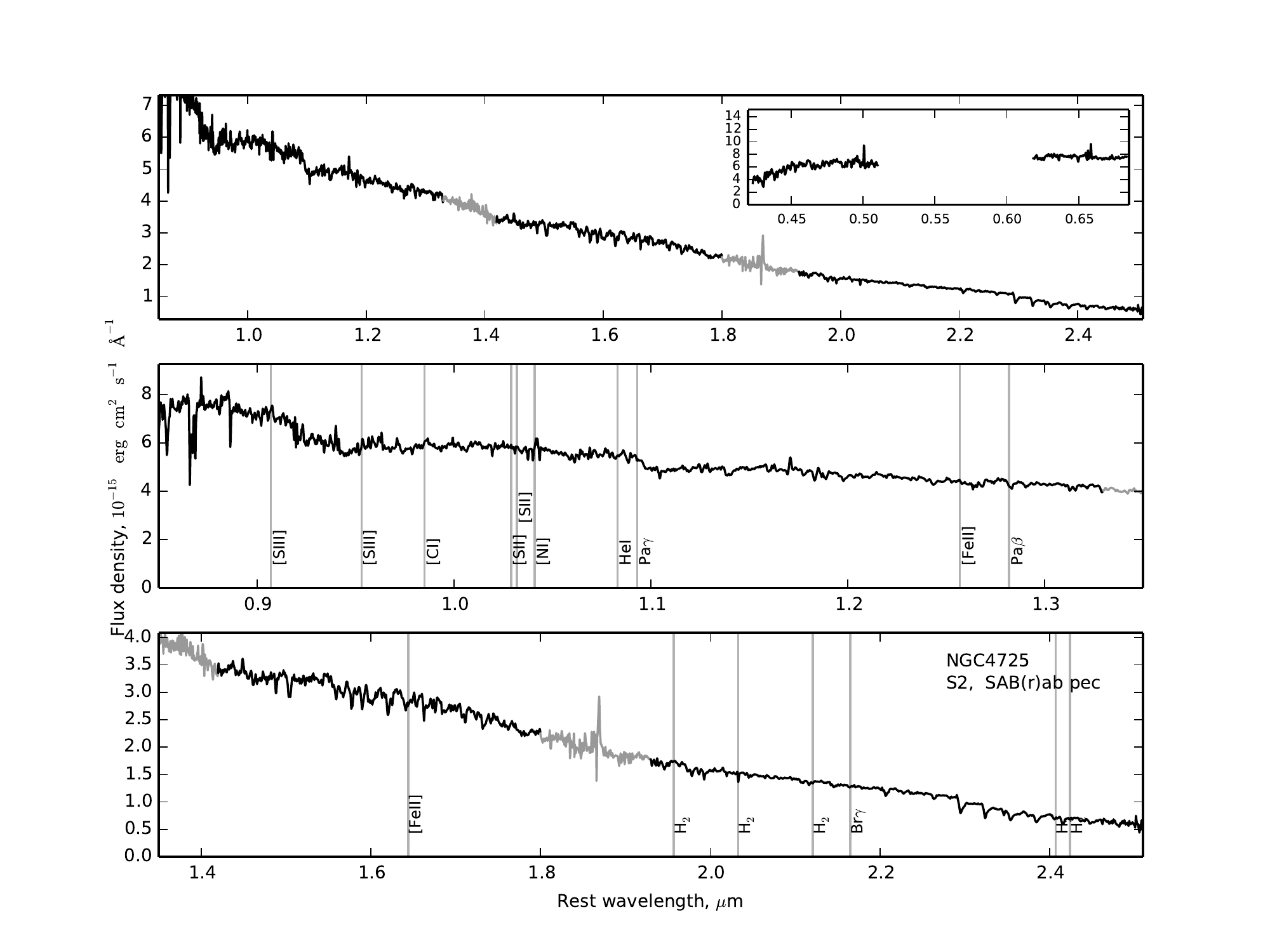}
\caption{ {\small As for Fig. \ref{f1}, for NGC~4725.}}
\label{f16}
\end{figure*}

NGC~4725 is a double-barred galaxy in which a handful of supernovae have occurred in recent decades. The optical classification of NGC~4725 as a Seyfert 2 is somewhat uncertain \citep{Ho97} and although a hard X-ray source is detected at the nucleus, its luminosity is comparable to that of surrounding, off-nuclear sources \citep{Ho01}. Although younger than the outer regions of the galaxy, the central stellar population is fairly old, with a luminosity-weighted mean age in the region of 4-5 Gyr \citep{deLorenzo-Caceres13}. No emission lines are detected in the GNIRS spectrum (Fig. \ref{f16}), although numerous stellar absorption lines, from the $K$-band CO bandheads and 0.85 $\mu$m CaT to the plethora of CO and metallic lines in the $H$-band, are visible. As with NGC~3147, the good telluric line removal means that the S/N in the spectrum is good at the locations of the H$_2$O and C$_2$ bands that are important for testing NIR stellar population models (\S\ref{abs}). \\

{\bf NGC 5194 (S2)} \\

\begin{figure*}
\hspace*{-10mm}
\includegraphics[scale=0.9]{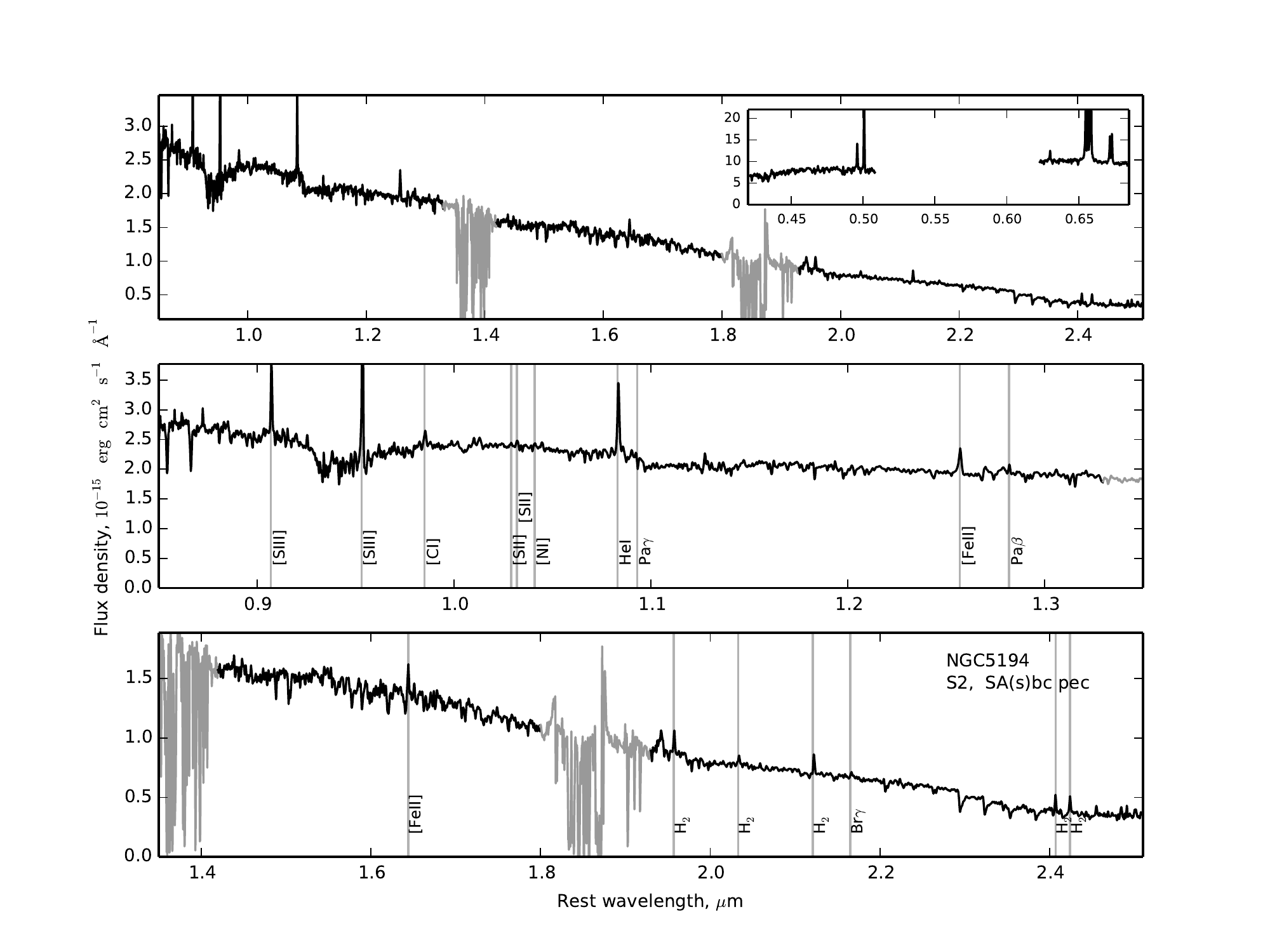}
\caption{ {\small As for Fig. \ref{f1}, for NGC~5194.}}
\label{f17}
\end{figure*}

NGC~5194, or M51, is a face-on spiral galaxy that is interacting with a neighboring system, NGC~5195 \citep[e.g.][]{BSmith10}. H$_2$O maser activity is detected in this galaxy, although it is unclear whether it is associated with an obscuring torus, jet, or star-forming region \citep{Ho87,Hagiwara01,Hagiwara07}. The stellar population in the inner few hundred parsecs of the galaxy is generally old, but this underlying distribution is punctuated by groups of massive stars \citep{Lamers02}. Lines of [SIII] are the strongest in the GNIRS spectrum (Fig. \ref{f17}), but emission from HeI, [FeII] and H$_2$ is also fairly prominent. Despite the line emission, many stellar absorption features are clearly visible in the spectrum.  \\

{\bf NGC 7743 (S2)} \\

\begin{figure*}
\hspace*{-10mm}
\includegraphics[scale=0.9]{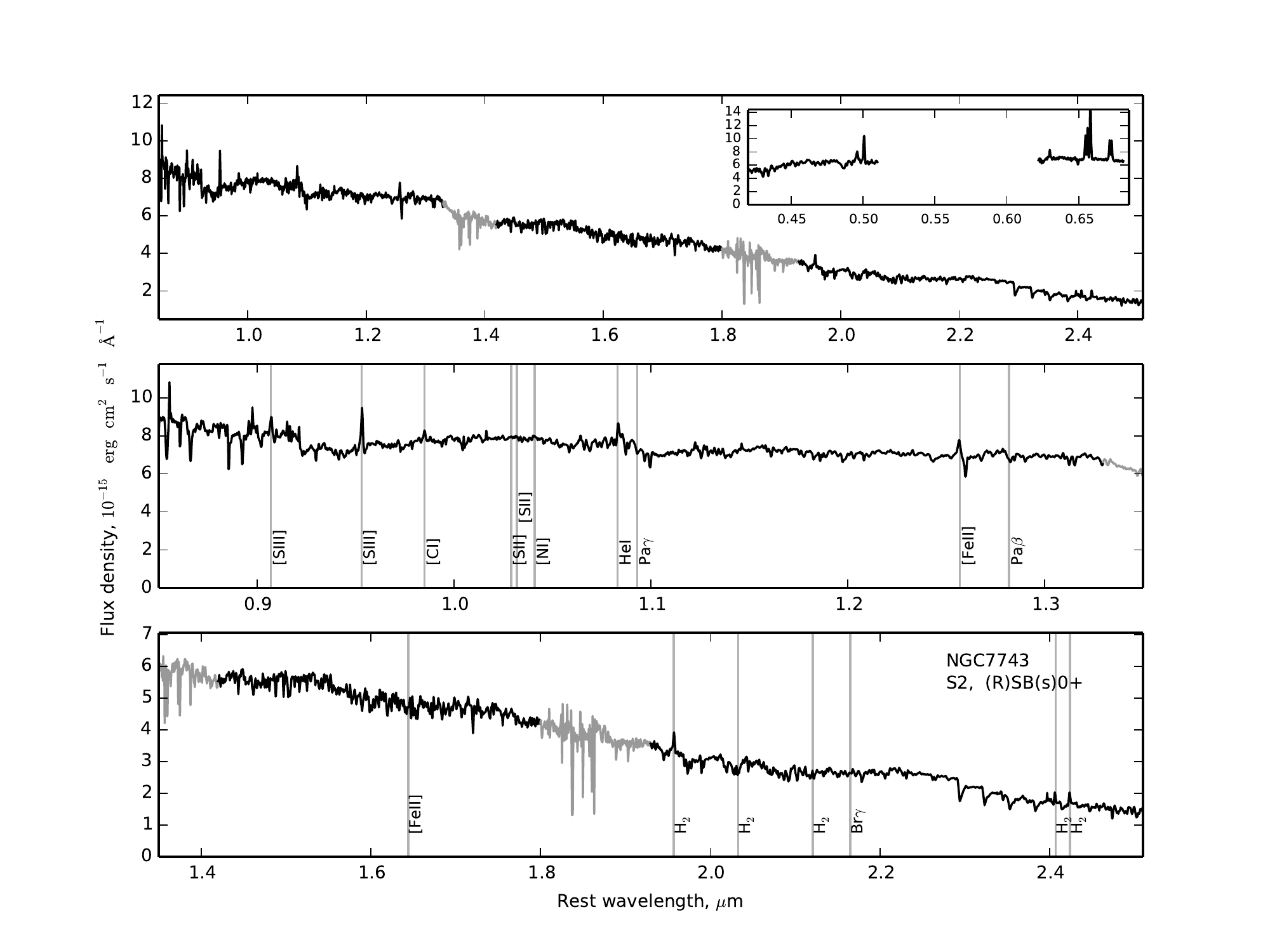}
\caption{ {\small As for Fig. \ref{f1}, for NGC~7743.}}
\label{f18}
\end{figure*}

NGC~7743 is a barred lenticular galaxy whose active nucleus appears to be highly obscured \citep{Gonzalez-Martin09b}.  \citet{Raimann01} find that the nuclear optical light is emitted by three stellar components, $\sim$40\% each from an old, metal-rich and a 1 Gyr component, with the remaining $\sim$20\% from a 100 Myr population. According to \citet{Silchenko06}, the luminosity-weighted mean age is $<$2 Gyr. The IR emission-line spectrum (Fig. \ref{f18}) is fairly similar to that of NGC~2655, another Seyfert 2 object, and a number of H$_2$ lines are present. Br$\gamma$ is detected in the spectrum presented by \citet{Mould12}, taken through a 1\arcsec\ slit, but is not visible in that of \citet[][ 8$\times$8\arcsec\ aperture]{Hicks13} or in Figure \ref{f18} (0.3\arcsec\ slit), suggesting spatial structure in that line. \\

\pagebreak

\subsection{LINER 1.9}

{\bf NGC 266 (L1.9)} \\

\begin{figure*}
\hspace*{-10mm}
\includegraphics[scale=0.9]{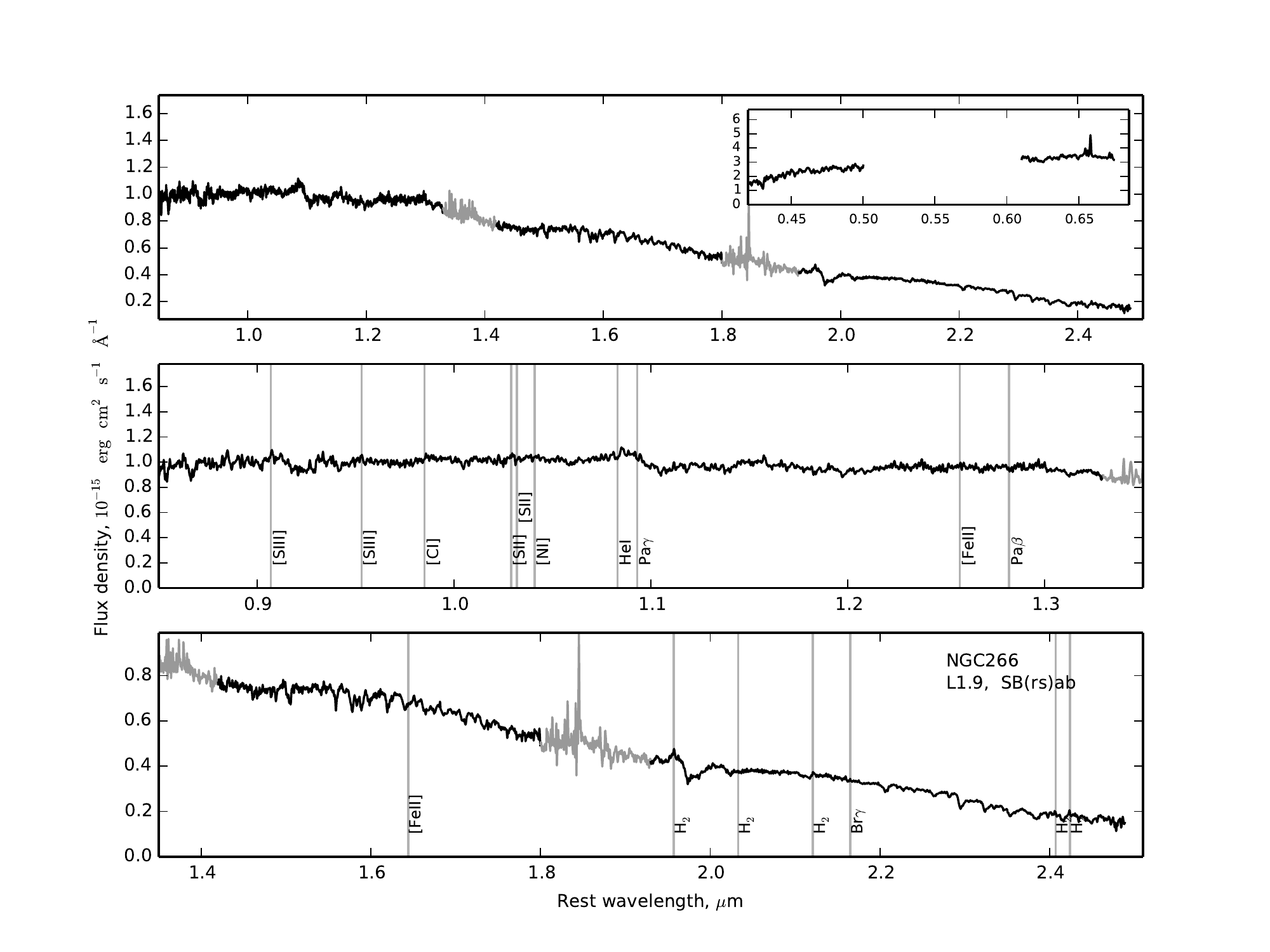}
\caption{ {\small As for Fig. \ref{f1}, for NGC~266.}}
\label{f19}
\end{figure*}

This LINER is hosted in strongly barred galaxy \citep{Veron-Cetty06} which has undergone an interaction with neighboring UGC\,499 \citep{Noordermeer05}. NGC\,266 has been studied from X-rays \citep{Younes12} to radio frequencies \citep{Nagar05}. In its optical spectrum, a broad H$\rm{\alpha}$ component is obvious even before profile decomposition, with a FWHM of the broad component of 1350 km/s \citep{Ho97}. With the possible exception of a weak, broad HeI 1.08 $\mu$m feature, no emission lines are detected in the GNIRS spectrum (Fig. \ref{f19}). Numerous stellar absorption lines, however, are visible in the spectrum. The nuclear stellar population of NGC~266 is found to be predominantly old, with just a 5\% contribution of intermediate-age stars to the optical light \citep{CidFernandes04}. \\

{\bf NGC 315 (L1.9)} \\

\begin{figure*}
\hspace*{-10mm}
\includegraphics[scale=0.9]{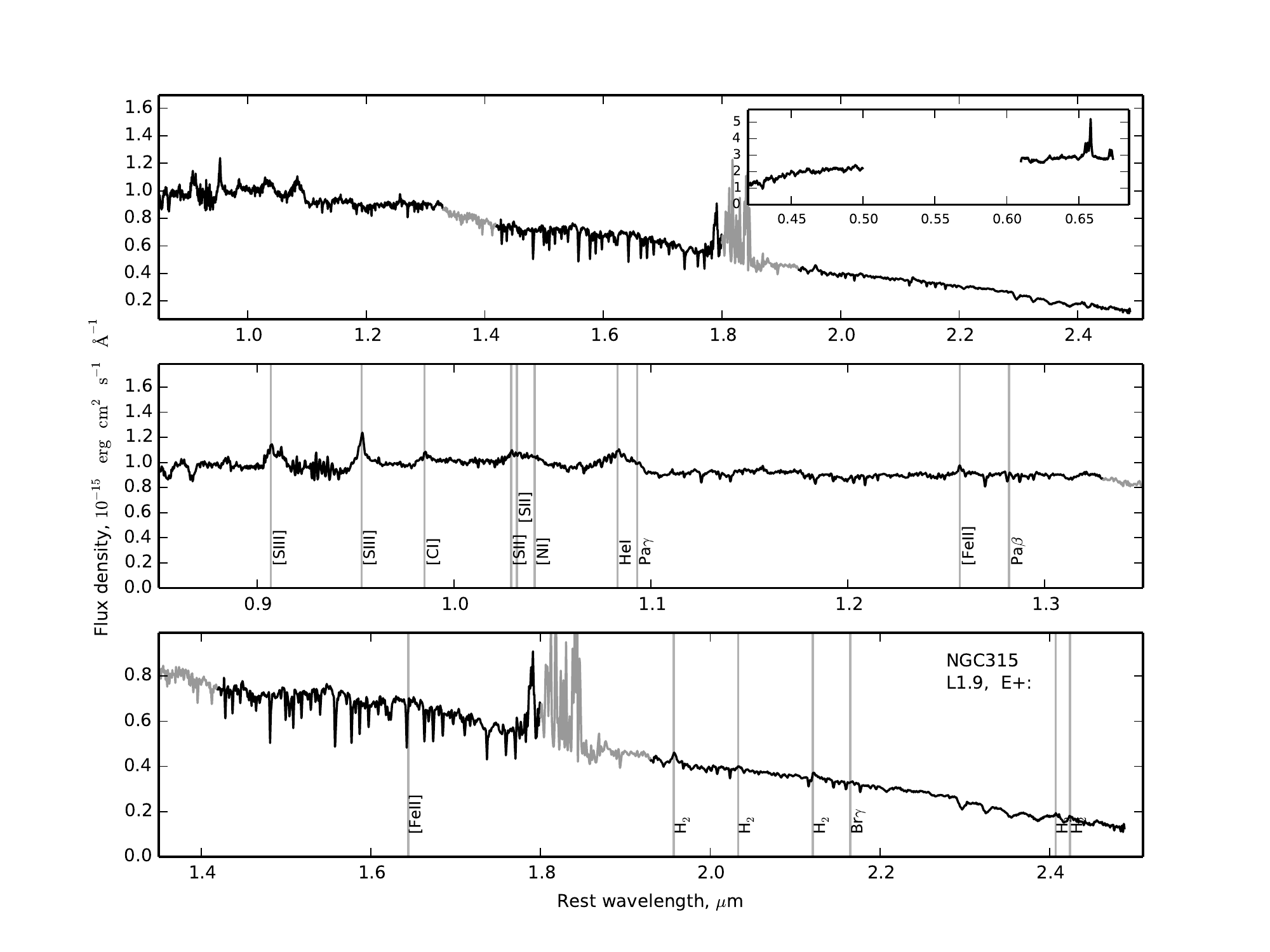}
\caption{ {\small As for Fig. \ref{f1}, for NGC~315. Removal of atmospheric OH emission is unusually poor for this source, as shown by the strong residuals appearing as negative ``spikes'' between $\sim$1.4 - 1.8 $\mu$m.}}
\label{f20}
\end{figure*}

The host galaxy of NGC~315 is a giant cD radio galaxy located in the Zwicky cluster 0107.5+3212 \citep{Zwicky96}. Two-sided, well-resolved radio jets are shown in both VLA and VLBI observations \citep{Cotton99,Venturi93}. \citet{Nagar05} reported an unresolved core in addition to the radio jet at VLBI resolutions. The high spatial resolution provided by \emph{Chandra} imaging allowed the detection of X-ray jets, the most striking being the one extending $\rm{\sim}$10\arcsec\ to the NW \citep{Donato04,Gonzalez-Martin09a}. This elliptical galaxy has a nuclear dust disk and an unresolved optical nucleus \citep{Capetti00}. At optical wavelengths the emission-line spectrum of NGC\,315 has LINER characteristics \citep{Ho95}, and the detection of a broad component to H$\rm{\alpha}$ \citep{Ho97} supports the identification of this object as a LLAGN.  The GNIRS spectrum (Fig. \ref{f20}) shows prominent CaT and CO absorption lines. Low ionisation lines of [SIII], [CI], and [Fe II] are detected, together with He I 1.083 $\mu$m, and a hint of a broad Pa$\rm{\gamma}$ emission line.  The nuclear spectrum also includes at least one weak $\rm{H_{2}}$ line in the $K$-band.  As with NGC~266, the stellar population of NGC~315 is  old, with intermediate-age stars contributing only 10\% of the optical light \citep[][; see also Zhang et al. 2008]{CidFernandes04}. \\

{\bf NGC 1052 (L1.9)} \\

\begin{figure*}
\hspace*{-10mm}
\includegraphics[scale=0.9]{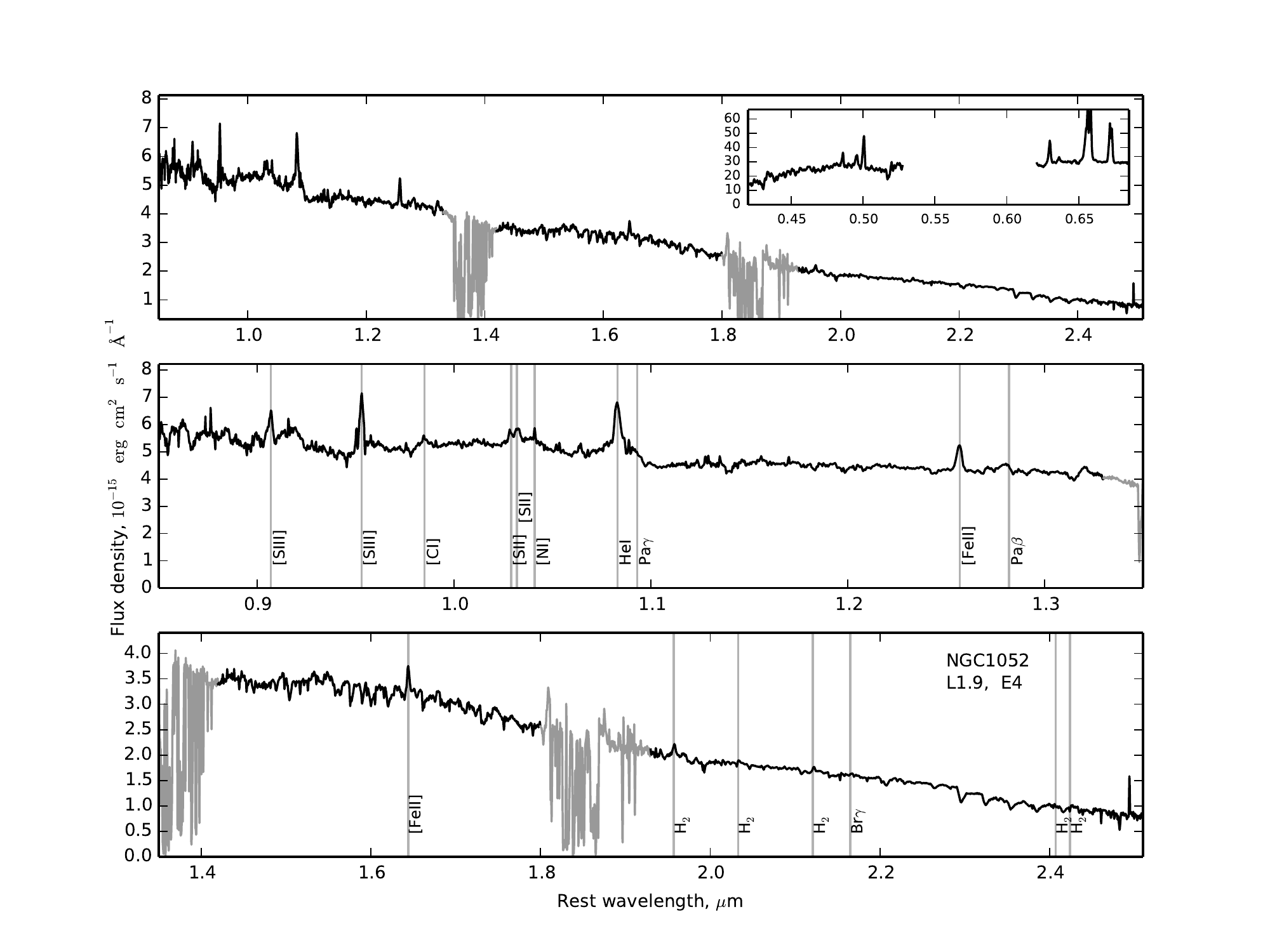}
\caption{ {\small As for Fig. \ref{f1}, for NGC~1052.}}
\label{f21}
\end{figure*}

NGC~1052 is the brightest member of a small group which, together with the NGC\,1069 group, makes up the Cetus I cluster \citep{Wiklind95}. Broad lines are clearly detected in the optical spectrum \citep{Barth99,Ho97,Balmaverde14}. NGC\,1052 also shows  $\rm H_2O$ megamaser emission \citep{Claussen98}, and variability in the radio, \citep{Vermeulen03}, ultraviolet \citep{Maoz05}, and X-rays \citep{Hernandez-Garcia13}.The X-ray data clearly indicate the presence of an unresolved nuclear source in the hard bands \citep{Satyapal05,Gonzalez-Martin06}, and  \citet{Gonzalez-Martin14} recently used artificial neural networks to show that this object might be more similar to a Seyfert galaxy than other LINERs, from the X-ray point of view. The GNIRS spectrum (Fig. \ref{f21}) contains low ionization lines of [SIII], [SII], HeI, and [Fe II]. [CI] is also present, along with possible weak, broad Pa$\rm{\gamma}$ and Pa$\rm{\beta}$. The strongest lines are from He I and [S III]. Moreover, He\,I shows some broadening. \citet{Mould12} also presented a NIR spectrum of NGC~1052, showing prominent Pa$\rm{\beta}$ and [Fe II] emission lines. In the $K$-band, the nuclear spectrum also includes $\rm{H_{2}}$ lines. The continuum shows CaT and CO absorption lines around 0.85 and 2.3 $\mu$m, as well as numerous stellar absorption features in the $H$-band.  \citep{CidFernandes04} and find that the optical emission of NGC~1052 is mostly from an old stellar population, and \citet{Zhang08} estimate a mean age of roughly 11 Gyr. \\

{\bf NGC 3718 (L1.9)} \\

\begin{figure*}
\hspace*{-10mm}
\includegraphics[scale=0.9]{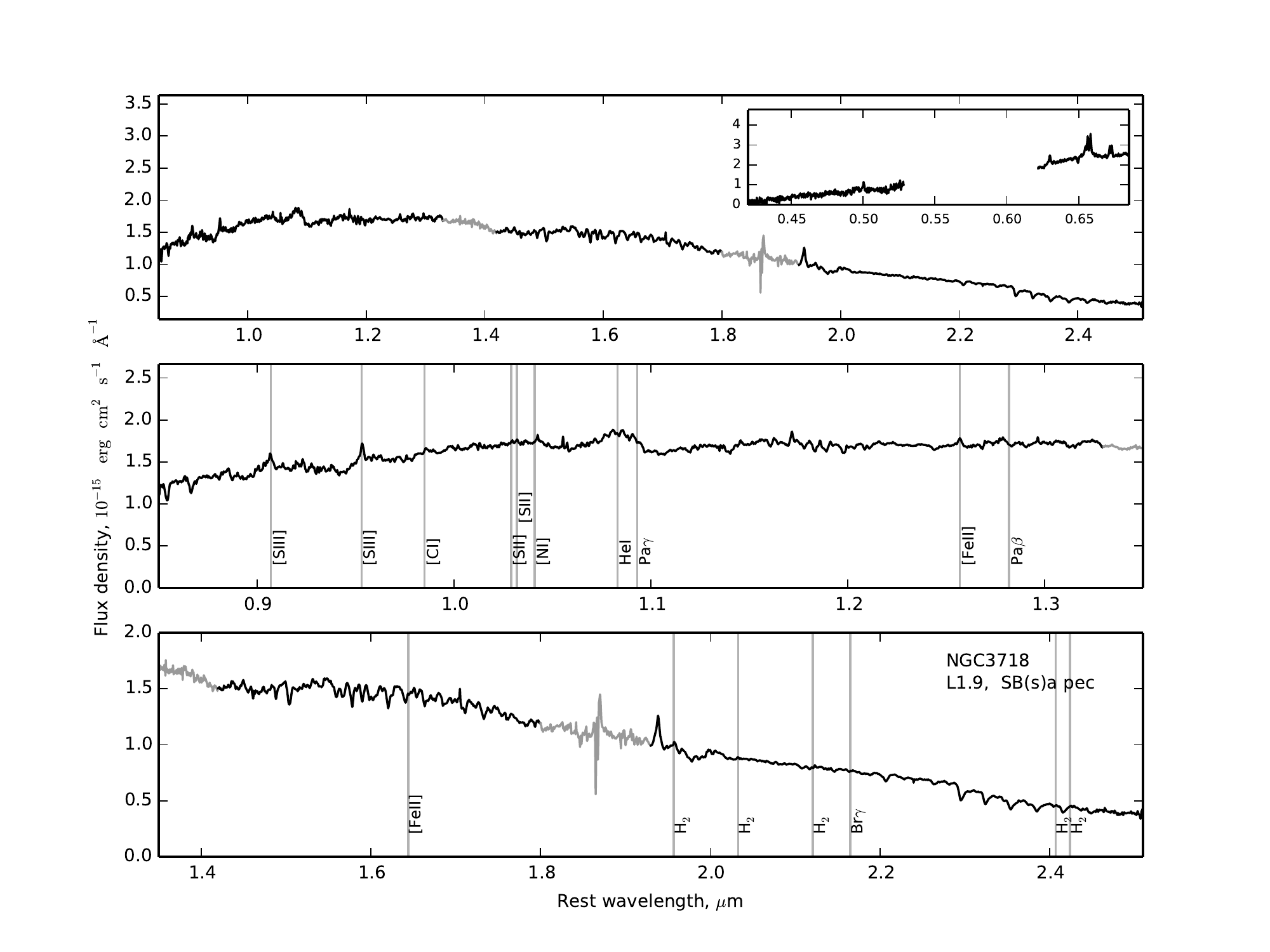}
\caption{ {\small As for Fig. \ref{f1}, for NGC~3718.}}
\label{f22}
\end{figure*}

Morphologically, NGC~3718 is an SBa galaxy with a prominent dust lane crossing its central bulge. On kpc scales it is also known for its strongly warped atomic and molecular gas disk \citep{Sparke09}. A type 1.9 LINER nucleus \citep[FWHM$_{\rm{H\alpha}}$ = 2350 km s$^{-1}$; ][]{Ho97} lies behind this dust lane, with a compact jet extending to the northwest \citep{Krips07,Nagar02}. On large scales, NGC\,3718 has a dwarf companion, NGC\,3729 \citep{Tully96}. Asymmetries found in CO(1-0) and CO(2-1) observations might be evidence for a tidal interaction with its companion at large scales and gas accretion onto the nucleus at small scales \citep{Krips05}.  
The GNIRS spectrum (Fig. \ref{f22}) is continuum-dominated, with weak emission lines of [S III], [Fe II], and He I. Stellar CaT and CO absorption bands are clearly detected near 0.85 and 2.3 $\mu$m, but little information about the stellar population of the nucleus of NGC~3718 is available in the literature. \\

{\bf NGC 3998 L1.9} \\

\begin{figure*}
\hspace*{-10mm}
\includegraphics[scale=0.9]{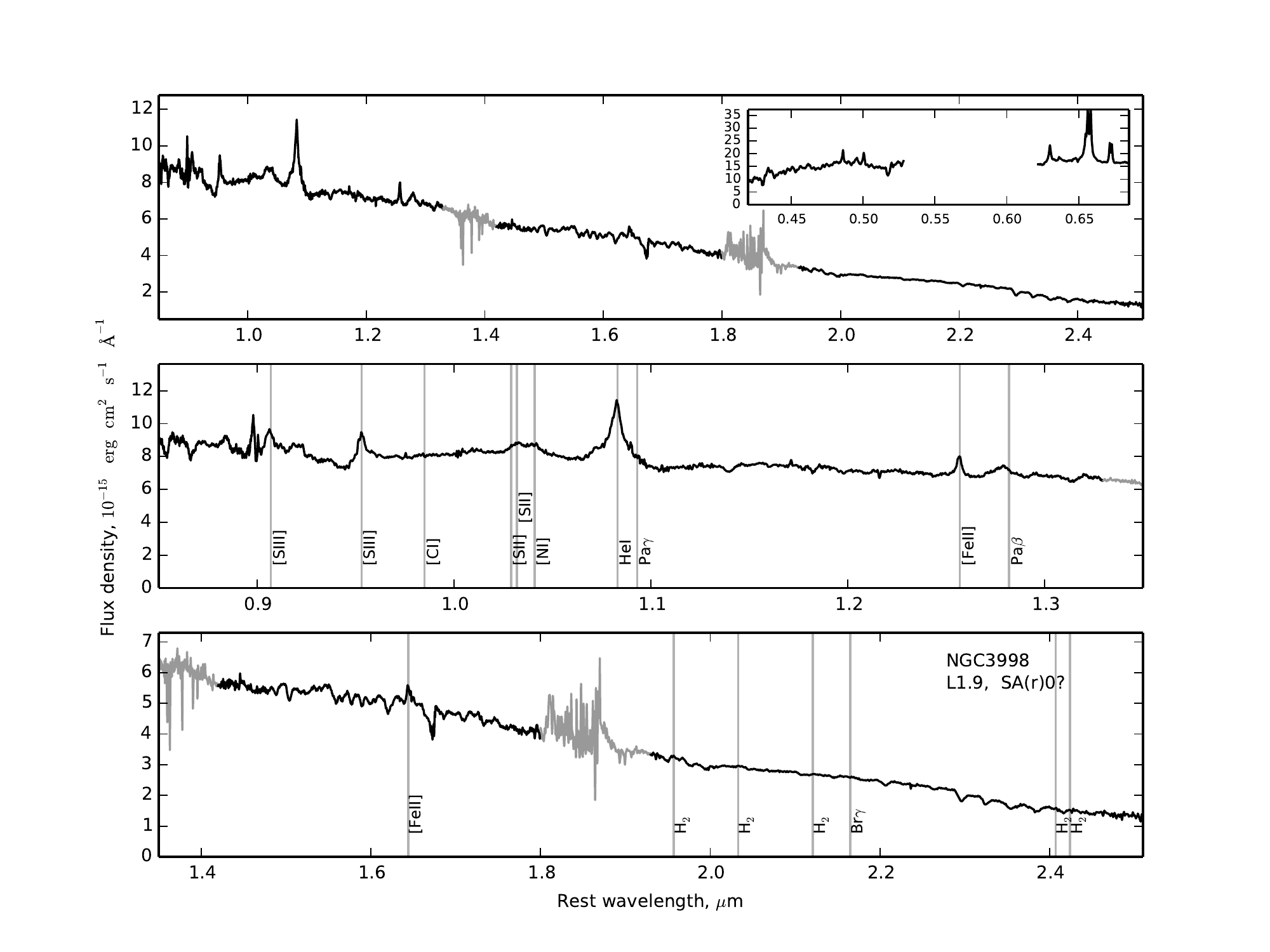}
\caption{ {\small As for Fig. \ref{f1}, for NGC~3998.}}
\label{f23}
\end{figure*}

Five galaxies (NGC\,3990, NGC\,3977, NGC\,3972, NGC\,3982, and UGC\,6919) are seen within a projected distance of 250 kpc of this S0 galaxy, all but one at redshifts consistent with sharing the same physical association \citep{Gonzalez-Martin09a}. A clear detection of a broad H$\rm{\alpha}$ line was reported by \citet{Heckman80} and \citet{Ho97}. Long-term variability is widely reported at radio \citep{Hummel84,Filho02}, UV \citep{Maoz05}, and X-ray \citep{Younes12,Hernandez-Garcia13} frequencies. The nuclear stellar population of NGC~3998 appears very young, being dominated by stars of 10$^{6}$ -- 10$^{7}$ yr, although the blue continuum emission of the AGN in the high spatial resolution data used for the analysis renders this uncertain \citep{GonzalezDelgado04}. The GNIRS spectrum of NGC\,3998 (Fig. \ref{f23}) is full of emission lines, with the most prominent being He I 1.08 $\mu$m (showing a clear broad component), [S III], and [Fe II]. [S II] and [NI] are also detected, and Pa$\rm{\gamma}$ and Pa$\rm{\beta}$ emission lines may also be present. At high angular resolution, the IR SED of this object is red and can be explained by emission from warm dust \citep{Mason13}, but in these lower-resolution data the continuum shape is characteristic of stellar emission. The continuum is imprinted with the CaT and CO absorption bands, as well as numerous other weak, stellar features. \\

{\bf NGC 4203 (L1.9)} \\

\begin{figure*}
\hspace*{-10mm}
\includegraphics[scale=0.9]{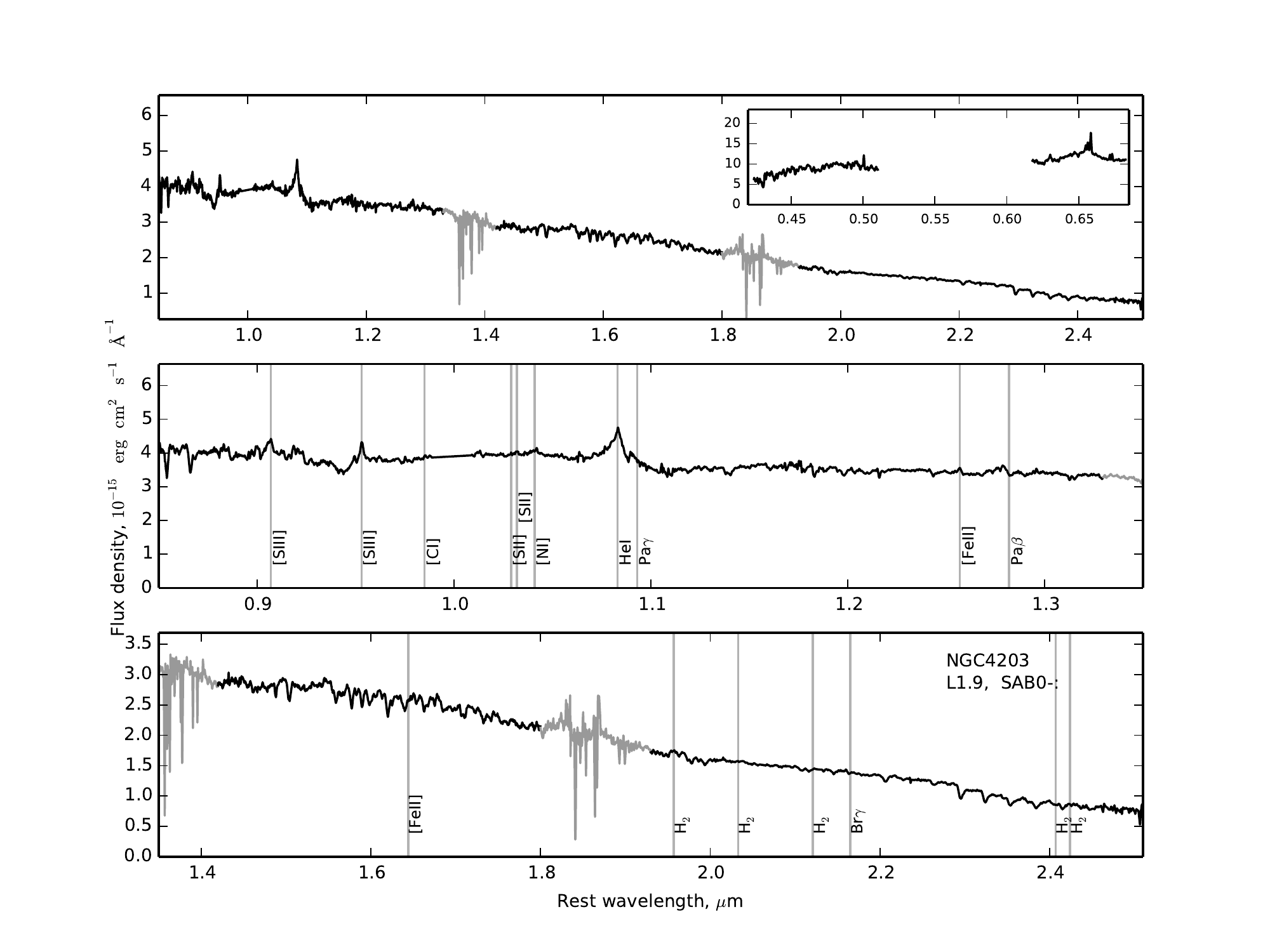}
\caption{ {\small As for Fig. \ref{f1}, for NGC~4203. A prominent detector artefact near 1.0 $\mu$m has been interpolated over in this spectrum.}}
\label{f24}
\end{figure*}

This AGN is hosted by a SAB0 galaxy seen nearly face-on, with a few dust lanes visible in the bulge. The LINER has prominent, double-peaked broad H$\alpha$ lines \citep{Shields00} and fairly symmetric optical emission lines \citep{Ho97}. The AGN is also detected at various other wavelengths: as a point source in X-rays \citep{Ho01}, as a variable UV source \citep{Maoz05}, and as a point source with an inverted radio spectrum \citep{Nagar00}. \citet{Mould12} obtained an infrared spectrum with TripleSpec on the 5-m telescope at Palomar, and report that no emission lines were detected despite a S/N near 100. Our spectrum (Fig. \ref{f24}) has a broad HeI line and [SIII] emission lines, probably made visible by the lesser dilution of the AGN spectrum in the 0.3\arcsec\ slit used for this work (cf. 1\arcsec\ used by Mould et al.). CaT and CO absorption bands are also visible, as well as numerous stellar absorption lines in the $H$-band. The nuclear stellar population of NGC~4203 has been examined by \citet{Sarzi05}. They find roughly equal ($\sim$45\%) contributions from young ($\le 10^{7.5}$ yr) and old ($10^{10}$ yr) components, but they note that the AGN featureless continuum in their high angular resolution data could be an alternative explanation for the apparent presence of large numbers of young stars. \\

{\bf NGC 4450 (L1.9)} \\

\begin{figure*}
\hspace*{-10mm}
\includegraphics[scale=0.9]{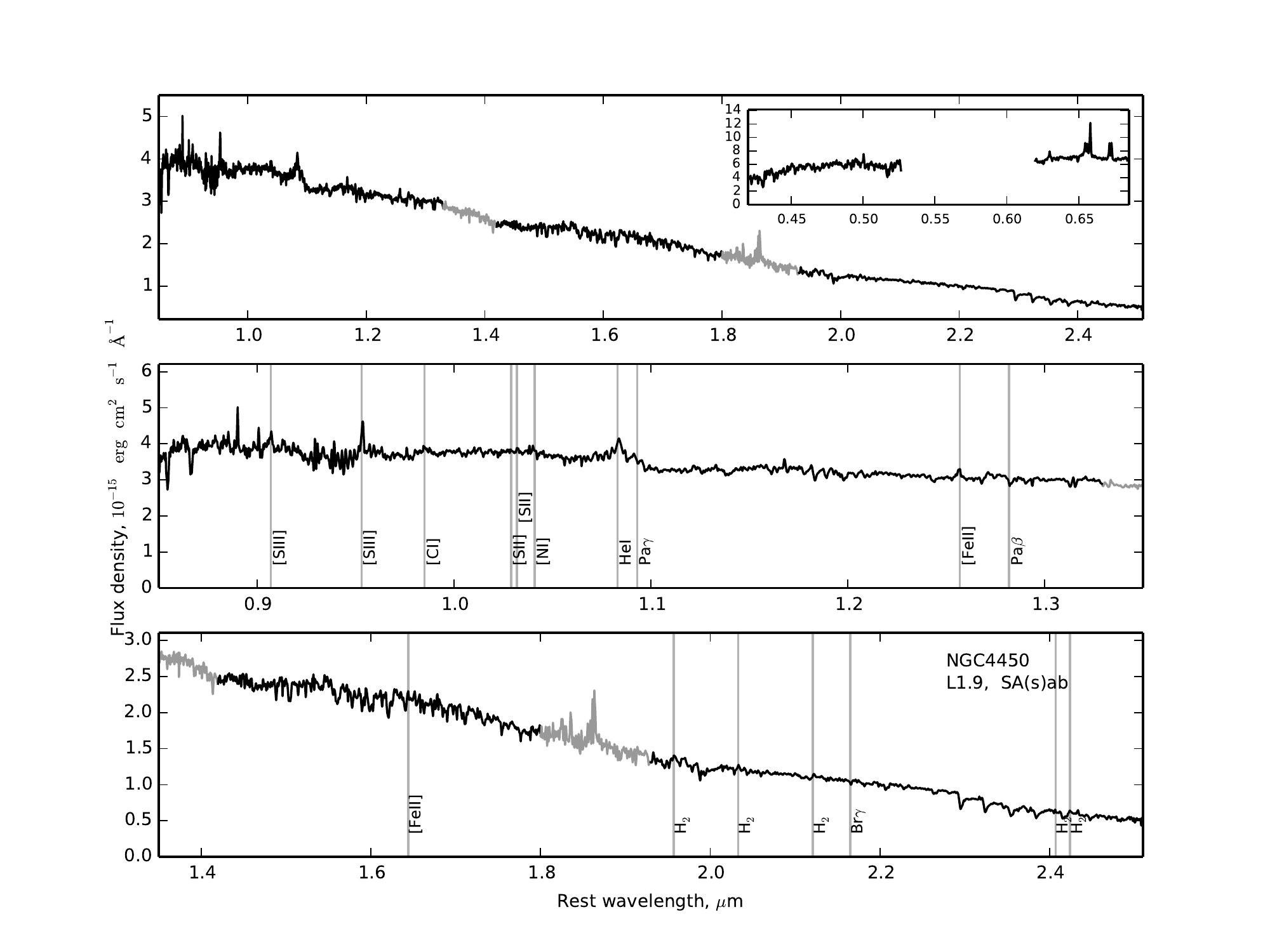}
\caption{ {\small As for Fig. \ref{f1}, for NGC~4450.}}
\label{f25}
\end{figure*}

Hosted in an SA(s)ab galaxy with medium inclination and soft spiral arms, this LINER has asymmetric narrow optical lines and a double-peaked, broad H$\alpha$ line \citep{Ho00}. VLA images reveal extended emission around the core \citep{Anderson05}, whose radio spectrum is flat or mildly inverted \citep{Ho01c}. The GNIRS spectrum (Fig. \ref{f25}) has a few emission lines ([SIII], HeI, [FeII]) and stellar absorption bands including CaT and CO.  Like NGC~4203, the nuclear stellar population of NGC~4450 has been examined by \citet{Sarzi05}. They find that young ($\le 10^{7.5}$ yr) and old ($\ge 10^{10}$ yr) stars contribute roughly 40\% and 60\% of the optical light, respectively, but they note that the AGN featureless continuum in their high angular resolution data could be an alternative explanation for the apparent presence of large numbers of young stars. \\

{\bf NGC 4750 (L1.9)} \\

\begin{figure*}
\hspace*{-10mm}
\includegraphics[scale=0.9]{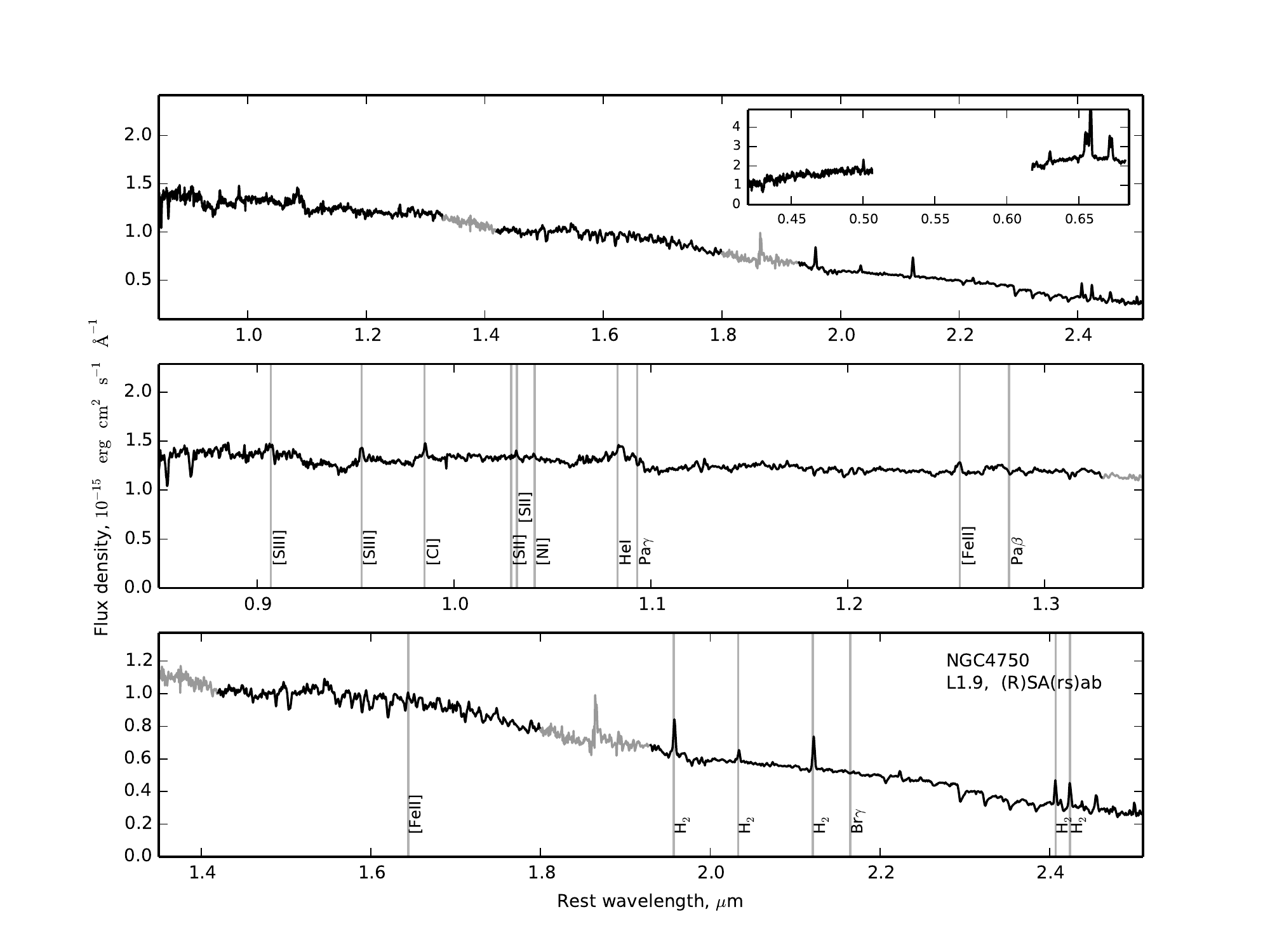}
\caption{ {\small As for Fig. \ref{f1}, for NGC~4750.}}
\label{f26}
\end{figure*}

\begin{figure*}
\hspace*{-10mm}
\includegraphics[scale=0.9]{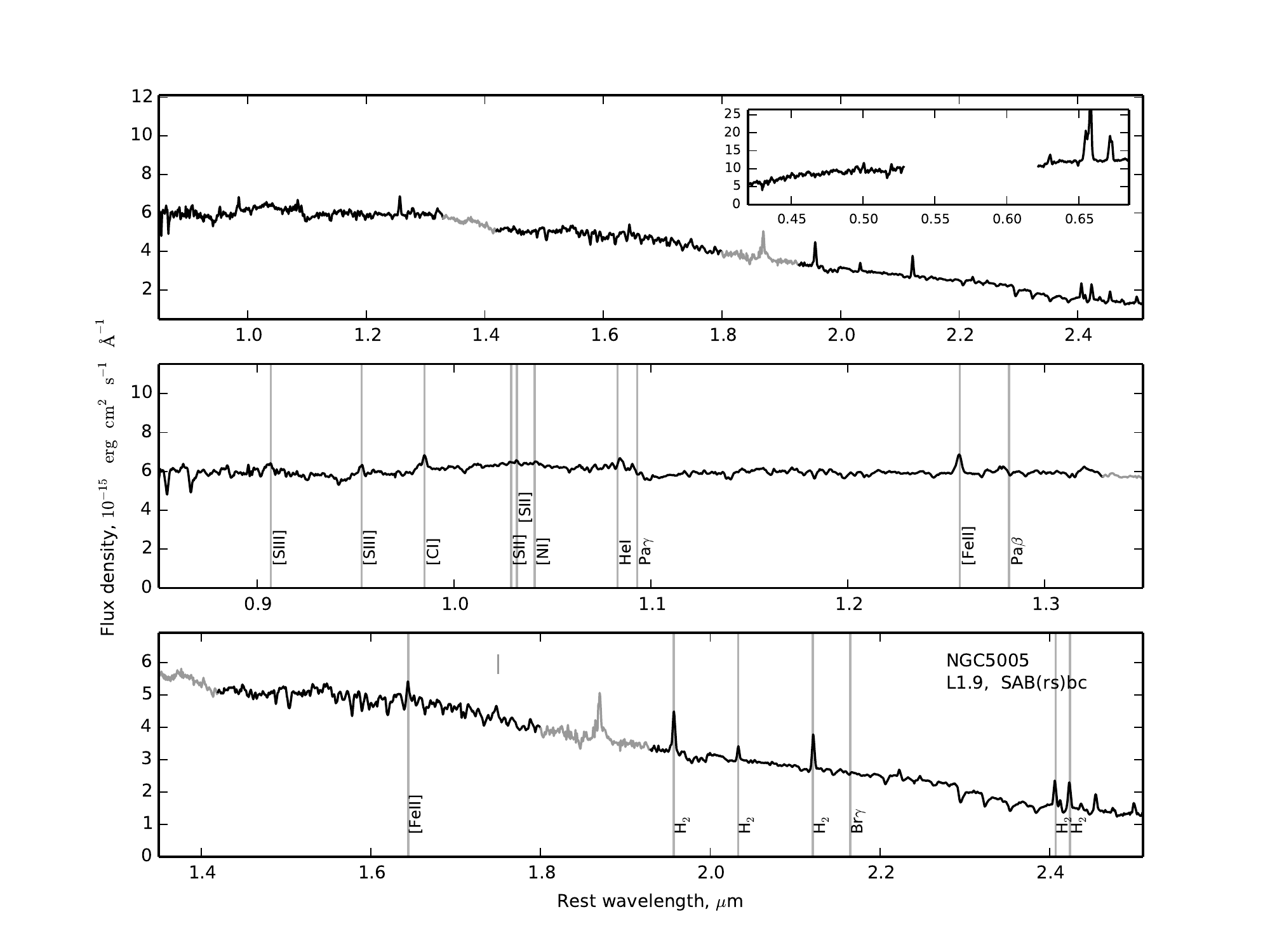}
\caption{ {\small As for Fig. \ref{f1}, for NGC~5005. The short vertical line near 1.73 $\mu$m indicates an artefact from division by the standard star that is common to several of the spectra.}}
\label{f27}
\end{figure*}

The host of this AGN has an unusual morphology and is classified as a (R)SA(rs)ab galaxy. It has a very small, very bright nucleus with a nuclear bar. Additionally, images of the galaxy show a ring completely detached from the nucleus and probably consisting of two overlapping spiral arms \citep{Erwin04}. \citet{Ho97b} detect faint wings in the H$\alpha$ profile of this LINER and otherwise very symmetric narrow emission lines. Several emission lines (including [SIII, [CI], [FeII], and H$_2$) are detected in the GNIRS spectrum (Fig. \ref{f26}), which rather closely resembles that of NGC~4450 except for the presence of the fairly strong H$_2$ emission in the $K$-band. CaT, CO, and many other stellar absorption features are also present. \\

{\bf NGC 5005 (L1.9)} \\

NGC~5005 is a weakly barred spiral galaxy and part of a pair that also includes NGC~5033 \citep{Helou82}. It is optically classified as a type 1.9 LINER \citep{Ho97}, and it has an intermediate-age stellar population contributing about 45\% of the nuclear optical light \citep{CidFernandes04}. The presence of a weak AGN is confirmed by the detection of a compact hard X-ray source \citep{Gonzalez-Martin06,Dudik09}. $H$- and $K$-band spectra of NGC~5005 were analysed by \citet{Bendo04}, who noted the presence of H$_2$ and [FeII] emission and the absence of detectable Br$\gamma$. They interpreted the H$_2$ and [FeII] lines as indicative of shocks, which could be caused by supernovae, interaction with NGC~5033, or gas outflow from the AGN. All of the lines noted by \citet{Bendo04} are visible in the GNIRS spectrum of NGC~5005, along with additional H$_2$ and [FeII] lines outside their spectral coverage. Emission from [SIII] and HeI is also detected, along with CaT and CO absorption, and numerous stellar absorption lines in the $H$-band. \\

\pagebreak

\subsection{LINER 2}

{\bf NGC 404 (L2)} \\

\begin{figure*}
\hspace*{-10mm}
\includegraphics[scale=0.9]{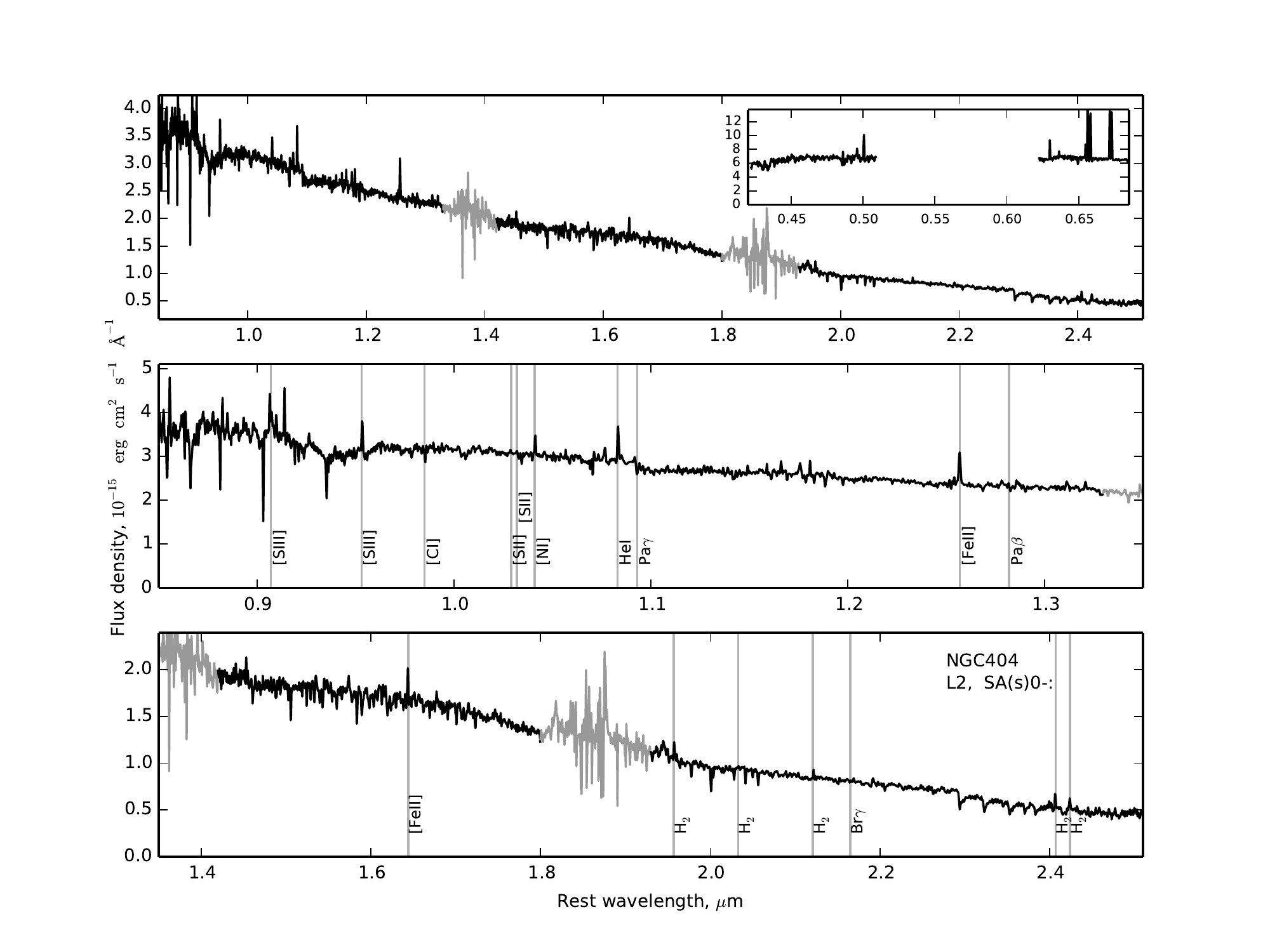}
\caption{ {\small As for Fig. \ref{f1}, for NGC~404.}}
\label{f28}
\end{figure*}

While the hard X-ray point source at the center of NGC~404 is faint enough that it could be an X-ray binary, the spectral shape of the X-ray emission and the radio characteristics of this object suggest the presence of an accreting nuclear black hole \citep{Binder11,Nyland12}. Modelling of the stellar kinematics implies that the mass of the black hole is relatively low, $\rm M_{BH} <1 \times 10^{5} \rm M_\odot$ \citep{Seth10}. The nuclear star cluster of NGC~404 is fairly young, with stars $\sim$1 Gyr old accounting for over half of the optical light in a 1\arcsec\ slit \citep[][see also Cid Fernandes et al. 2004]{Seth10}. The GNIRS spectrum of NGC~404 (Fig. \ref{f28}) shows emission from [SIII], [NI], HeI, [FeII], and H$_2$, with stronger and more numerous lines than many of the other type 2 LINERs in this sample. This is probably a result of the galaxy's proximity (D=2.4 Mpc), meaning that the nuclear emission is less diluted by stellar emission. Despite the presence of a 1 Gyr-old stellar cluster of roughly solar metallicity, deep molecular absorption features from TP-AGB stars, predicted by some stellar population models \citep{Maraston05}, are not visible in the IR spectrum.  \\

{\bf NGC 474 (L2::)} \\

\begin{figure*}
\hspace*{-10mm}
\includegraphics[scale=0.9]{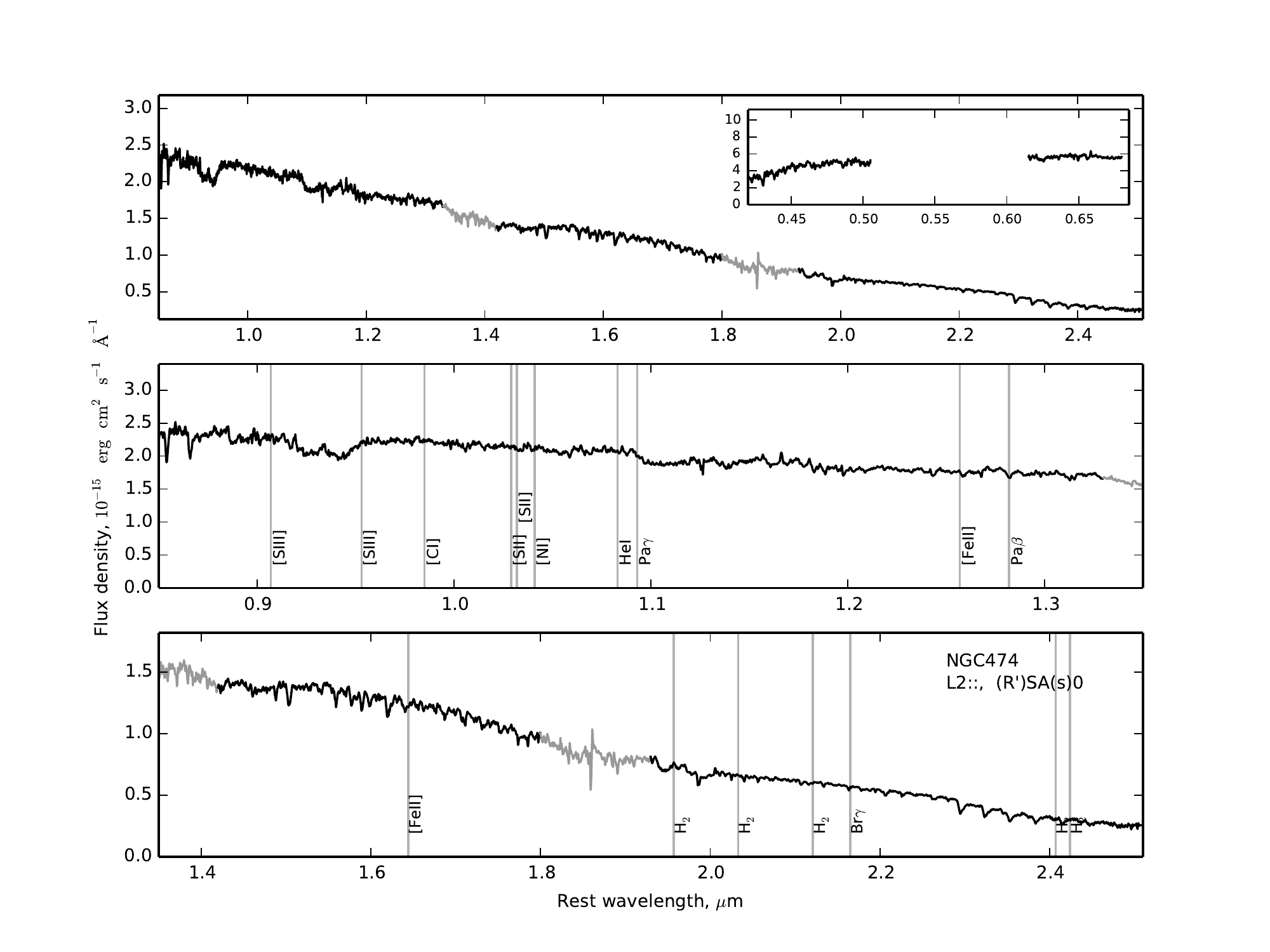}
\caption{ {\small As for Fig. \ref{f1}, for NGC~474.}}
\label{f29}
\end{figure*}

NGC~474 (Arp 227) is a lenticular galaxy surrounded by arcs or shells of stars \citep{Turnbull99} that were possibly formed in an interaction with neighboring NGC~470, or in a past merger event. The central regions of the galaxy are old ($\sim$9 Gyr), while the age of the stellar population decreases towards the galaxy's outskirts \citep{Kuntschner10}. The classification of NGC~474 as a type 2 LINER is highly uncertain, and its optical emission lines are weak \citep{Ho97}. If an AGN is present in NGC~474, it may be deeply obscured \citep{Gonzalez-Martin09b}. No emission lines are detected in the IR spectrum (Fig. \ref{f29}), and the good S/N and excellent inter-band telluric line removal in the data clearly reveal the nuclear stellar population. Molecular absorption features at $\sim$0.93 $\mu$m (probably a blend of TiO, ZrO and CN) and 1.1 $\mu$m (CN) are relatively strong in this object. \\

{\bf NGC 1961 (L2)} \\

\begin{figure*}
\hspace*{-10mm}
\includegraphics[scale=0.9]{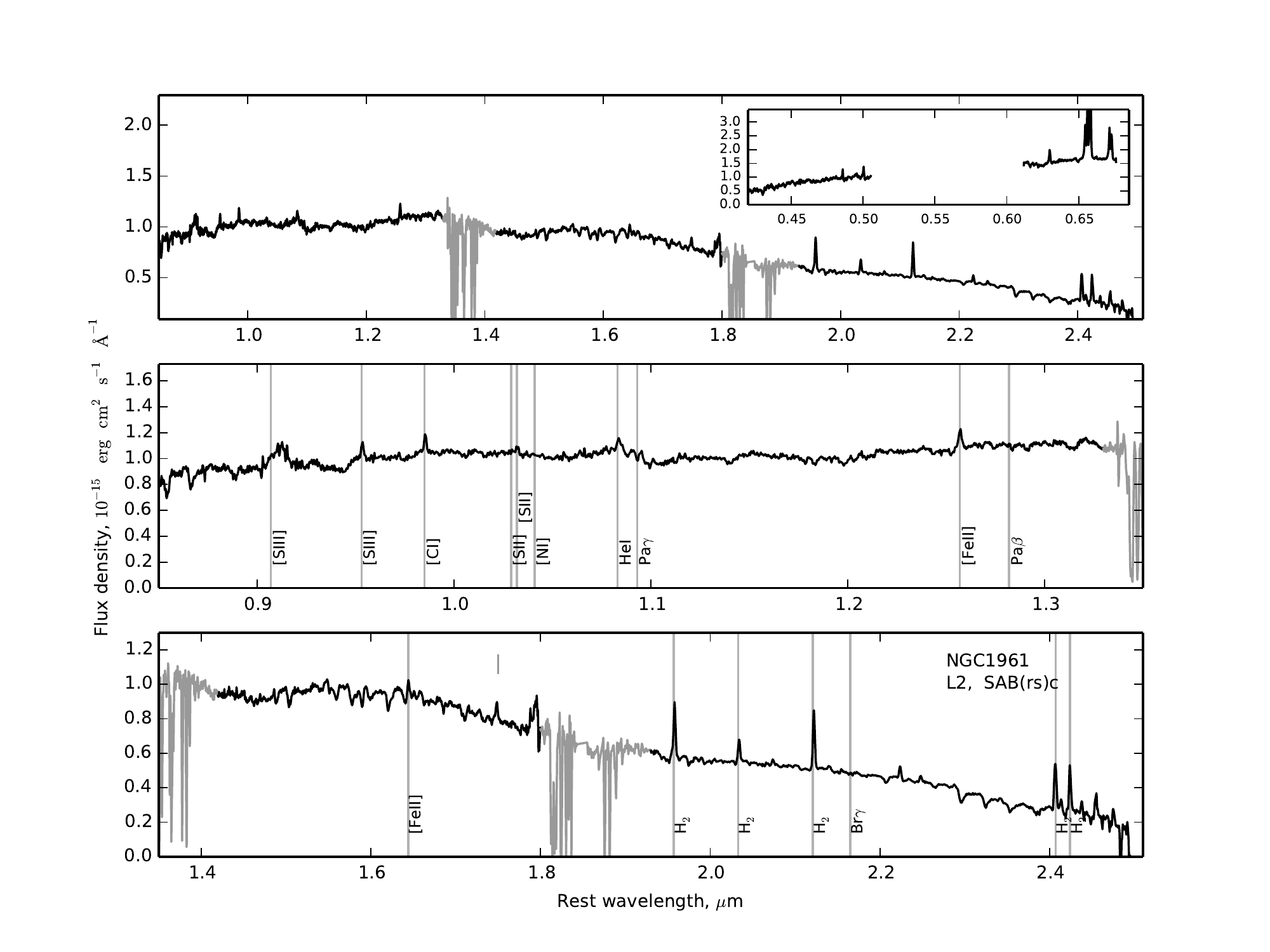}
\caption{ {\small As for Fig. \ref{f1}, for NGC~1961. The short vertical line near 1.73 $\mu$m indicates an artefact from division by the standard star that is common to several of the spectra.}}
\label{f30}
\end{figure*}

Studies of the CO emission in NGC~1961 (UGC~3334) suggest that the galaxy once collided with a companion, and that its disk has since become strongly warped \citep{Combes09}. Consideration of the radio and infrared properties of this galaxy led to its classification as a starburst by \citet{Condon91}, and it has been host to several recent supernovae. Dust lanes are visible in the optical image of this inclined, SABc galaxy \citep{Sandage94}, and nondetection of a nuclear UV source also suggests that it may be fairly dusty \citep{Barth98}. This is consistent with the downturn in flux at short wavelengths in the GNIRS spectrum (Fig. \ref{f30}). Lines from species such as [SIII], [FeII] and HeI are present, although fairly weak, and numerous, prominent emission lines of H$_{2}$ are present in the $K$-band. \\

{\bf NGC 2768 (L2)} \\

\begin{figure*}
\hspace*{-10mm}
\includegraphics[scale=0.9]{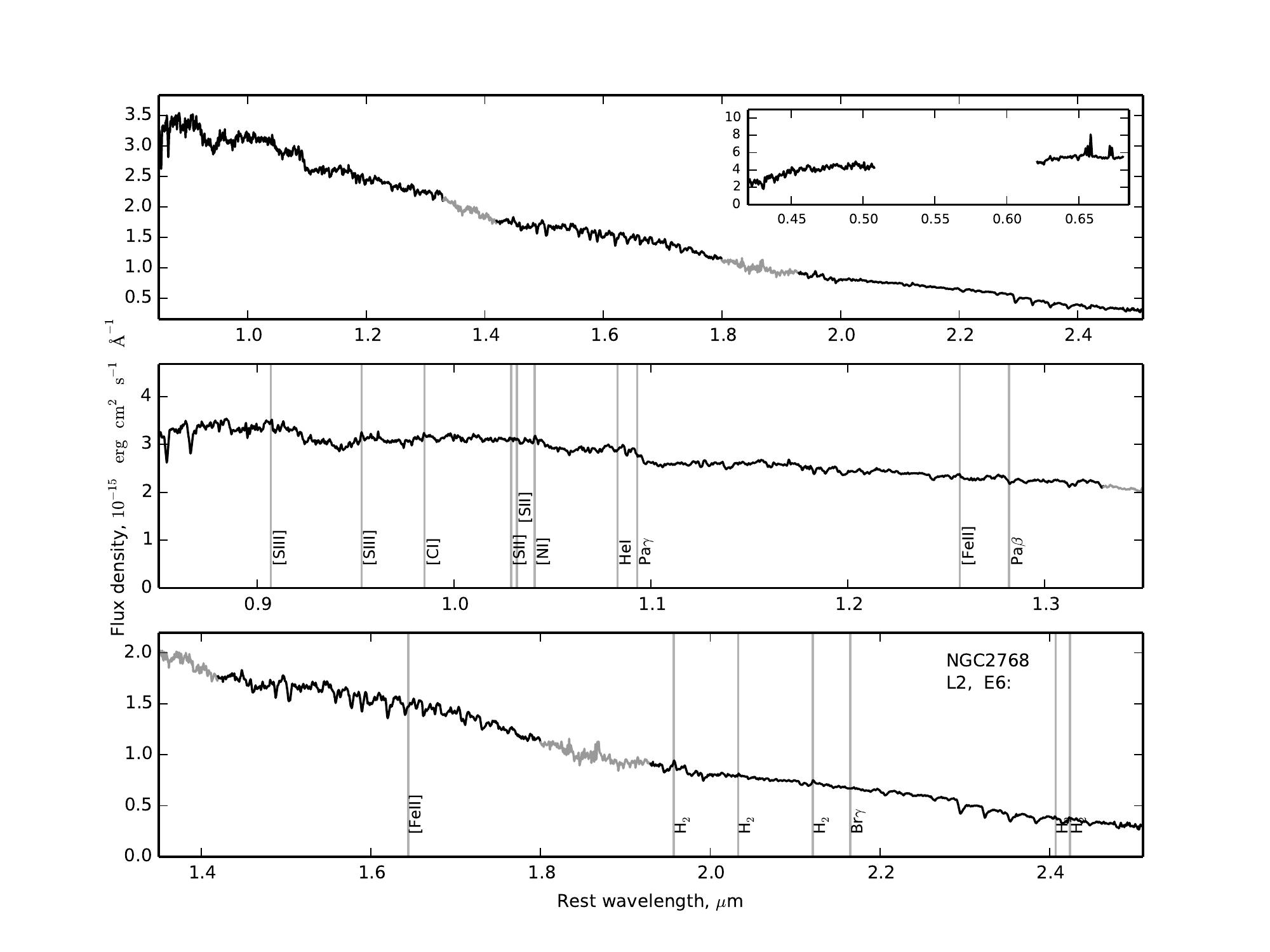}
\caption{ {\small As for Fig. \ref{f1}, for NGC~2768.}}
\label{f31}
\end{figure*}

This polar ring elliptical galaxy is detected in CO, and may have acquired material during a tidal accretion event \citep{Crocker08}. \citet{Silchenko06} derive a luminosity-weighted mean age of 11 Gyr for the galaxy's nucleus, although \citet{Crocker08} and \citet{Kuntschner10} find evidence of a star forming episode occurring a few Gyr ago, \citet{Zhang08} find a mean age of $\sim$6 Gyr, and \citet{Serra08} estimate a mean age of roughly 4 Gyr. In any case, there is little evidence of ongoing star formation. NGC~2768 is host to a compact radio source and variable X-ray nucleus \citep{Nagar05,Filho06,Komossa99}, and there are weak emission lines in its optical spectrum \citep{Ho95}. The GNIRS spectrum (Fig. \ref{f31}) is practically devoid of emission lines, with a couple of H$_2$ perhaps weakly detected. The continuum shape and absorption features are broadly similar to those of several of the other type 2 LINERs in this sample. \\

{\bf NGC 2832 (L2::)} \\

\begin{figure*}
\hspace*{-10mm}
\includegraphics[scale=0.9]{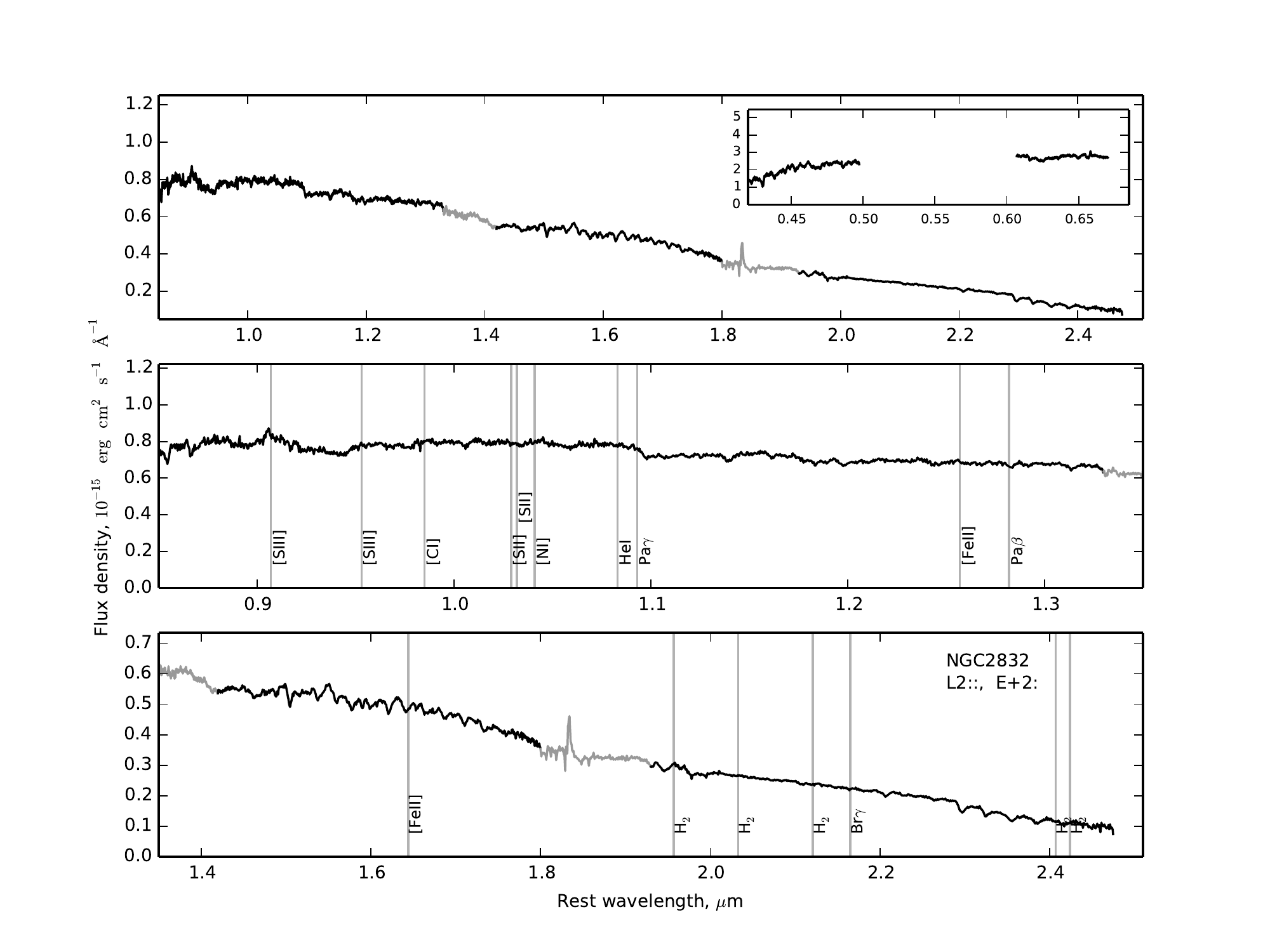}
\caption{ {\small As for Fig. \ref{f1}, for NGC~2832.}}
\label{f32}
\end{figure*}

NGC~2832 is an elliptical galaxy and the brightest galaxy of the Abell 779 cluster. It may be interacting with a satellite galaxy, NGC~2831 \citep{Lauer88}. The object has some weak optical line emission, leading to an uncertain LINER 2 classification \citep{Ho97}, but evidence of AGN activity has not been forthcoming \citep[for instance, the nucleus has not been detected in VLBA and VLA observations;][]{Liuzzo10,Nagar05}.
The central stellar population of NGC~2832 is found to be old, with derived ages of 7.5 --  13 Gyr \citep{Proctor02,Sanchez-Blazquez06,Zhang08,Loubser09}. No emission lines are detected in the IR spectrum (Fig. \ref{f32}), which otherwise broadly resembles those of the other type 2 LINERs presented here. \\

{\bf NGC 3169 (L2)} \\

\begin{figure*}
\hspace*{-10mm}
\includegraphics[scale=0.9]{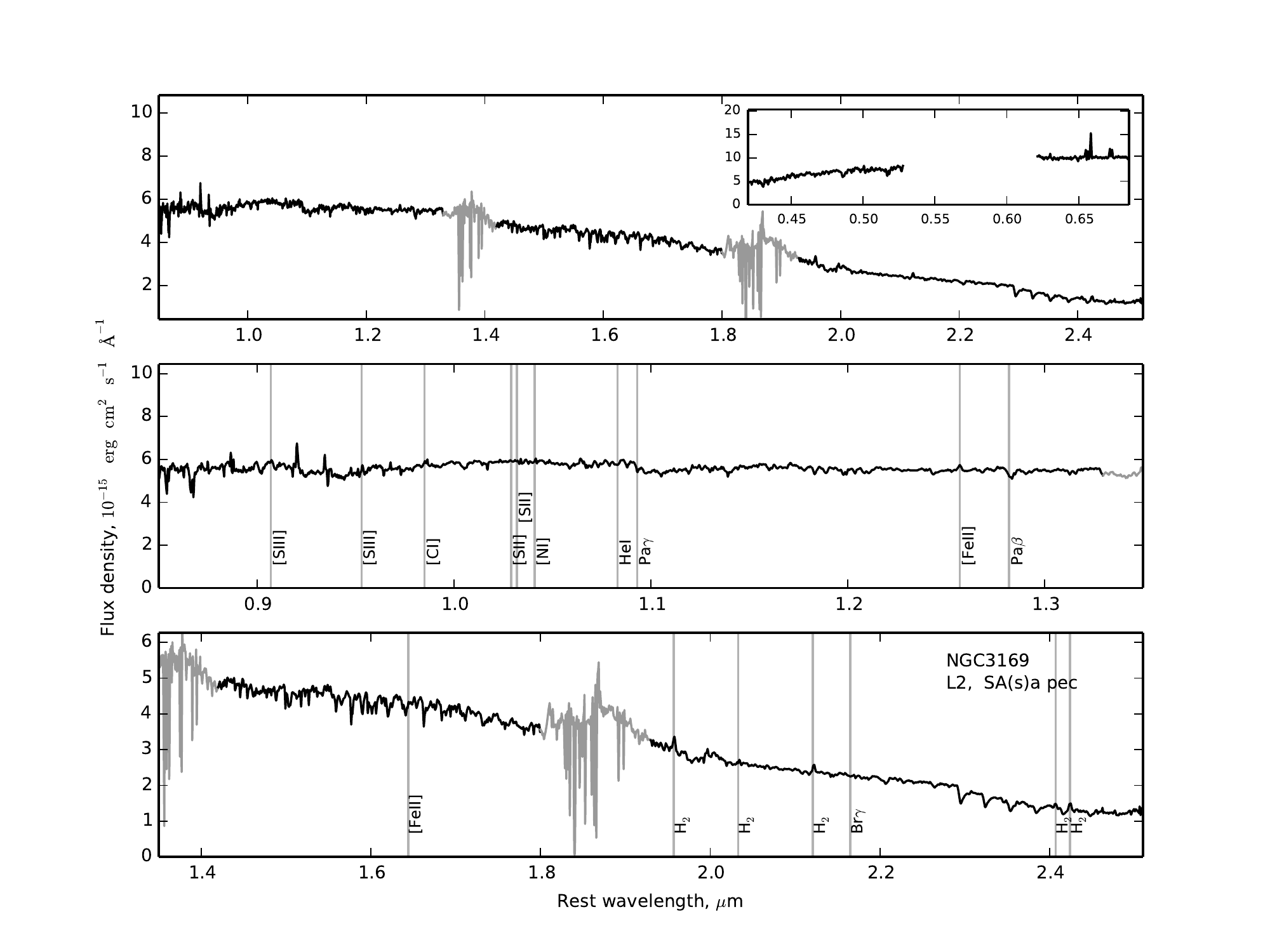}
\caption{ {\small As for Fig. \ref{f1}, for NGC~3169.}}
\label{f33}
\end{figure*}

The Sa galaxy NGC~3169 is part of a gas-rich interacting group that also includes NGC~3166 and NGC~3156. NGC~3169 hosts a low-luminosity AGN, revealed by the presence of a hard X-ray point source \citep{Satyapal05} and compact, high brightness temperature radio source \citep{Falcke00,Nagar05}. According to \citet{CidFernandes04}, the optical emission is predominantly from old stars, although with a minor contribution (21\%) from an intermediate-age (10$^8$ - 10$^9$ yr) population \citep[see also][]{Silchenko06}. 
There is weak H$_2$ emission in the GNIRS spectrum of NGC~3169 (Fig. \ref{f33}), and possibly [FeII] 1.257 $\mu$m as well. The continuum shows absorption bands of both atomic and molecular origin, qualitatively similar to those observed in many of the other type 2 LINERs in this sample. The flattening at the short wavelength end is probably due to extinction, as the nucleus of this object contains chaotic, dusty structures \citep{GonzalezDelgado08}. \\

{\bf NGC 3190 (L2)} \\

\begin{figure*}
\hspace*{-10mm}
\includegraphics[scale=0.9]{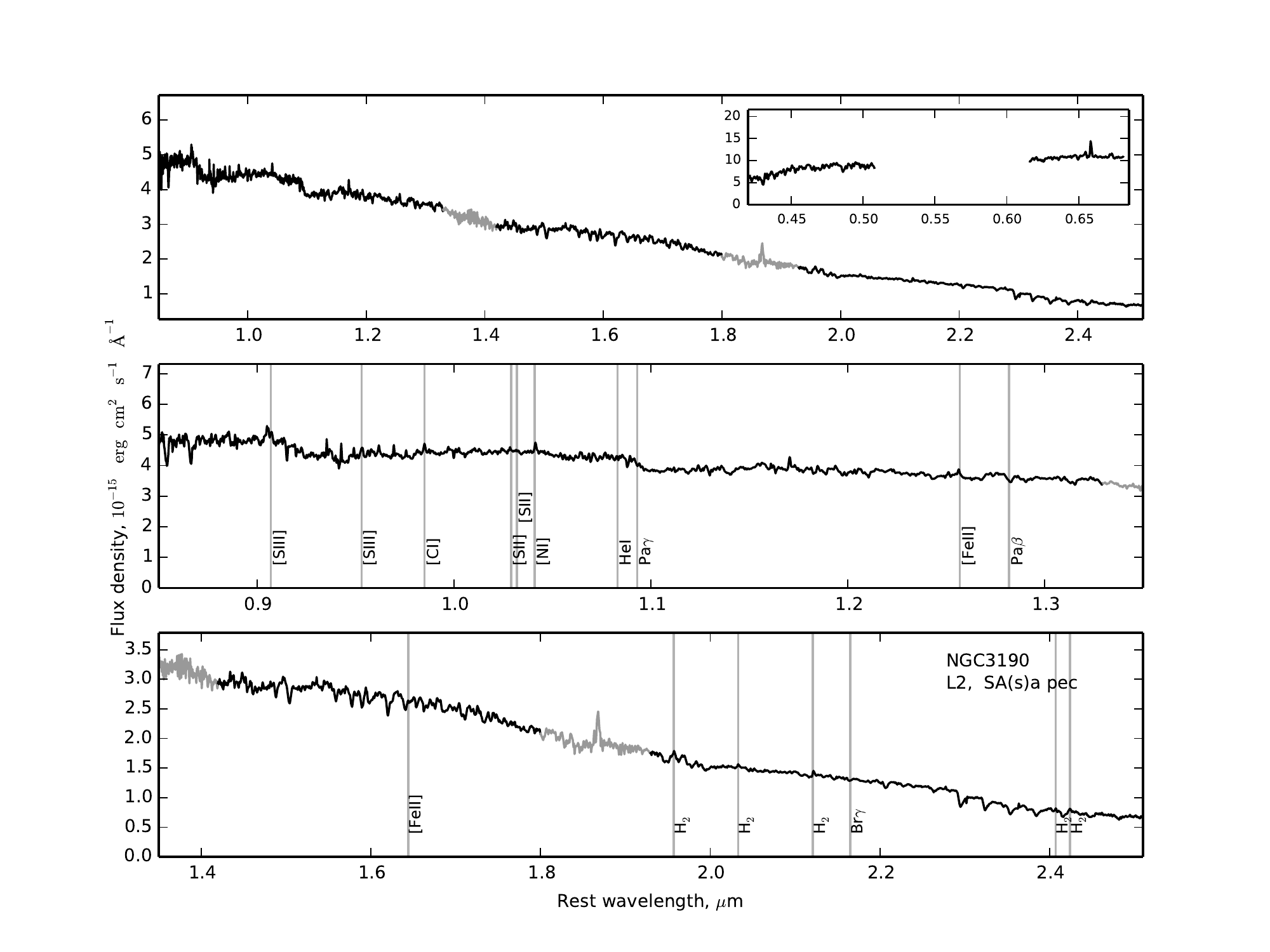}
\caption{ {\small As for Fig. \ref{f1}, for NGC~3190.}}
\label{f34}
\end{figure*}

Very little-studied to date, NGC~3190 is a highly-inclined spiral galaxy and part of the HGC44 group.
In common with several other LINER 2s in this sample, the IR spectrum of this object (Fig. \ref{f34}) shows just a handful of very weak emission lines, in this case from [CI], [NI], FeII, and H$_2$. \\

{\bf NGC 3607 (L2)} \\

\begin{figure*}
\hspace*{-10mm}
\includegraphics[scale=0.9]{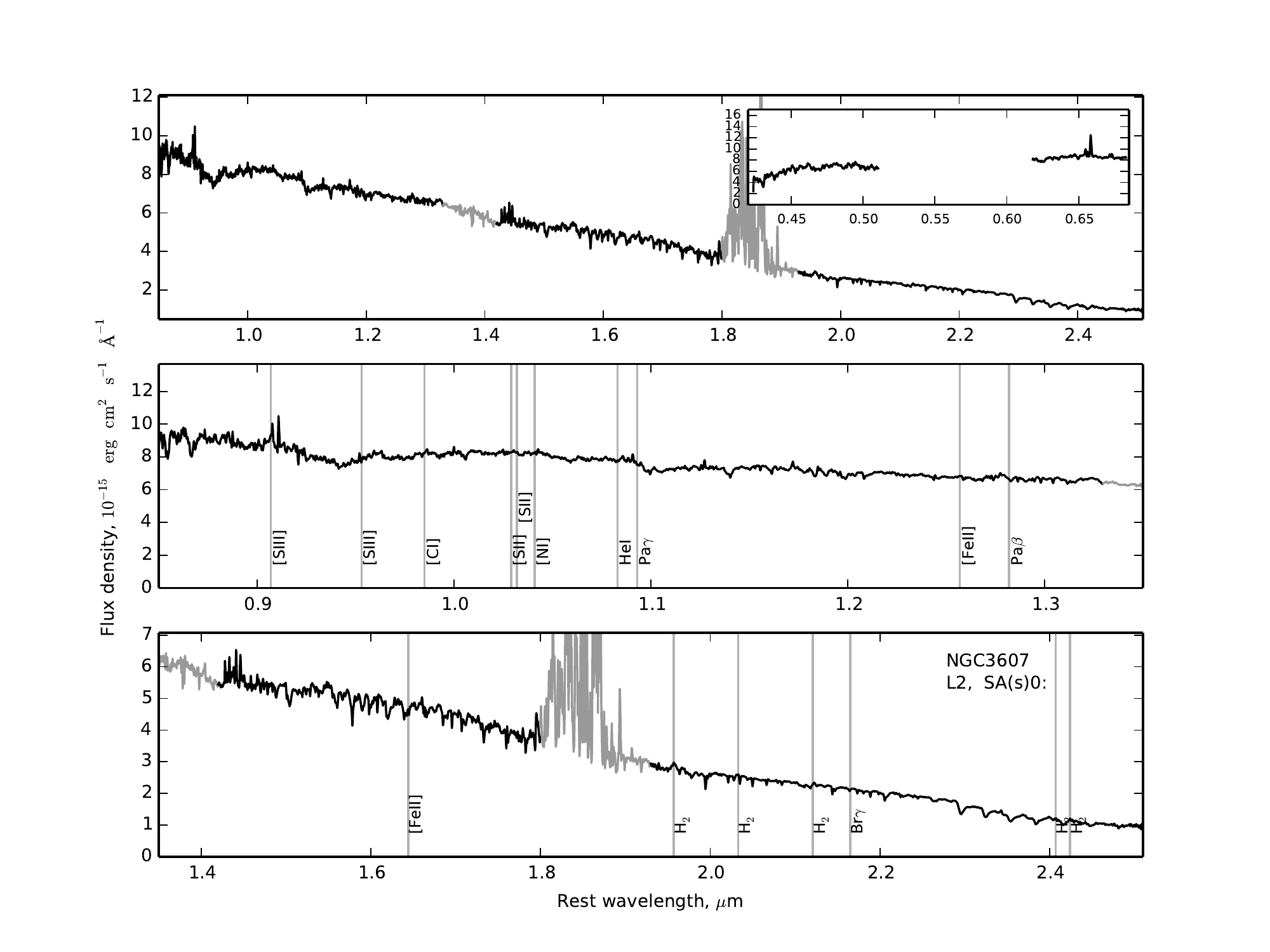}
\caption{ {\small As for Fig. \ref{f1}, for NGC~3607.}}
\label{f35}
\end{figure*}

Along with NGC~3605 and NGC~3608, NGC~3607 forms part of the G49 (Leo II) group \citep{deVaucouleurs75}. X-ray observations show no clear evidence for the presence of an AGN \citep{Terashima02,Flohic06,Gonzalez-Martin09a}.
Estimates of the age of the central stellar population vary.  \citet{Rickes09} find that most of the optical emission arises in components of $\ge$5 Gyr, and \citet{Proctor02} find a mean age of 5.6 Gyr.  \citet{Annibali07} and \citet{Terlevich02} find younger average ages of $\sim$3 Gyr, but \citet{Silchenko06} estimate a luminosity-weighted age of 12 Gyr. The IR spectrum of NGC~3607 (Fig. \ref{f35}) is qualitatively similar to those of many of the other type 2 LINERs in this sample, and only one or two weak emission lines (from [SIII] and perhaps H$_2$) are detected. \\
 
{\bf NGC 4346 (L2::)} \\

\begin{figure*}
\hspace*{-10mm}
\includegraphics[scale=0.9]{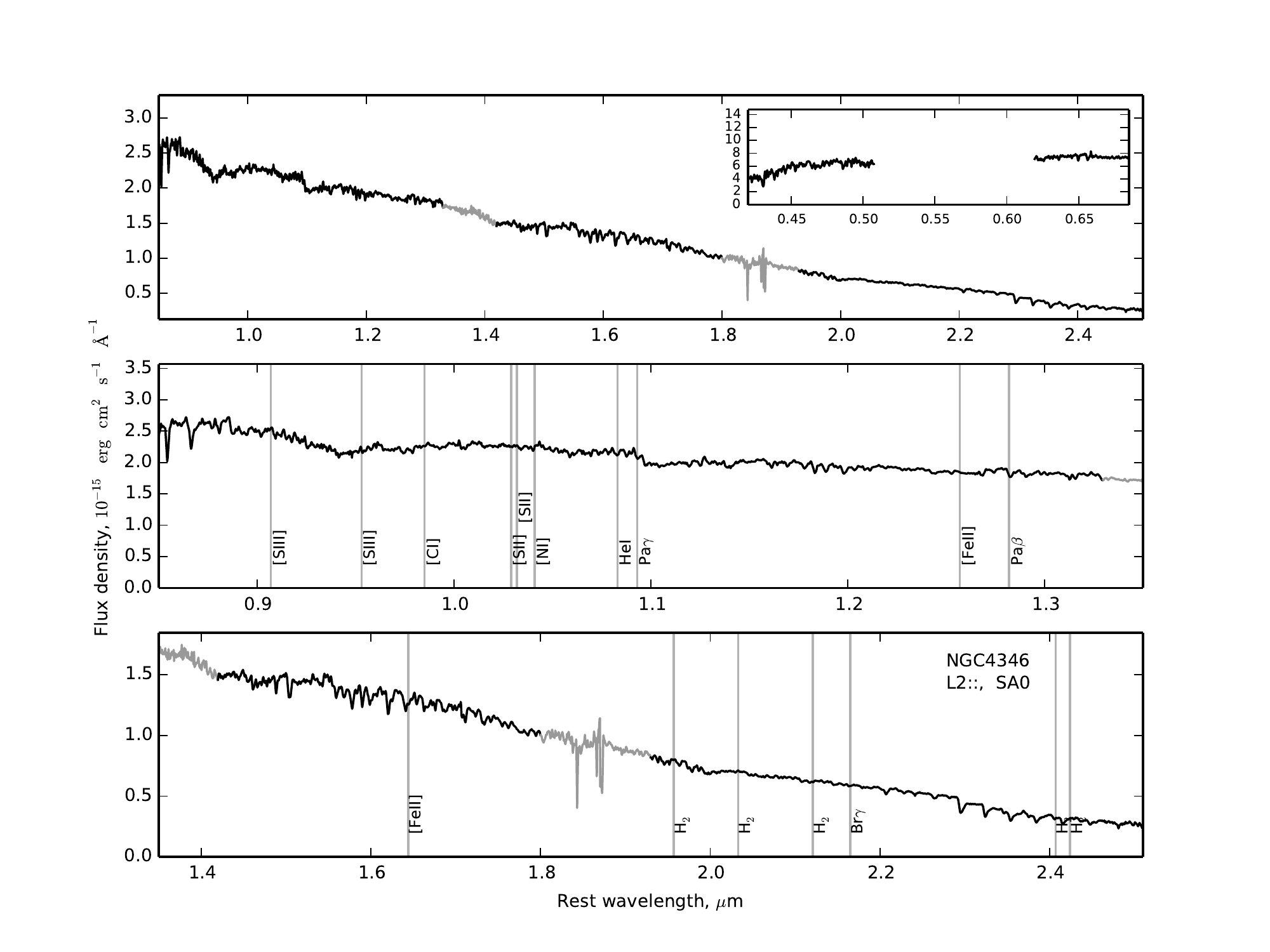}
\caption{ {\small As for Fig. \ref{f1}, for NGC~4346.}}
\label{f36}
\end{figure*}

NGC~4346 appears to be an unremarkable S0 galaxy, with few, weak optical emission lines, and a tentative LINER 2 classification. To date, it is undetected in X-rays \citep{Halderson01} and in radio surveys aimed at searching for nuclear activity \citep{Nagar05}. The good telluric line removal, high S/N, and lack of detectable emission lines in the GNIRS spectrum (Fig. \ref{f36}) make this object useful for detailed examination of stellar features in the IR. \\

{\bf NGC 4548 (L2)} \\

\begin{figure*}
\hspace*{-10mm}
\includegraphics[scale=0.9]{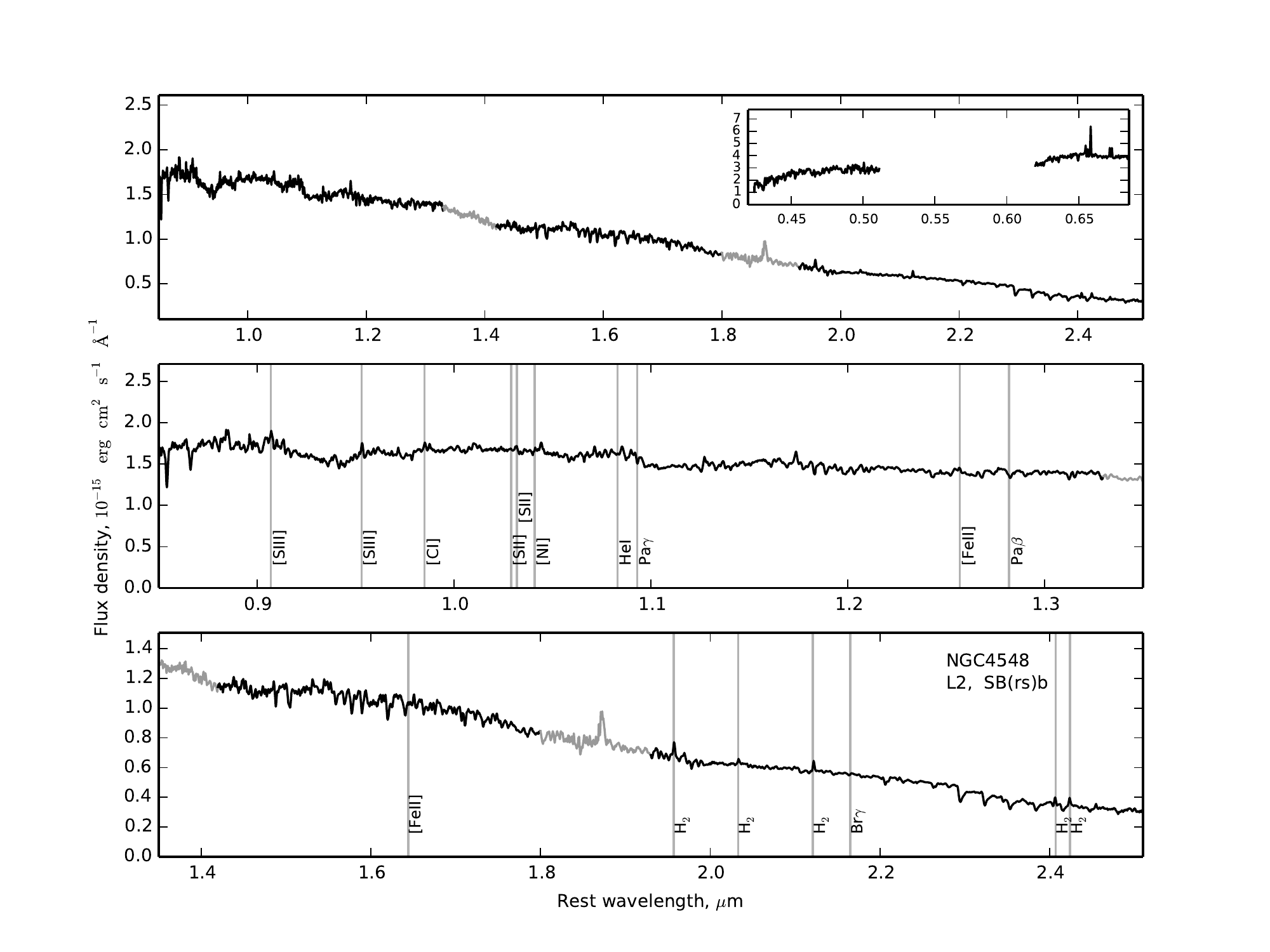}
\caption{ {\small As for Fig. \ref{f1}, for NGC~4548.}}
\label{f37}
\end{figure*}

NGC~4548 is an ``anemic'' spiral galaxy whose HI gas deficiency may have been caused by ram pressure stripping as it passed close to the centre of the Virgo cluster \citep{Vollmer99}. The nuclear stellar population is old, $\sim$10 Gyr \citep{Sarzi05}, and X-ray emission from a low-luminosity and somewhat obscured AGN is detected \citep{Terashima03,Eracleous10a}. The GNIRS spectrum (Fig. \ref{f37}) contains only a few, very weak emission lines, and is generally rather similar to most of the other LINER 2 galaxies in this sample. \\

{\bf NGC 4594 (L2)} \\

\begin{figure*}
\hspace*{-10mm}
\includegraphics[scale=0.9]{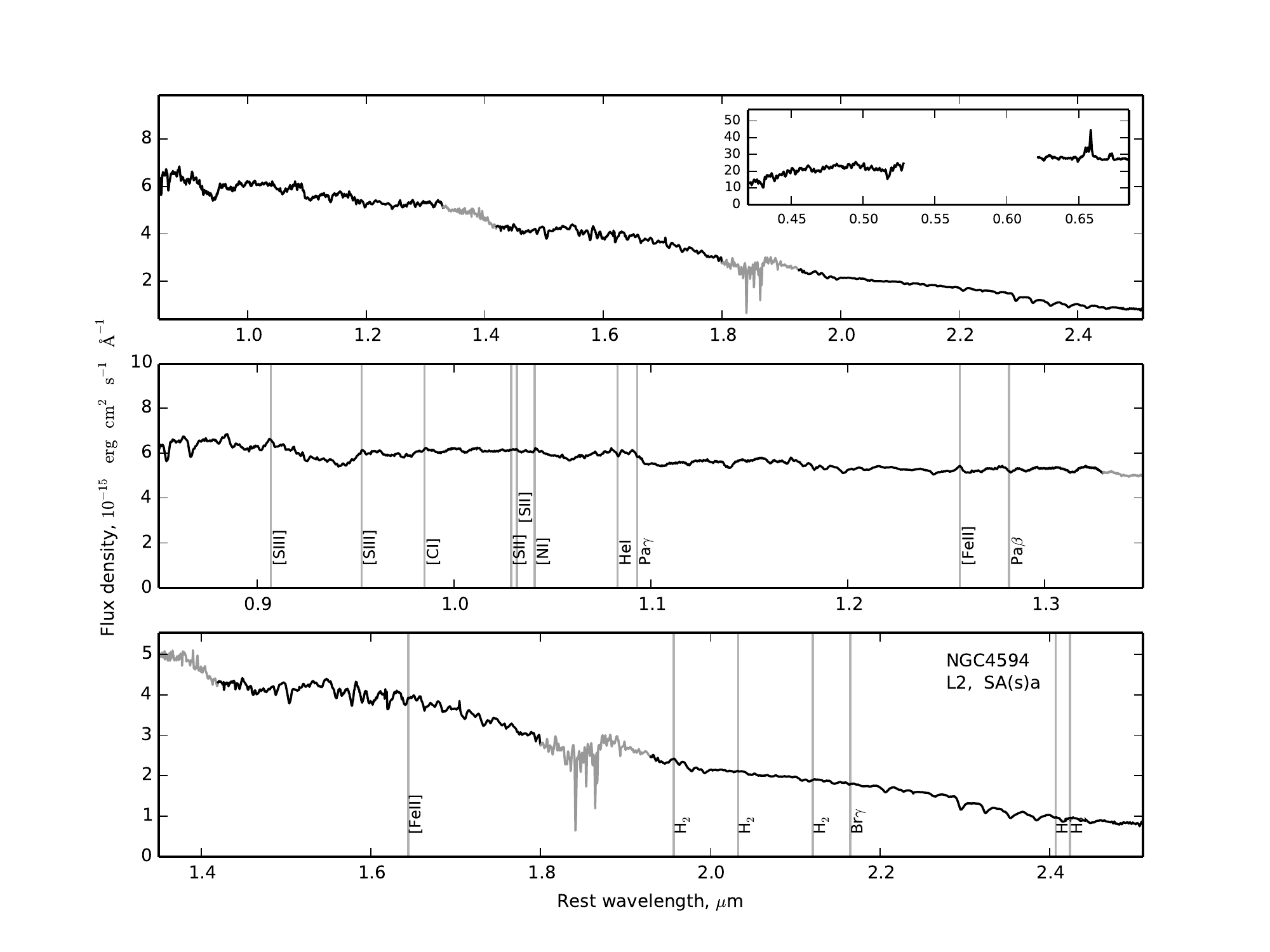}
\caption{ {\small As for Fig. \ref{f1}, for NGC~4594.}}
\label{f38}
\end{figure*}

NGC~4594 (M104; the Sombrero galaxy) has been extensively studied from radio to gamma-ray wavelengths. The edge-on host contains a weak AGN accompanied by parsec-scale jets \citep{Hada13,Mezcua14}, and at sub-arcsecond angular resolution the IR part of the SED appears to be dominated by synchrotron emission \citep{Mason12}. The lower spatial resolution GNIRS spectrum (Figure \ref{f38}) shows no clear sign of this, being similar in shape to the other type 2 LINERs in the sample. Only a few, weak emission lines of [SIII] and [FeII] are tentatively detected. The central stellar population of NGC~4594 has been studied by \citet{Sanchez-Blazquez06}, who find ages ranging from 6 to 12 Gyr, depending on the optical line index used. \\

{\bf NGC 4736 (L2)} \\

\begin{figure*}
\hspace*{-10mm}
\includegraphics[scale=0.9]{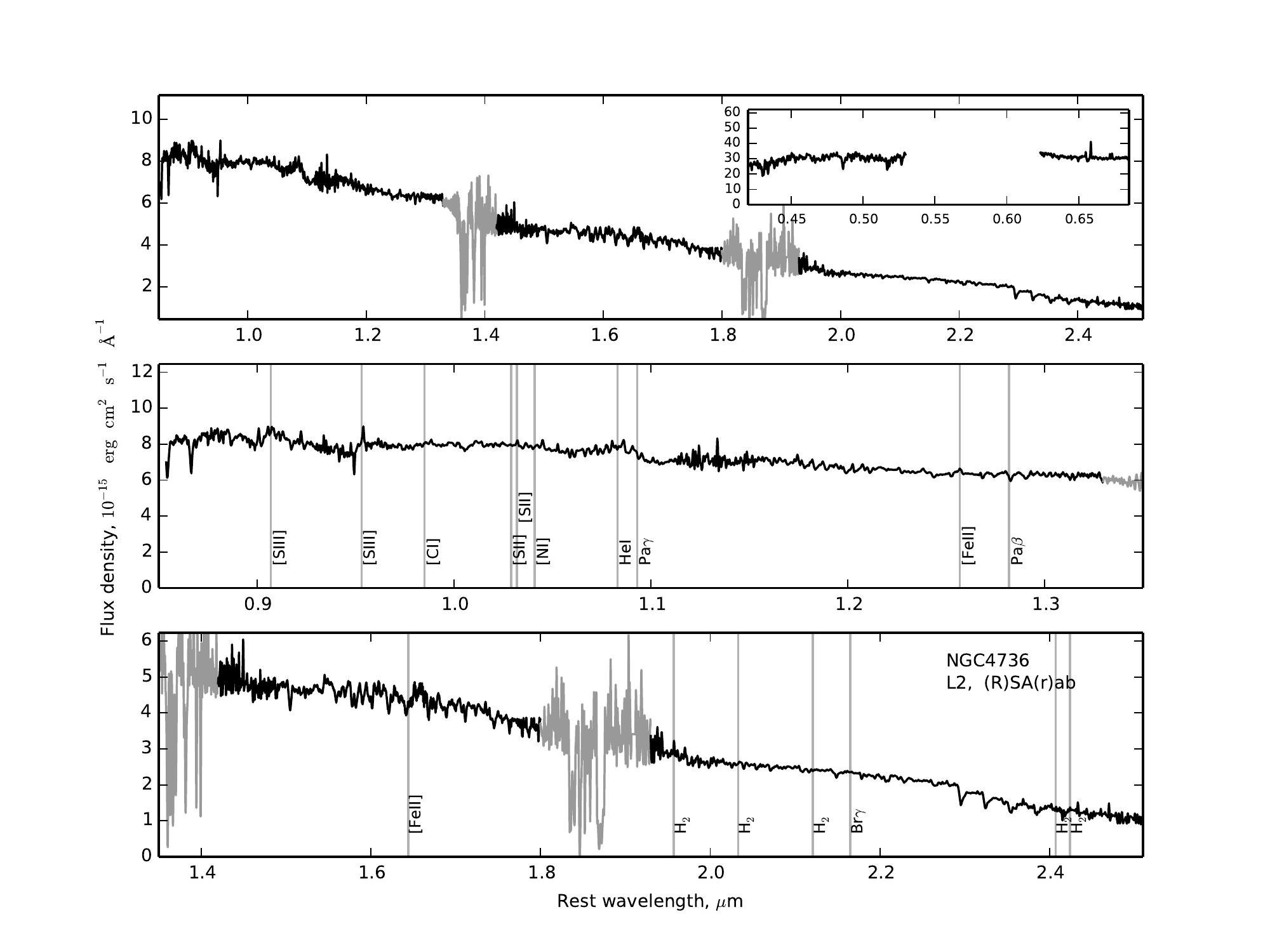}
\caption{ {\small As for Fig. \ref{f1}, for NGC~4736.}}
\label{f39}
\end{figure*}

NGC~4736 is a face-on, ringed galaxy also known as M94.
The galaxy hosts a multitude of nuclear X-ray sources, likely the consequence of an intense period of star formation \citep{Eracleous02}. However, the presence of a variable UV source and compact radio core suggest that it also hosts a low-luminosity AGN \citep{Maoz05,Nagar05}.
\citet{CidFernandes04} find that just under half of the nuclear 4020\AA\ emission of NGC~4736 is from a 10$^8$ -- 10$^9$ year-old stellar population, and \citet{Taniguchi96} classify the galaxy as a post-starburst system on the basis of its optical Balmer absorption lines. The GNIRS spectrum (Fig. \ref{f39}) is of somewhat lower S/N than most of the other spectra in this paper. A couple of weak emission lines may be present in the data, and the continuum shape is broadly similar to that of many of the other types 2 LINERs discussed here.  \\

{\bf NGC 5371 (L2)} \\

\begin{figure*}
\hspace*{-10mm}
\includegraphics[scale=0.9]{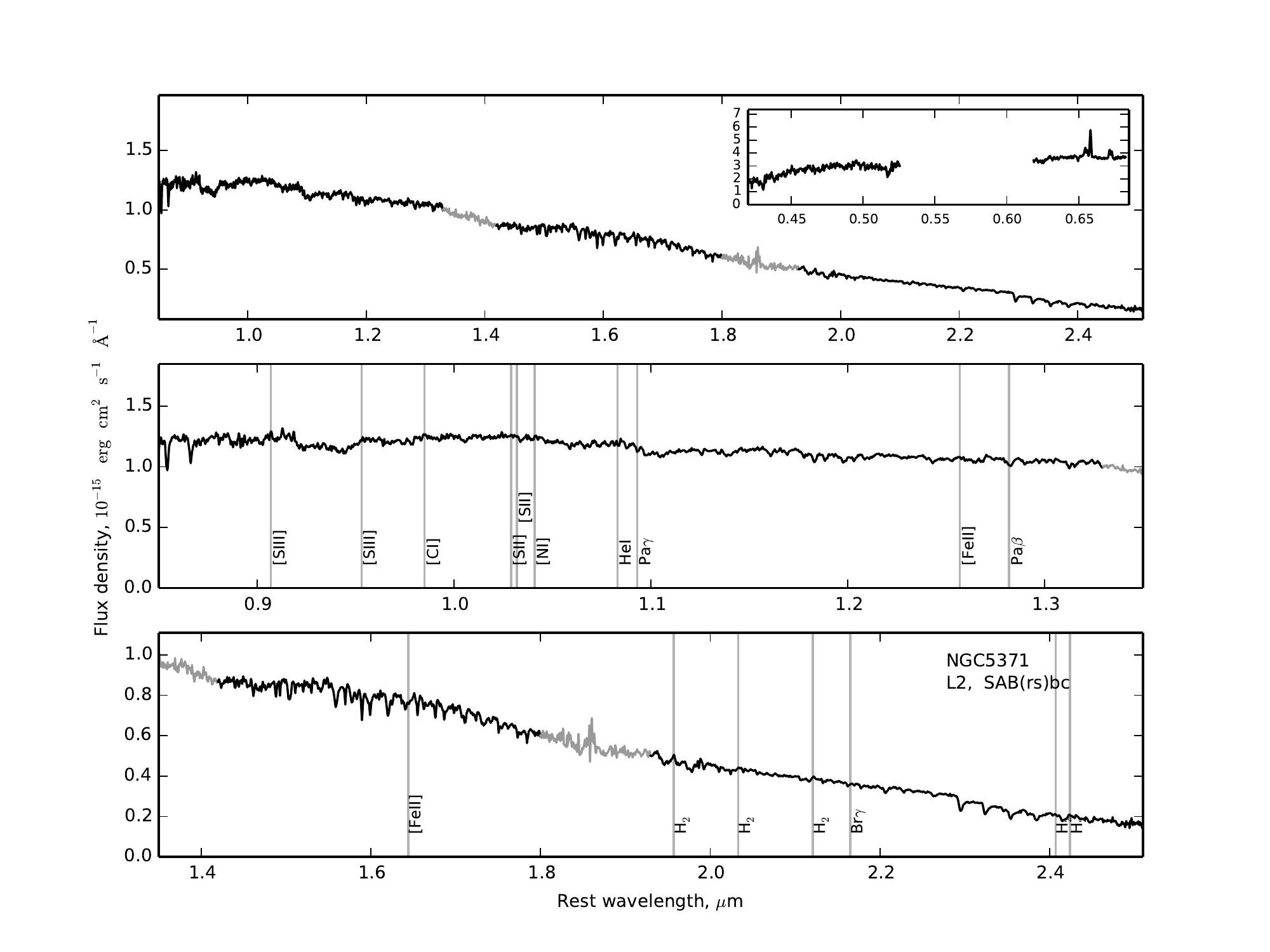}
\caption{ {\small As for Fig. \ref{f1}, for NGC~5371.}}
\label{f40}
\end{figure*}

NGC~5371, a weakly-barred spiral galaxy, hosts a low-luminosity hard X-ray point source \citep{Cisternas13} that has not been detected in MERLIN or VLA radio observations \citep{Nagar05,Filho06}. No emission lines are detected in the IR spectrum (Fig. \ref{f40}), in agreement with the $H$- and $K$-band data of \citet{Bendo04}. \\

{\bf NGC 5850 (L2)} \\

\begin{figure*}
\hspace*{-10mm}
\includegraphics[scale=0.9]{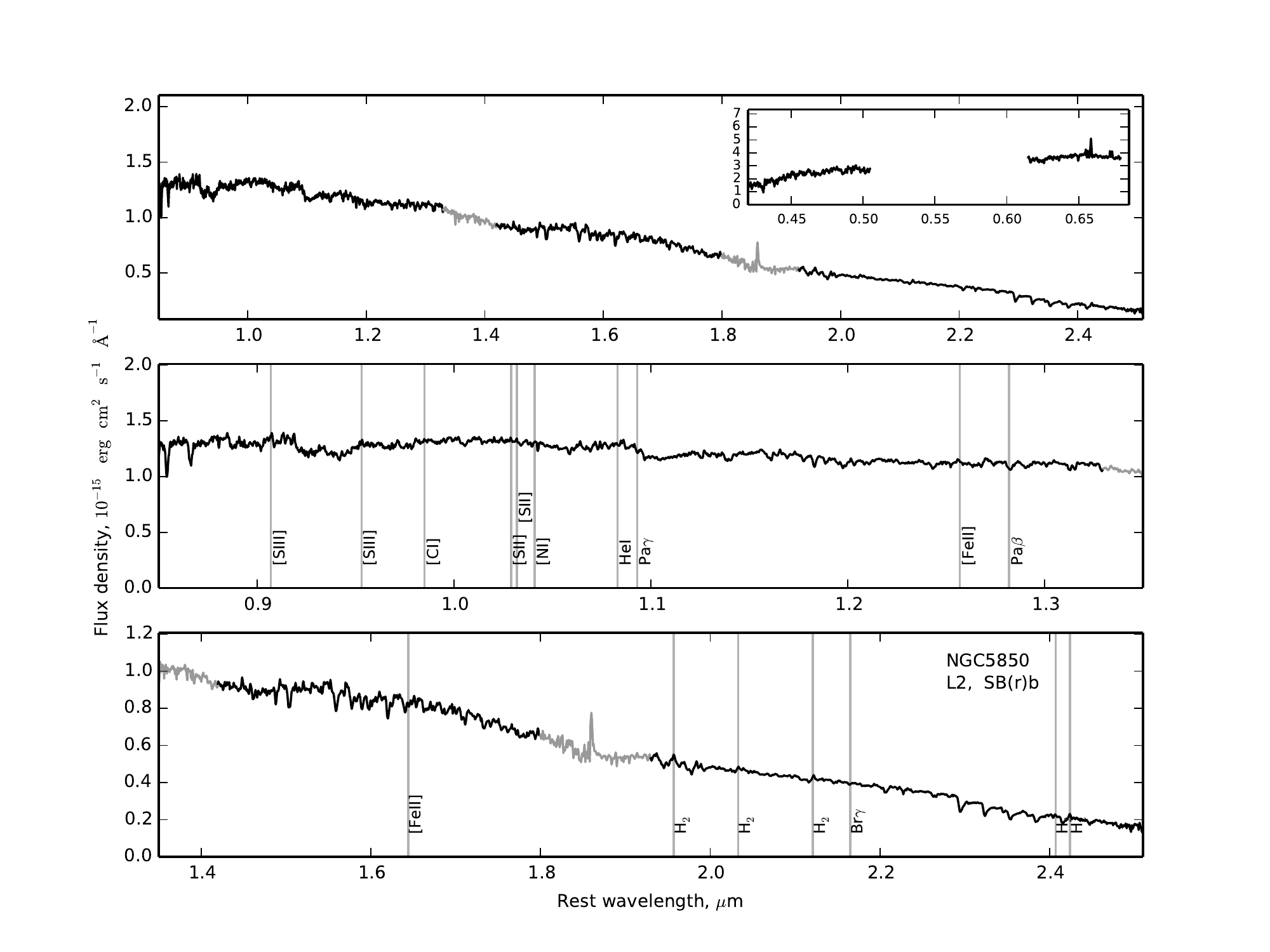}
\caption{ {\small As for Fig. \ref{f1}, for NGC~5850.}}
\label{f41}
\end{figure*}

NGC~5850 is a double-barred spiral galaxy that may have survived an encounter with a nearby massive elliptical galaxy, NGC~5846 \citep{Higdon98}. The optical line emission from NGC~5850 is spatially extended, and sources of ionisation may include post-AGB stars and low-level star formation \citep{Bremer13}. \citet{deLorenzo-Caceres13} observe spiral structures and twisted velocity fields that suggest the presence of gas streaming towards the nucleus, and find that the central stellar population of NGC~5850 has a luminosity-weighted age in the region of 5 Gyr.  In contrast, the IR spectrum of NGC~5850 (Fig. \ref{f41}) is almost devoid of lines, with only H$_{2}$ emission being weakly detected. \\

{\bf NGC 6500 (L2)} \\

\begin{figure*}
\hspace*{-10mm}
\includegraphics[scale=0.9]{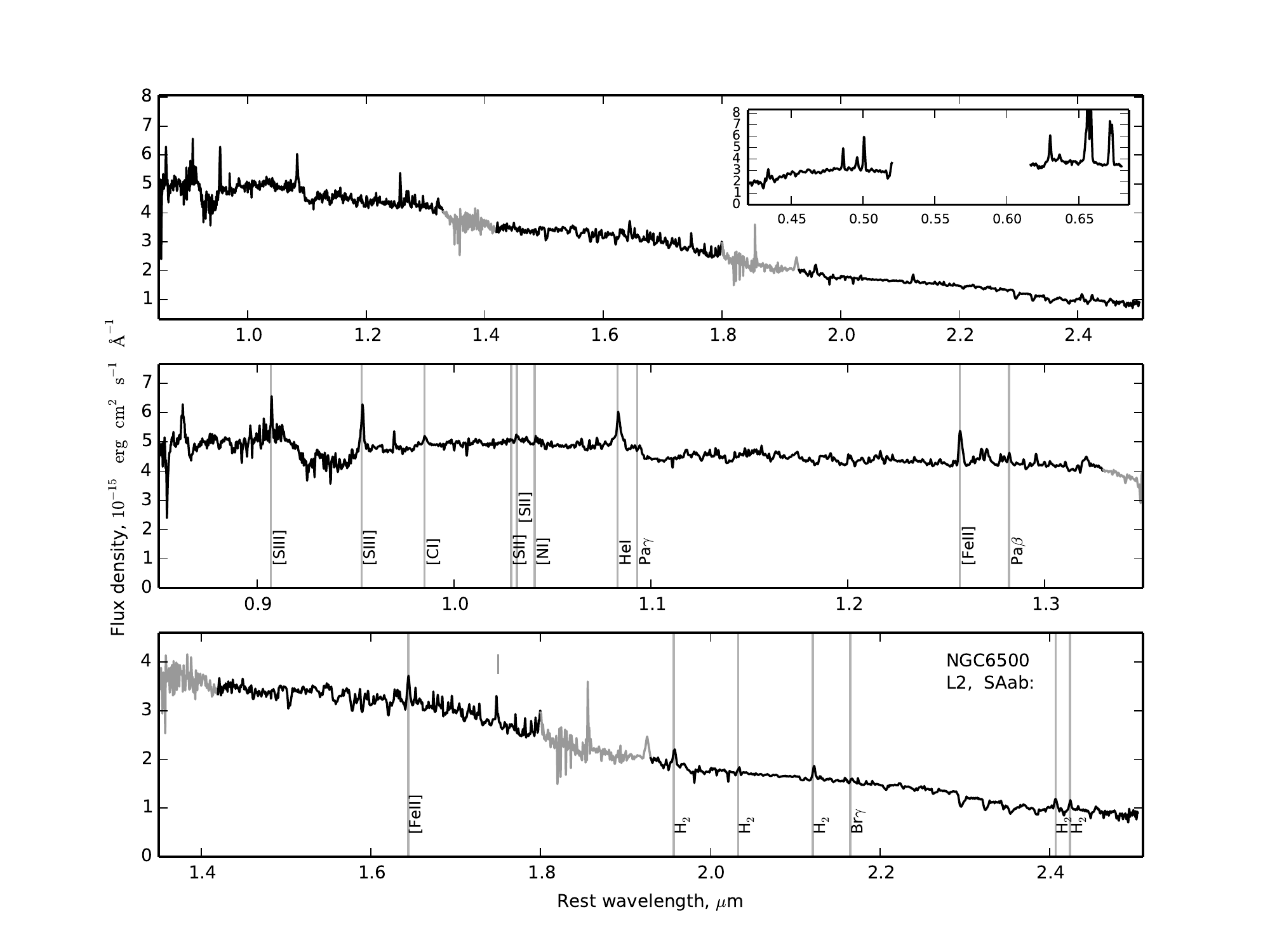}
\caption{ {\small As for Fig. \ref{f1}, for NGC~6500. The short vertical line near 1.73 $\mu$m indicates an artefact from division by the standard star that is common to several of the spectra.}}
\label{f42}
\end{figure*}

NGC~6500 contains a compact, flat-spectrum radio core \citep{Falcke00,Filho02}, variable UV nucleus \citep{Maoz05}, and nuclear hard X-ray source \citep{Terashima03,Dudik05} that indicate the presence of an AGN in this spiral galaxy. The optical LINER emission is extended, with kinematics suggestive of an outflowing wind  \citep{GonzalezDelgado96}. Most of the nuclear optical light in NGC~6500 comes from an old stellar population, although about 25\% appears to arise in young ($<$10$^7$ yr) stars \citep{CidFernandes04}. Unlike most of the LINER 2 nuclei in this sample, the GNIRS spectrum of NGC~6500 (Fig. \ref{f42}) contains numerous, fairly prominent emission lines, from species including [SIII], HeI and [FeII]. \\

{\bf NGC 7217 (L2)} \\

\begin{figure*}
\hspace*{-10mm}
\includegraphics[scale=0.9]{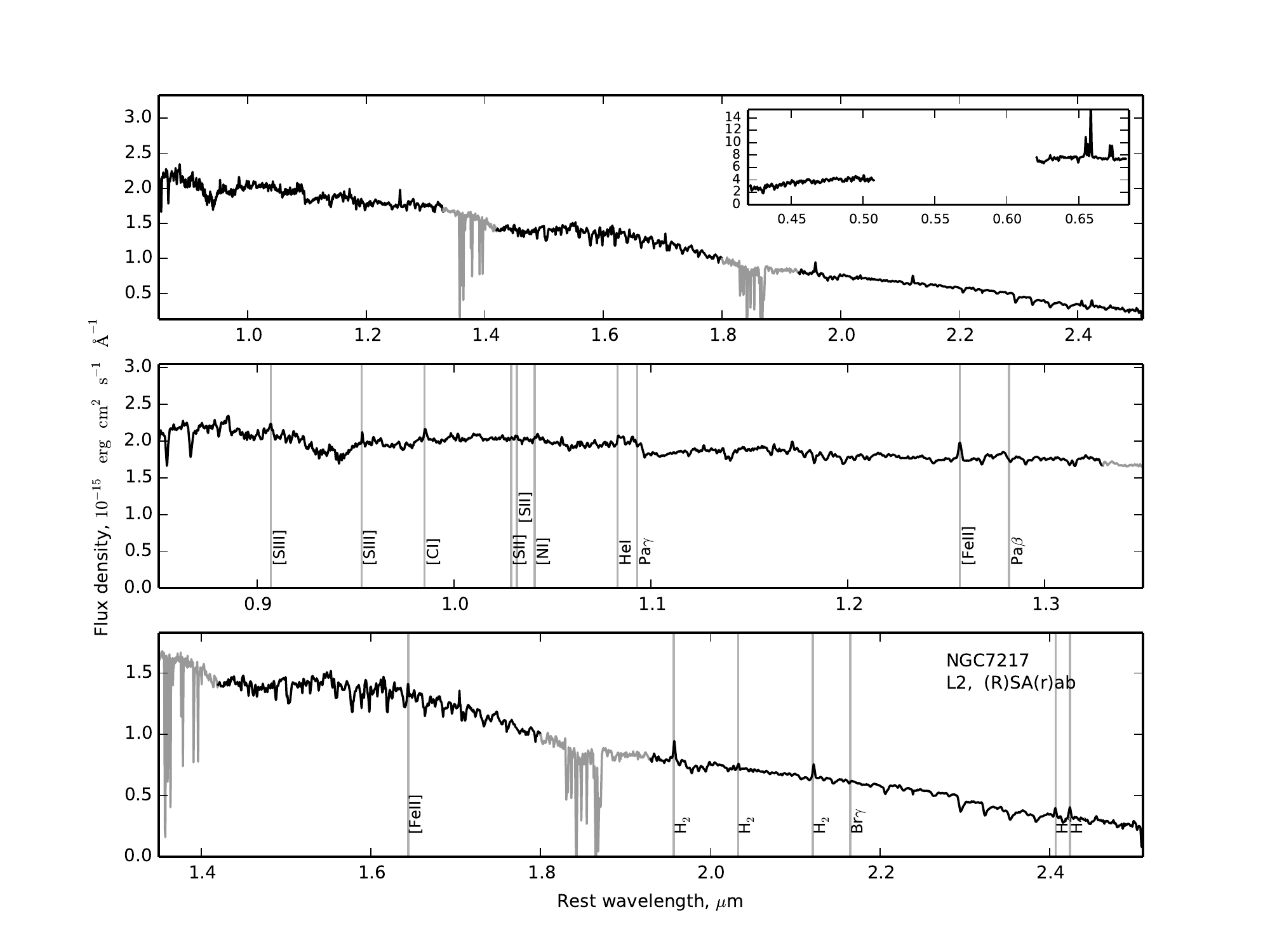}
\caption{ {\small As for Fig. \ref{f1}, for NGC~7217.}}
\label{f43}
\end{figure*}

\begin{figure*}
\hspace*{-10mm}
\includegraphics[scale=0.9]{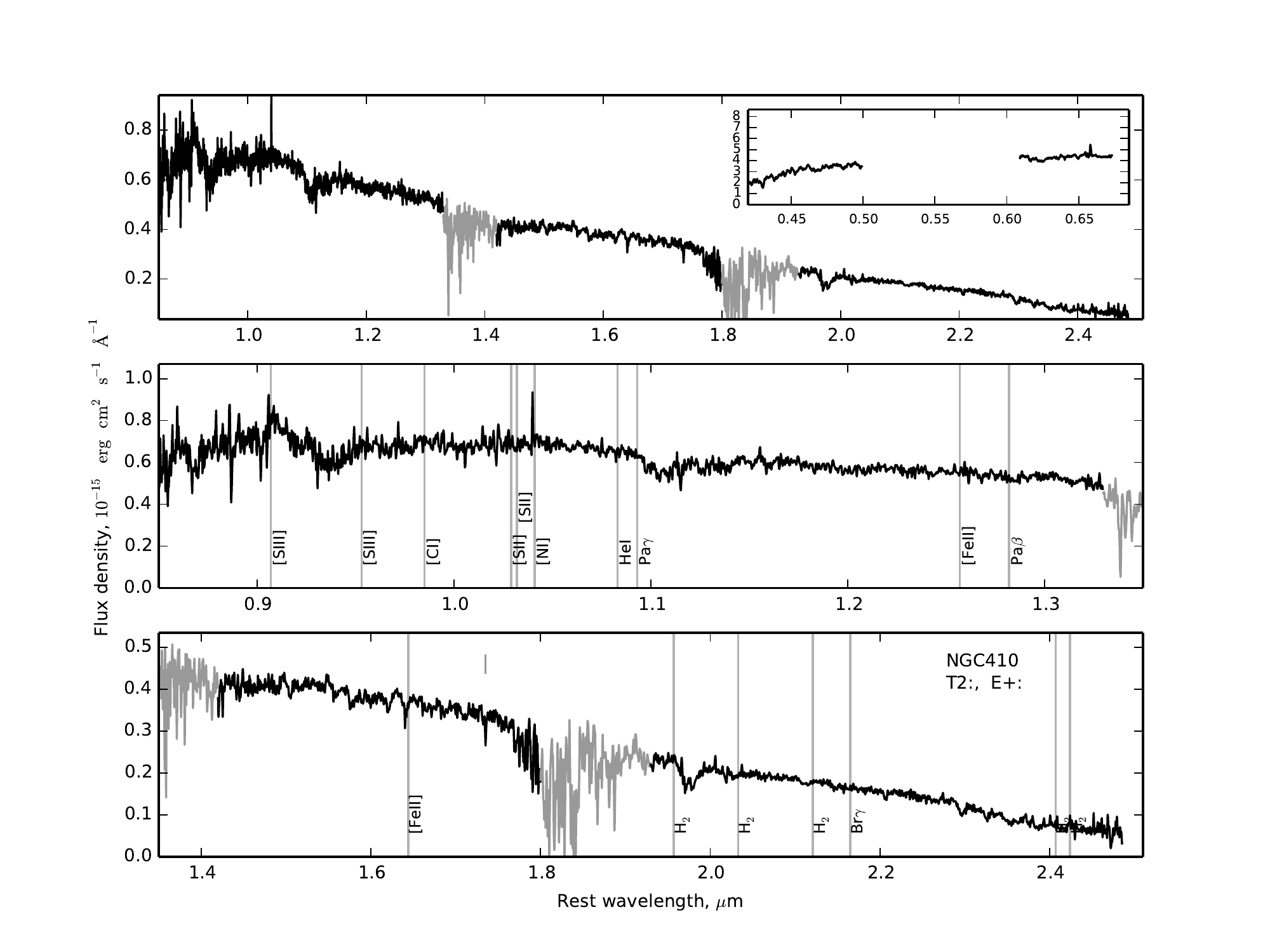}
\caption{ {\small As for Fig. \ref{f1}, for NGC~410. The short vertical line near 1.73 $\mu$m indicates an artefact from division by the standard star that is common to several of the spectra.}}
\label{f44}
\end{figure*}

This galaxy contains a number of star-forming rings \citep[e.g.][]{Buta95,Sarzi07}, which lie outside the slit used for these observations, and the gas within the innermost ring is currently outflowing \citep{Combes04}. The nuclear stellar population is predominantly old, with just a few per cent of the optical light being emitted by intermediate-age (10$^8$ -- 10$^9$ yr) stars \citep{CidFernandes04}. The GNIRS spectrum (Fig. \ref{f43}) contains several emission lines, of which those of [FeII] and H$_2$ are strongest. The overall continuum shape qualitatively resembles that of most of the other type 2 LINERs in this work. \\

\pagebreak

\subsection{Transition Objects, Inactive Galaxies}

{\bf NGC 410 (T2:)}  \\

A relatively little-studied elliptical galaxy, NGC~410 is the central galaxy of the Pisces cluster \citep{Hudson01}. The S/N in the GNIRS spectrum is relatively low, but fairly strong stellar absorption bands at 1.1 $\mu$m and 0.93 $\mu$m are clearly visible. No identifiable emission lines are reliably detected in the NIR spectrum, although weak H$\beta$, [NII] $\lambda$6583, and [SII] $\lambda \lambda$6716,6731 are present in the optical spectrum of \citet{Ho97}. Stellar population synthesis indicates that the optical light of NGC~410 comes almost entirely from old stars \citep[$>10^9$ yrs; ][]{CidFernandes04,GonzalezDelgado04}. \\

{\bf NGC 660 (T2/H:)} \\

\begin{figure*}
\hspace*{-10mm}
\includegraphics[scale=0.9]{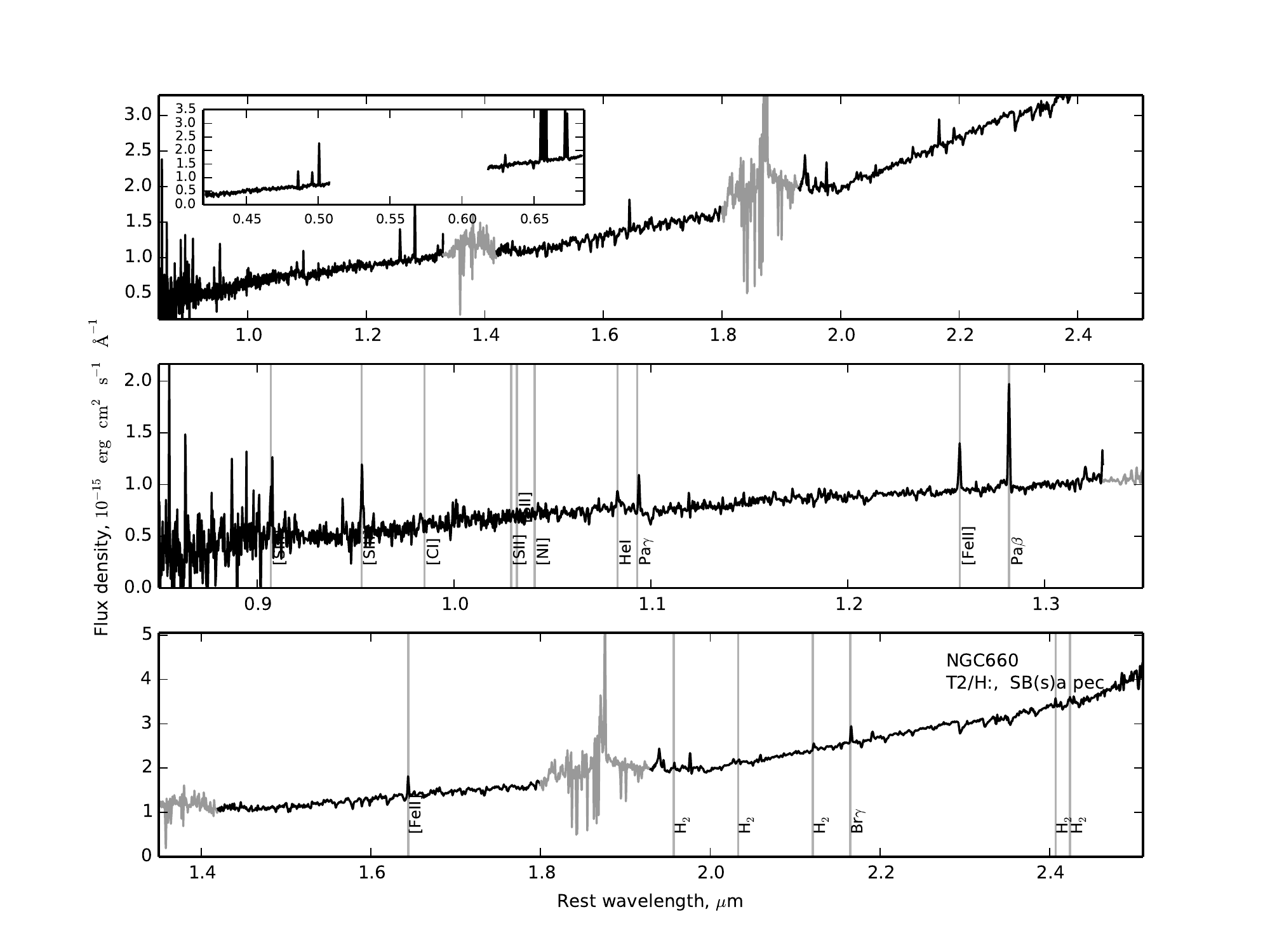}
\caption{ {\small As for Fig. \ref{f1}, for NGC~660.}}
\label{f45}
\end{figure*}

NGC~660 is a highly-inclined, barred spiral galaxy with a polar ring \citep{Whitmore90,Combes92} and a rather high IR luminosity \citep[log L$_{IR}$=10.5 L$_\odot$; ][]{Sanders03} that indicates ongoing star formation. Based on optical spectroscopy, \citet{CidFernandes04} find a significant intermediate-age population in the nuclear regions of this galaxy, with over half of the 4020\AA\ flux coming from stars of 10$^8$ -- 10$^9$ yr. The detection of the [NeV]$\lambda$14.32 line confirms that NGC~660 hosts an AGN \citep{Goulding09}, with a deep silicate absorption feature that shows it to be highly obscured \citep{Roche91}. The GNIRS spectrum of NGC~660 (Fig. \ref{f45}) is unique among the 50 sources  observed. It displays a prominent excess of NIR emission,
particularly in the K-band, where the continuum  flux increases steeply with increasing wavelength. This spectral shape is very similar to that observed in NGC~1068 and Mrk 1239 \citep{Rodriguez-Ardila06,Martins10a}, so we attribute it to emission from hot dust. Strong hydrogen recombination lines are observed, along with a series of fairly prominent [FeII] and [SIII] lines. \\

{\bf NGC 4569 (T2)} \\

\begin{figure*}
\hspace*{-10mm}
\includegraphics[scale=0.9]{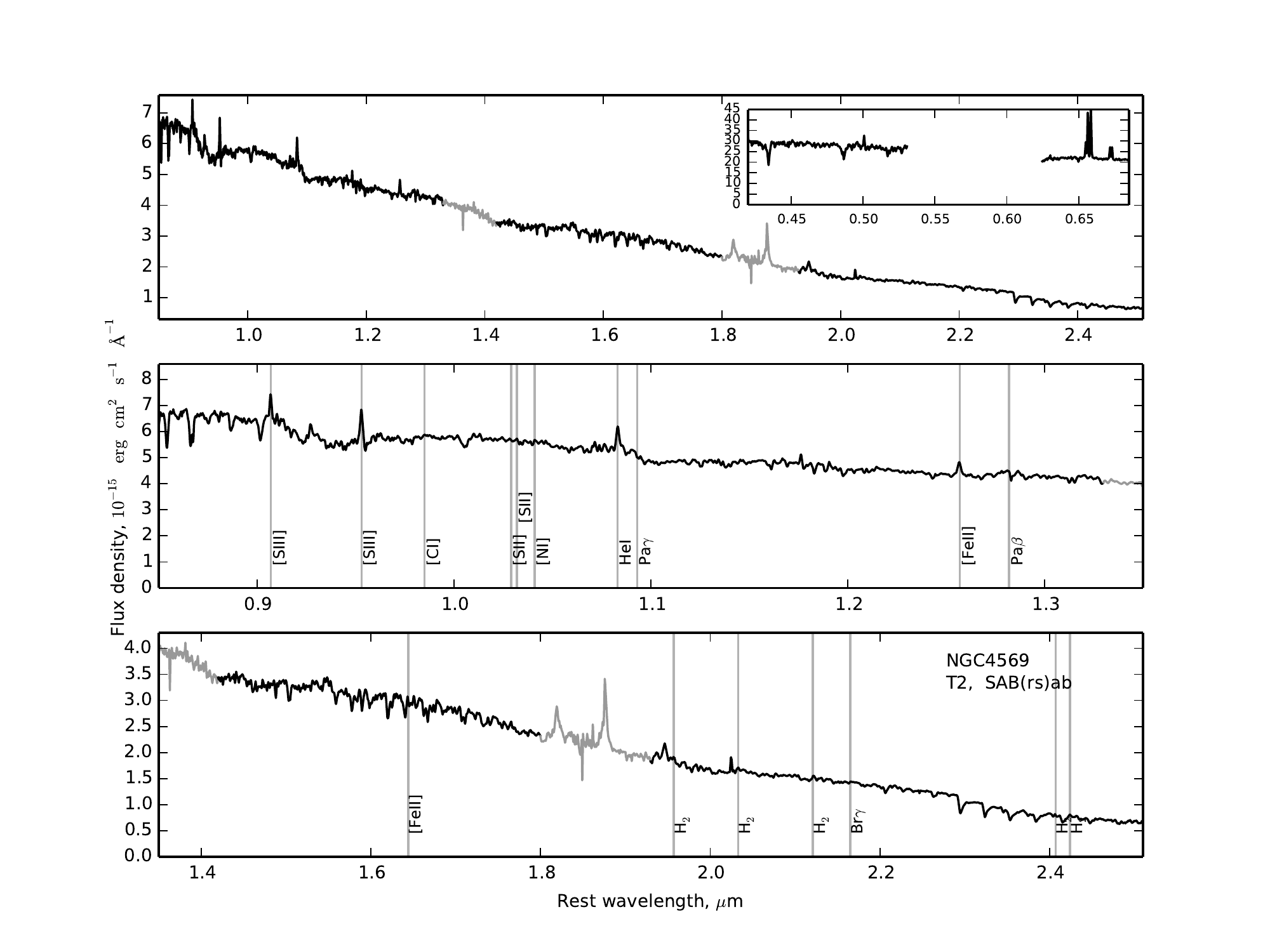}
\caption{ {\small As for Fig. \ref{f1}, for NGC~4569.}}
\label{f46}
\end{figure*}

NGC~4569 is a large,  anaemic (gas-poor) spiral galaxy in the Virgo cluster \citep{vandenBergh76}, and observations of its HI content suggest that a ram pressure stripping event occurred around 300 Myrs ago \citep{Vollmer04}. The nuclear UV spectrum of NGC~4569 indicates the presence of a young (5-6 Myr) starburst \citep{Maoz98,Gabel02}, while the optical spectrum shows strong, narrow Balmer lines that reveal a population of A-type supergiants \citep{Keel96} and a generally young-intermediate age population \citep{CidFernandes04}. Recent and/or ongoing star formation activity is also suggested by the detection of PAH emission in the mid-IR \citep{Dale06}. Any spectral signatures of a post-starburst population in NGC~4569 must fairly subtle, as the NIR absorption line spectrum of this nucleus (Fig. \ref{f46}) looks broadly similar to those of many other objects in this study. A handful of common emission lines are observed in the spectrum, from [SIII], HeI, and [FeII].\\

{\bf NGC 7331 (T2)} \\

NGC~7331 (Fig. \ref{f46}) is a highly-inclined spiral galaxy in the vicinity of Stephan's quintet, although it is not associated \citep{Gutierrez02}. The galaxy contains a ring of star formation (r$\sim$6 kpc) visible at infrared and longer wavelengths \citep[e.g.][]{Telesco82}, from which material appears to be flowing inwards \citep{Battaner03}. Mid-IR spectra covering the nucleus and inner ring show emission in several H$_2$ and fine structure lines, and the detection of [O IV] 25.9 $\mu$m emission at the nucleus implies the presence of either an AGN, Wolf-Rayet stars, or shocks \citep{Smith04}. PAH emission is also detected at the nucleus, as well as in the ring.

\begin{figure*}
\hspace*{-10mm}
\includegraphics[scale=0.9]{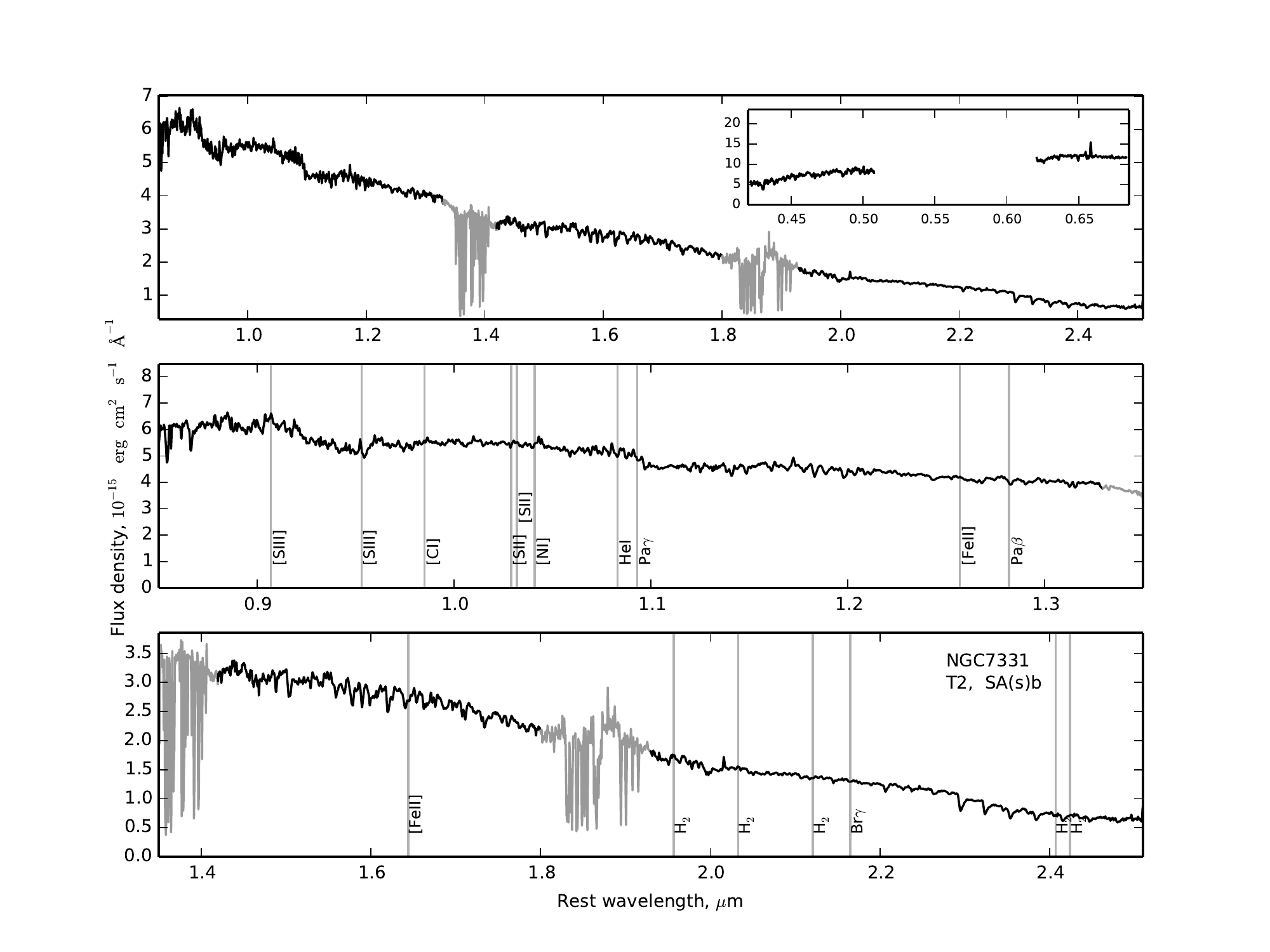}
\caption{ {\small As for Fig. \ref{f1}, for NGC~7331.}}
\label{f47}
\end{figure*}

In contrast, the nuclear optical spectrum of NGC~7331 shows only weak emission from H$\beta$, [OIII] $\lambda$5007, [NII] $\lambda$6583, and [SII] $\lambda \lambda$6716,6731. Its Balmer absorption lines led \citet{Ohyama96} to classify NGC~7331 as a post-starburst galaxy, with $\sim$40\% of the optical light being produced by stars with ages $\sim 5 \times 10^9$ yrs.  \citet{CidFernandes04} come to a different conclusion, finding only a 5\% contribution from intermediate-age stars. Spectral features from any post-starburst population are not obvious in the NIR spectrum of this object. The GNIRS spectrum (Fig. \ref{f47}) is almost devoid of emission lines; the energetic processes leading to the PAH and [O IV] lines in the mid-IR spectrum leave little trace in the NIR data. \\

{\bf NGC 205 (Inactive)} \\

\begin{figure*}
\hspace*{-10mm}
\includegraphics[scale=0.9]{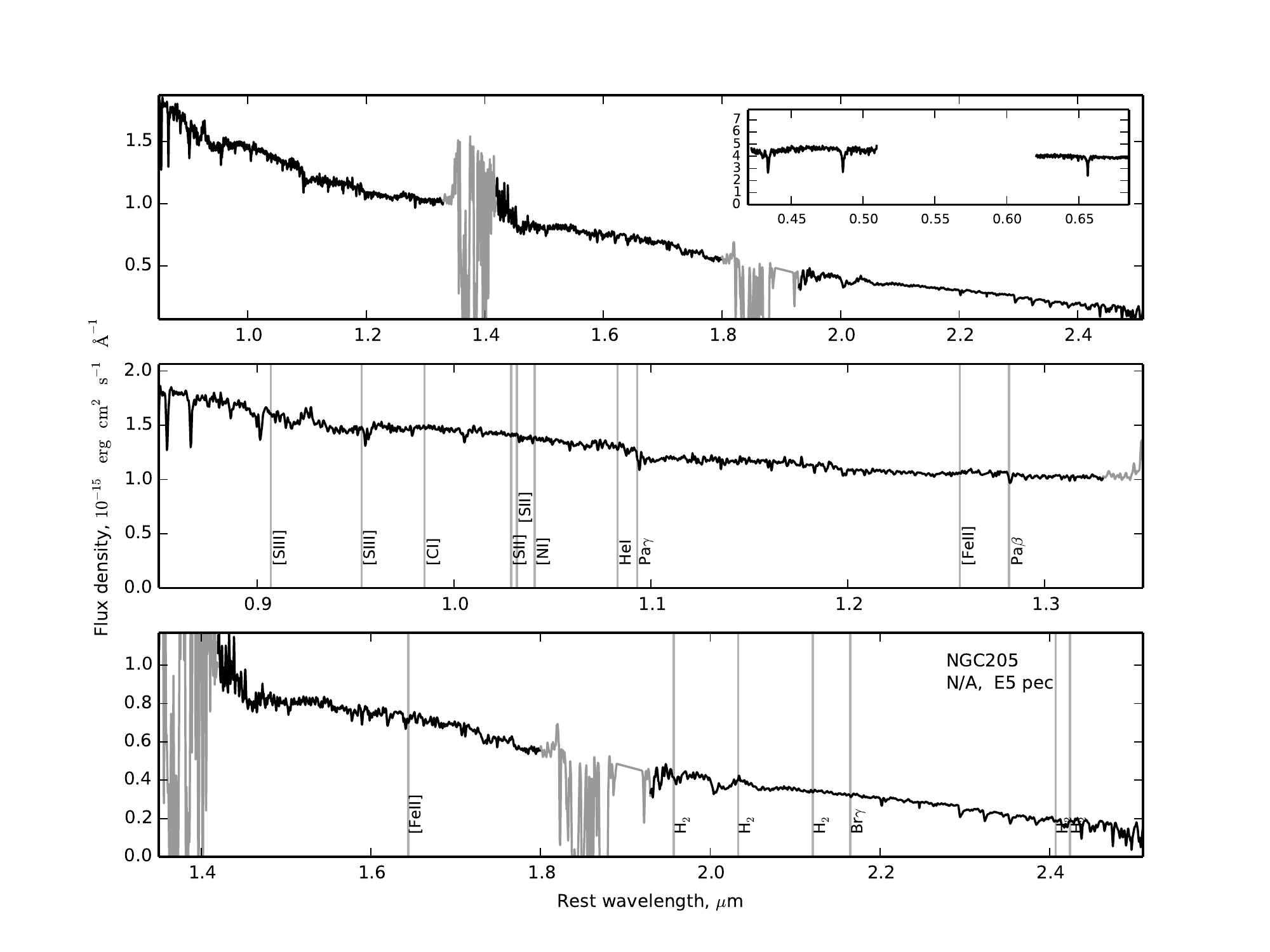}
\caption{ {\small As for Fig. \ref{f1}, for NGC~205.}}
\label{f48}
\end{figure*}

NGC~205 is a dwarf elliptical galaxy and a close satellite of M31. Given its proximity, it has been extensively studied at many wavelengths. No evidence of AGN activity has been detected to date \citep{Ho97,Colbert99}, and any black hole at the nucleus must have a mass of no more than a few $\times$ 10$^4\;$M$_{\odot}$ \citep{Valluri05}. No emission lines are visible in the optical spectrum of NGC~205 presented by \citet{Ho95}, but deep HST photometry reveals a population of young blue stars at its centre \citep{Monaco09}, and the optical spectrum displays a blue continuum and strong Balmer absorption lines indicative of A stars \citep{Bica90,GonzalezDelgado99,CidFernandes04}. The GNIRS spectrum (Fig. \ref{f48}) is similarly free from detectable emission lines. Qualitatively it is rather similar to those of the LINERs and transition objects in this sample, but many of the absorption lines are shallower and the broad H-band ``bump'' (\S\ref{cont}) appears somewhat weaker than in many other objects. The stellar population responsible for the optical Balmer lines leaves no gross signatures in the IR spectrum of NGC~205. \\

\pagebreak

\subsection{Non-Palomar Galaxies}
 
{\bf 1H\,1934-063 (NLS1)} \\

\begin{figure*}
\hspace*{-10mm}
\includegraphics[scale=0.9]{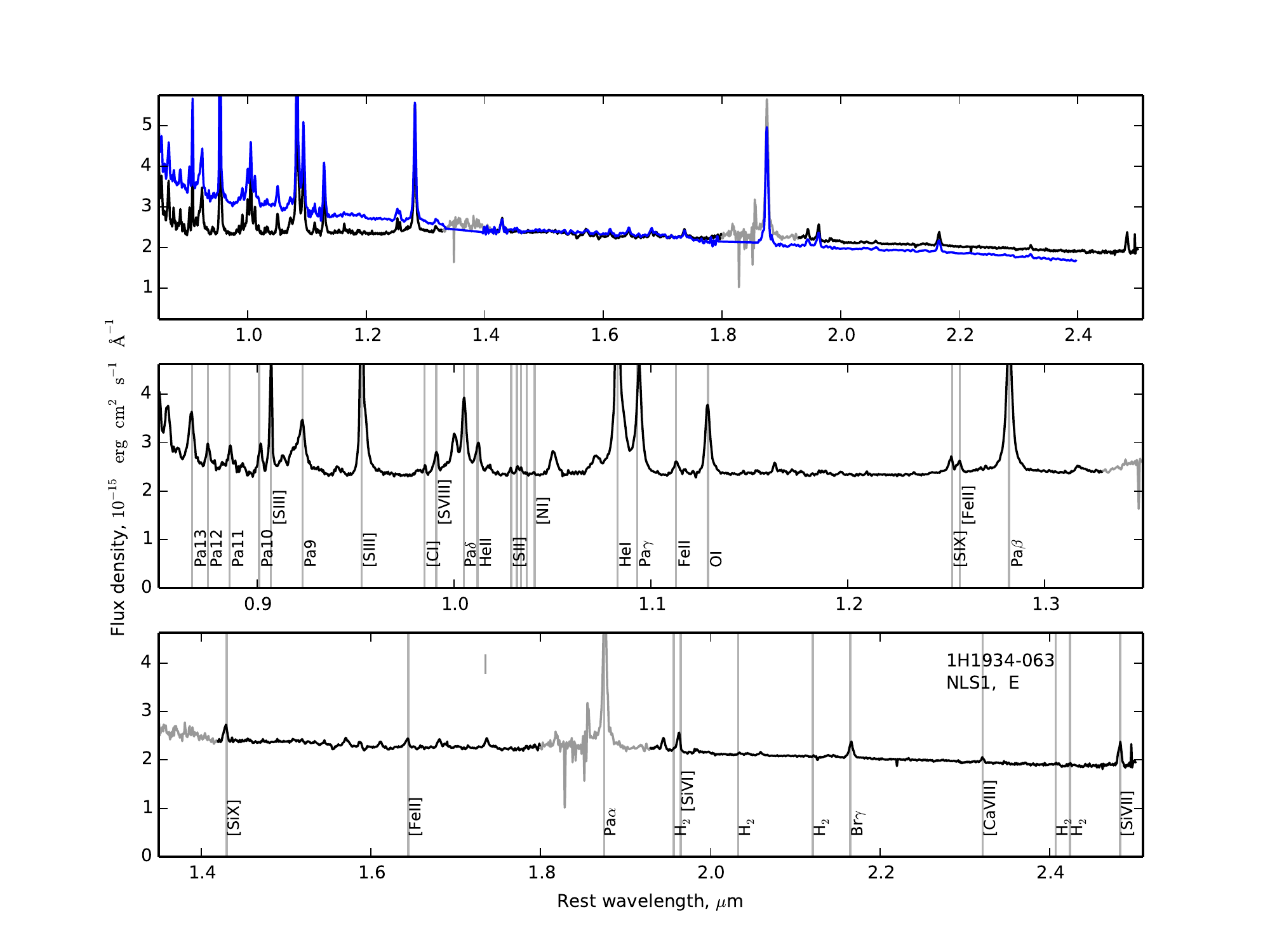}
\caption{ {\small As for Fig. \ref{f1}, for 1H~1934-063. The blue spectrum in the top panel is the IR spectrum of this object from \citet{Rodriguez-Ardila02}, scaled to match the GNIRS spectrum in flux in the relatively line-free spectral region around 1.5 $\mu$m. The short vertical line near 1.73 $\mu$m indicates an artefact from division by the standard star that is common to several of the spectra. See \citet{Rodriguez-Ardila02} and \citet{Riffel06} for identifications of additional weak lines.}}
\label{f50}
\end{figure*}

1H\,1934-063 is the only narrow-line Seyfert 1 AGN in the sample. It has been studied in the optical and NIR 
because of its prominent spectrum, rich in emission lines both from the NLR and
BLR. \citet{Rodriguez-Ardila02} were the first to publish a NIR spectrum of this source,
dominated by broad lines from Fe\,{\sc ii}, O\,{\sc i}, He\,{\sc i} and H\,{\sc i}. 
Narrow, bright forbidden lines from [S\,{\sc iii}] and prominent coronal lines
of [S\,{\sc viii}], [S\,{\sc ix}], [Si\,{\sc vi}], [Si\,{\sc x}] and [Ca\,{\sc viii}] are also easily 
seen in its NIR spectrum. The GNIRS spectrum (Fig. \ref{f50}) shows all these lines in addition to 
[Si\,{\sc vii}] at 2.48\,$\mu$m, which was not previously reported. The main difference 
between the present and previous NIR spectroscopy is that the continuum in the GNIRS spectrum is flatter. This is probably because the 0.3\arcsec\ slit used for these observations is significantly smaller than that of previous observations (0.8\arcsec) and therefore includes less emission from the host galaxy. Otherwise, the spectroscopy presented here matches very well previous observations covering a similar wavelength interval. \\

{\bf NGC 7469 (S1)} \\

\begin{figure*}
\hspace*{-10mm}
\includegraphics[scale=0.9]{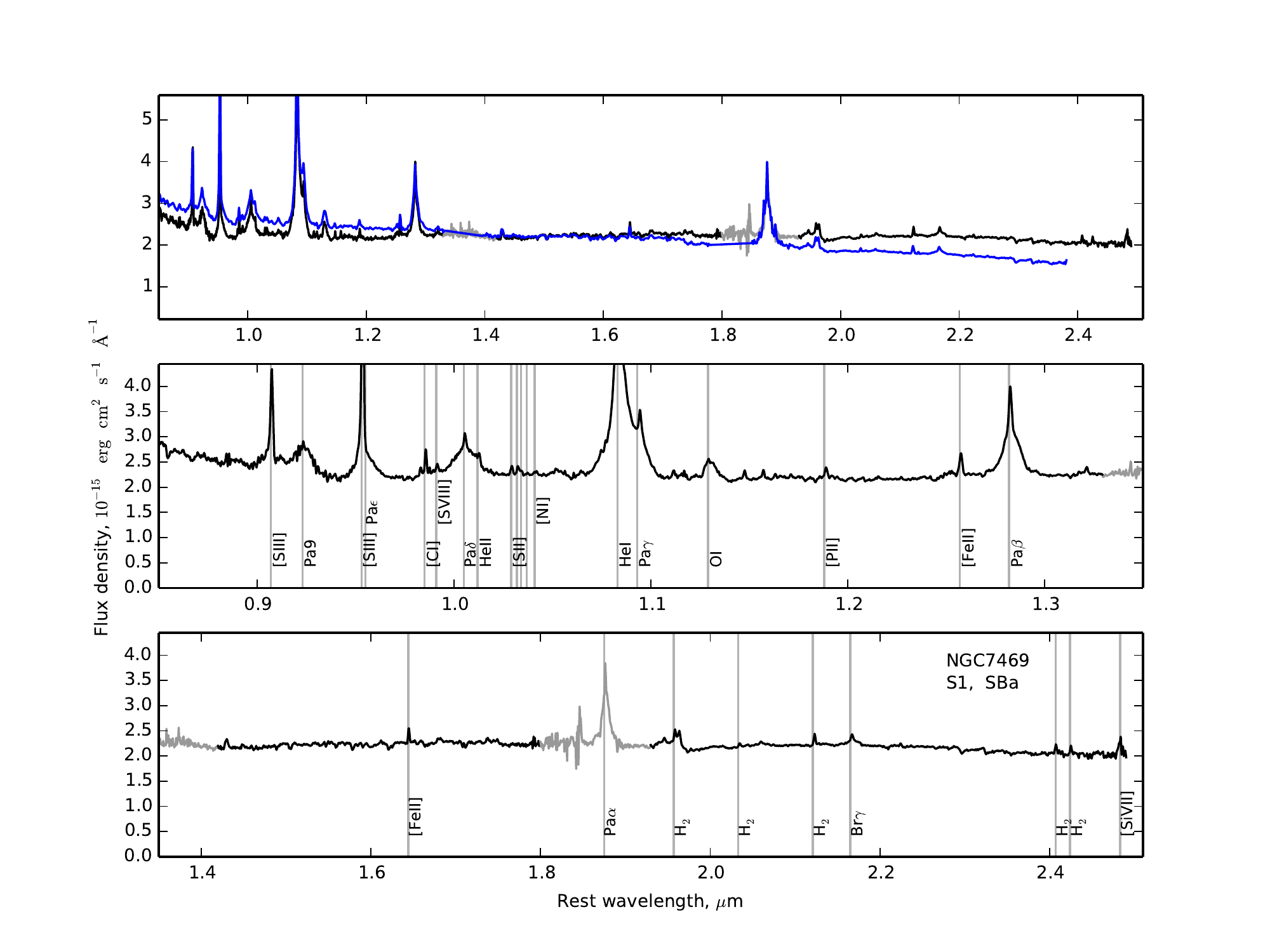}
\caption{ {\small As for Fig. \ref{f1}, for NGC~7469. The blue spectrum in the top panel is the IR spectrum of this object from \citet{Riffel06}, scaled to match the GNIRS spectrum in flux in the relatively line-free spectral region around 1.5 $\mu$m. See \citet{Riffel06} for identifications of additional weak lines.}}
\label{f49}
\end{figure*}

This Seyfert\,1 galaxy has been extensively studied in the literature. Its
most prominent characteristic is a circumnuclear ring (1.5$\arcsec-2.5\arcsec$) 
of powerful starburst activity observed at radio, optical, near-infrared, and mid-infrared
wavelengths \citep[][and references therein]{Davies04}, with a luminosity that amounts 
to two-thirds of the bolometric luminosity of the entire galaxy.  NIR spectroscopy 
of this source has been published by several authors including \citet{Genzel95,Thompson96,Sosa-Brito01,Riffel06} and \citet{Davies07}. The GNIRS emission line spectrum (Fig. \ref{f49}) agrees well
with that presented by \citet{Riffel06}. The main difference is found in the
continuum, with the GNIRS spectrum displaying an excess of continuum over the underlying power-law at the red end of the spectrum, peaking at $\sim 2.2 \; \mu$m. This feature can be interpreted as emission from dust heated
to its sublimation temperature ($\sim$ 1800~K) by the AGN. Although this excess
was also detected by \citet{Riffel06}, it is more conspicuous in the GNIRS
spectrum, probably due to the smaller slit width which isolates 
the NIR emission from the host galaxy and circumnuclear ring of star formation.
He\,{\sc i}\,1.083~$\mu$m and [S\,{\sc iii}]~0.907, 0.953\,$\mu$m are the most
conspicuous lines observed, and high-ionization lines of [S\,{\sc viii}], [S\,{\sc ix}], 
[Si\,{\sc vi}], [Si\,{\sc vii}] and [Si\,{\sc x}] are also identified.  
Stellar absorption features are detected mostly in the $H$- and $K$-bands, and the 2.3 $\mu$m CO bandheads are prominent. The CaT at 0.85~$\mu$m is also
present but is affected by the superimposed, broad O\,{\sc i}~0.845\,$\mu$m emission line. \\

\clearpage

\section{Appendix B: Atmospheric water vapor and inter-band signal-to-noise ratio}
\label{A2}

\begin{figure*}
\hspace*{-15mm}
\includegraphics[scale=0.95, clip, trim=0 200 0 0]{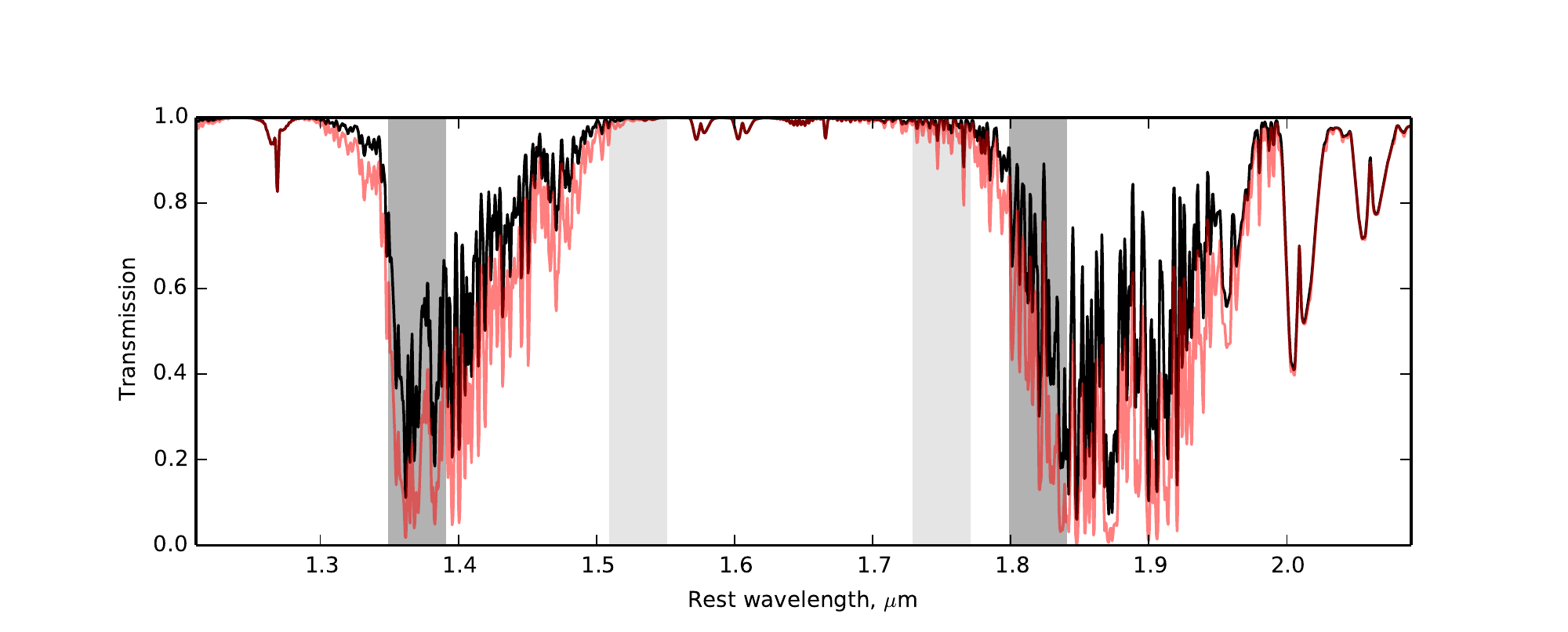}
\caption{ {\small R=1200 atmospheric transmission spectrum calculated for 1 (black) and 3 (red) mm water vapor at zenith, 4186 m altitude, using the ATRAN program \citep{Lord92}. ``Double-humped'' features around 1.6 and 2.0 $\mu$m are from atmospheric CO$_2$, while much of the remaining absorption is from H$_2$O. The shaded areas indicate the regions used to measure the S/N plotted in Figure \ref{snvtau}.}}
\label{trans}
\end{figure*}

The regions between the traditional NIR bands suffer from poor atmospheric transmission and strong absorption from telluric molecules (Fig. \ref{trans}). Access to these regions can be desirable, however, for studies of features such as H$_2$O and C$_2$ bands in evolved stellar populations, or emission lines in redshifted galaxies. The main atmospheric absorber in these inter-band regions is H$_2$O, while CO$_2$ also contributes around 1.4, 1.6, and (especially) 2.0 $\mu$m  \citep{Glass99,Seifahrt10}. Telluric water vapor is monitored at several observatory sites, giving observers the opportunity to adapt their observing strategy to the prevailing conditions. To the best of our knowledge, though, the empirical relationship between atmospheric water vapor and the S/N in the inter-band regions of the final spectrum, after dividing by a standard star to remove the telluric lines, has not yet been presented in an easily-accessible form. In the GNIRS data set we have many pairs of galaxies and standard stars observed under a wide range of observing conditions. We therefore use the galaxy spectra to explore the effect of telluric water vapor on data quality in the regions of the spectrum around 1.35 and 1.85 $\mu$m. 

\begin{figure*}
\hspace*{-15mm}
\includegraphics[scale=0.95]{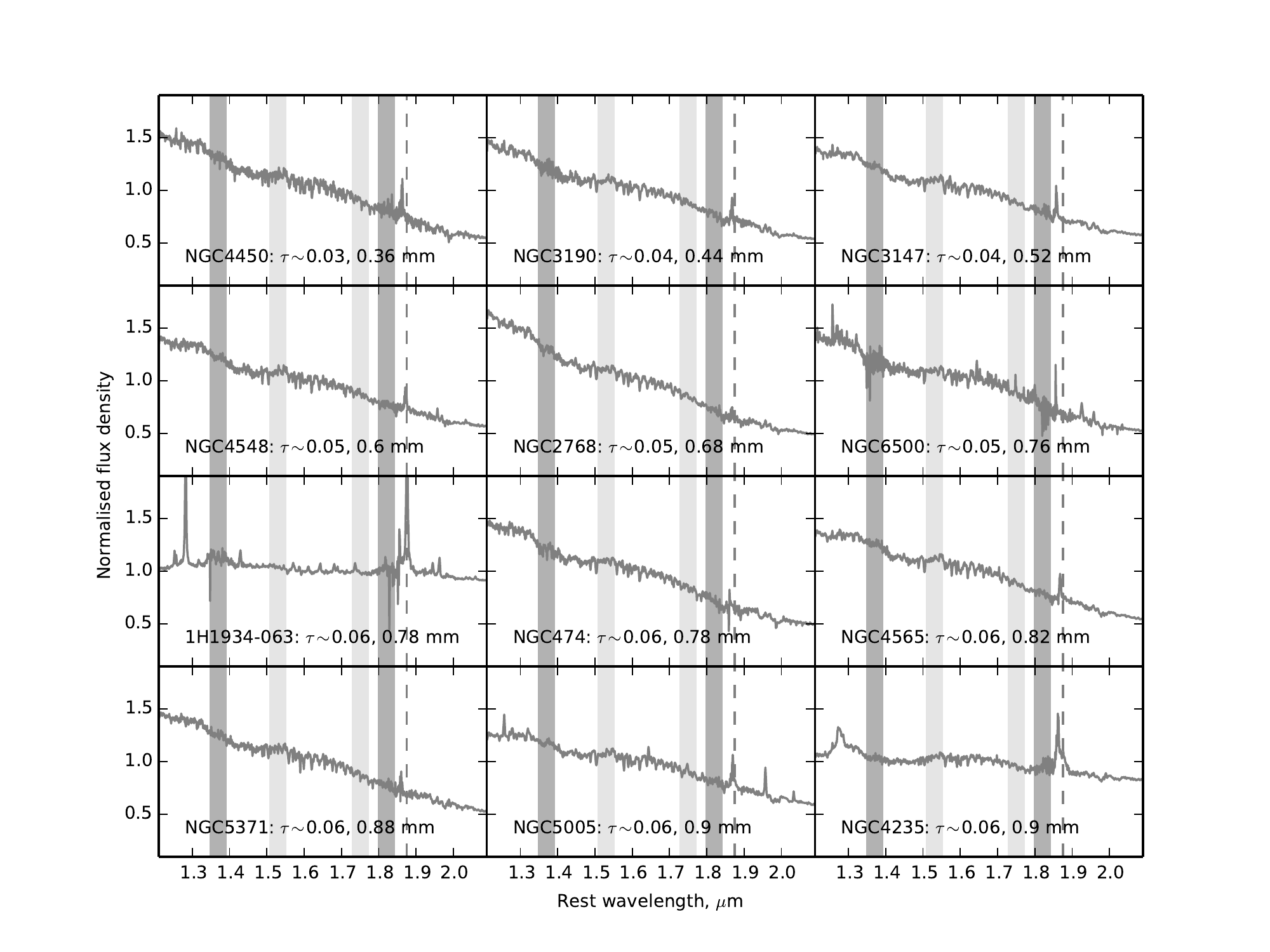}
\caption{ {\small Galaxy spectra arranged in order of increasing $\tau_{225 \; GHz}$. The shaded areas indicate regions used to measure the S/N plotted in Figure \ref{snvtau}. The dashed vertical line marks the wavelength of the Pa$\alpha$ emission line, which is not present in all the spectra but could be quite easily confused with a neighboring residual that appears in almost every data set. }}
\label{tau1}
\end{figure*}

\begin{figure*}
\hspace*{-15mm}
\includegraphics[scale=0.95]{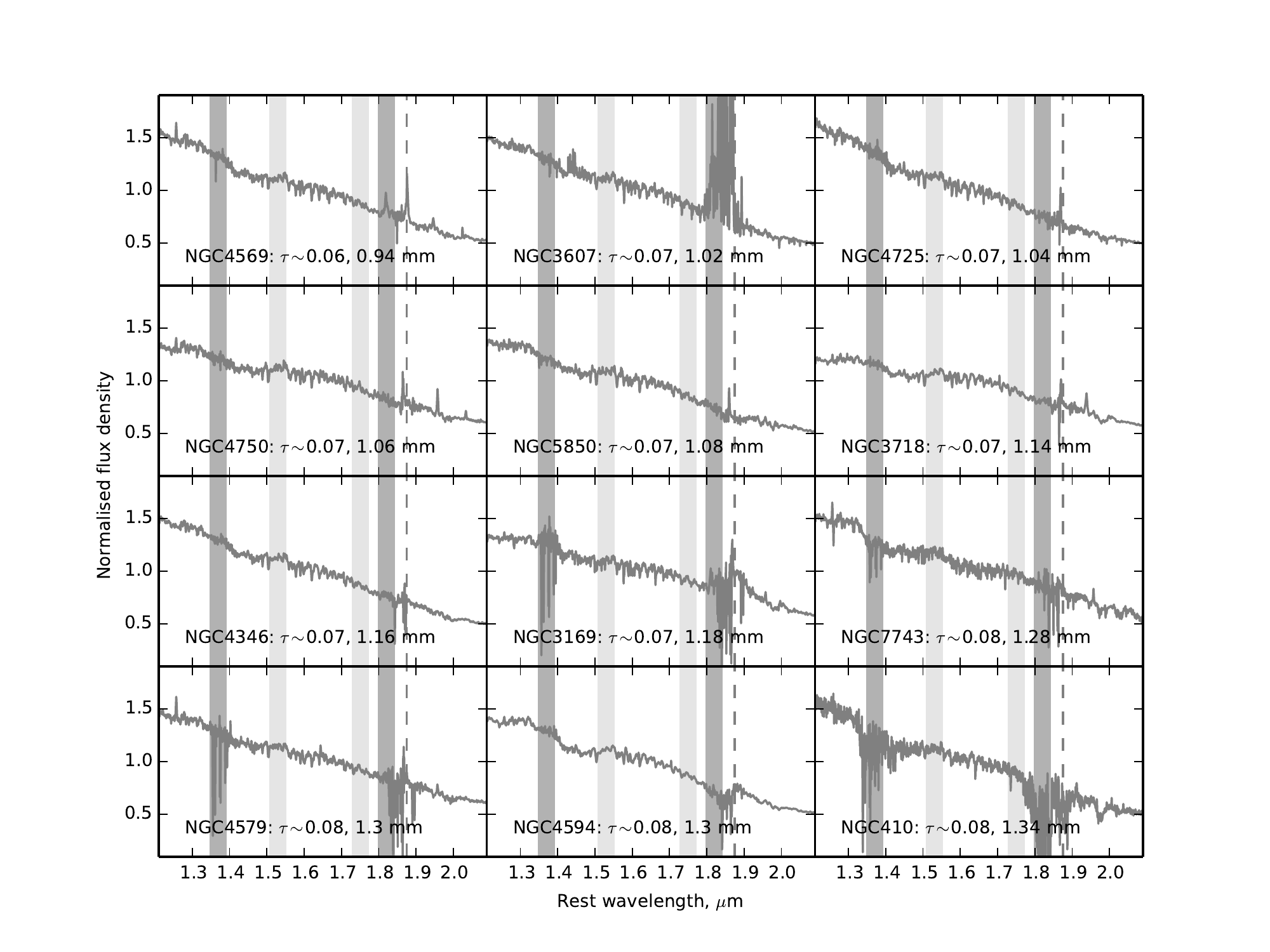}
\caption{ {\small As for Figure \ref{tau1}}}
\label{tau2}
\end{figure*}

\begin{figure*}
\hspace*{-15mm}
\includegraphics[scale=0.95]{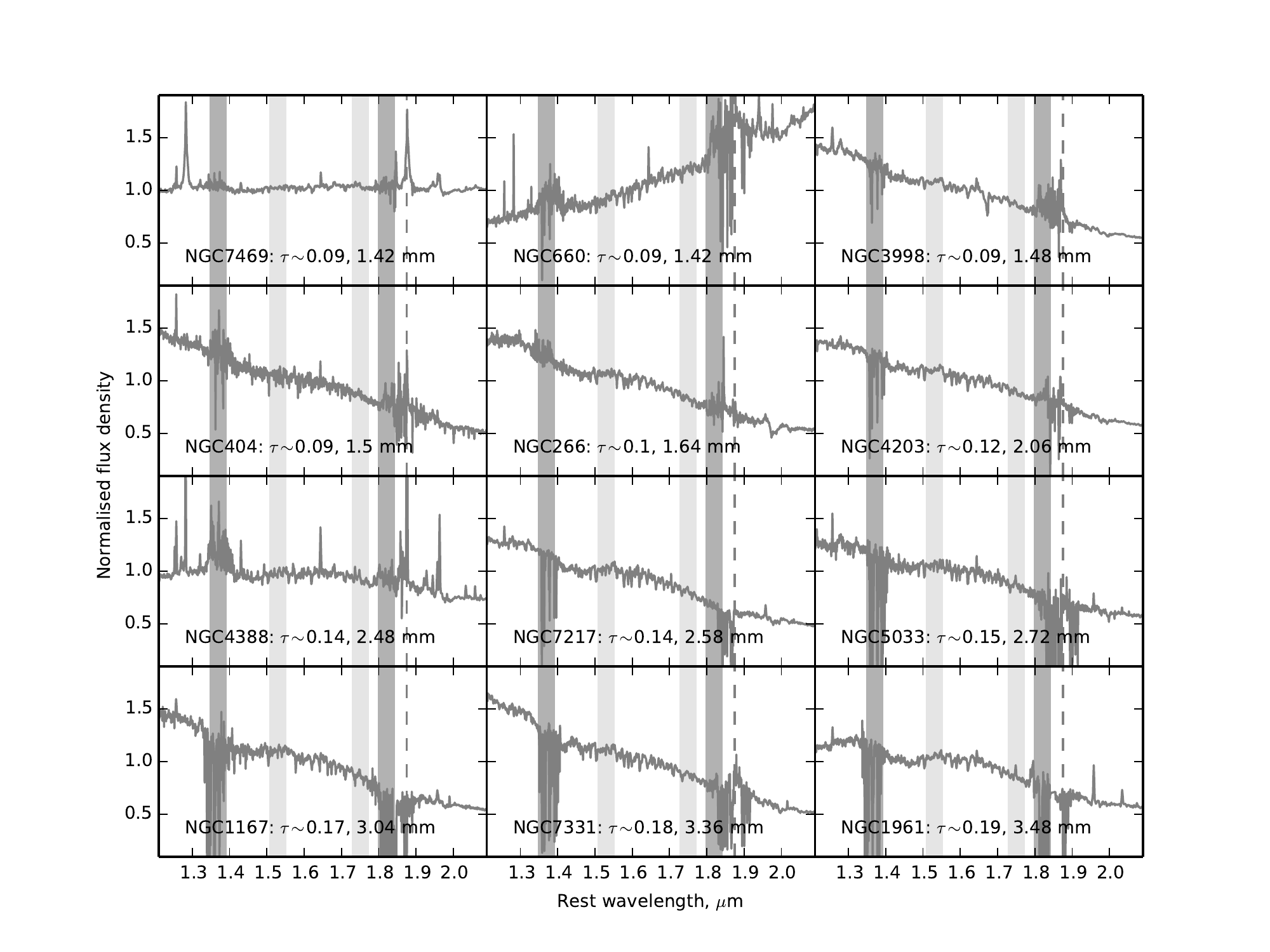}
\caption{ {\small As for Figure \ref{tau1}}}
\label{tau3}
\end{figure*}

\begin{figure*}
\hspace*{-15mm}
\includegraphics[scale=0.95, clip, trim=0 100 0 0]{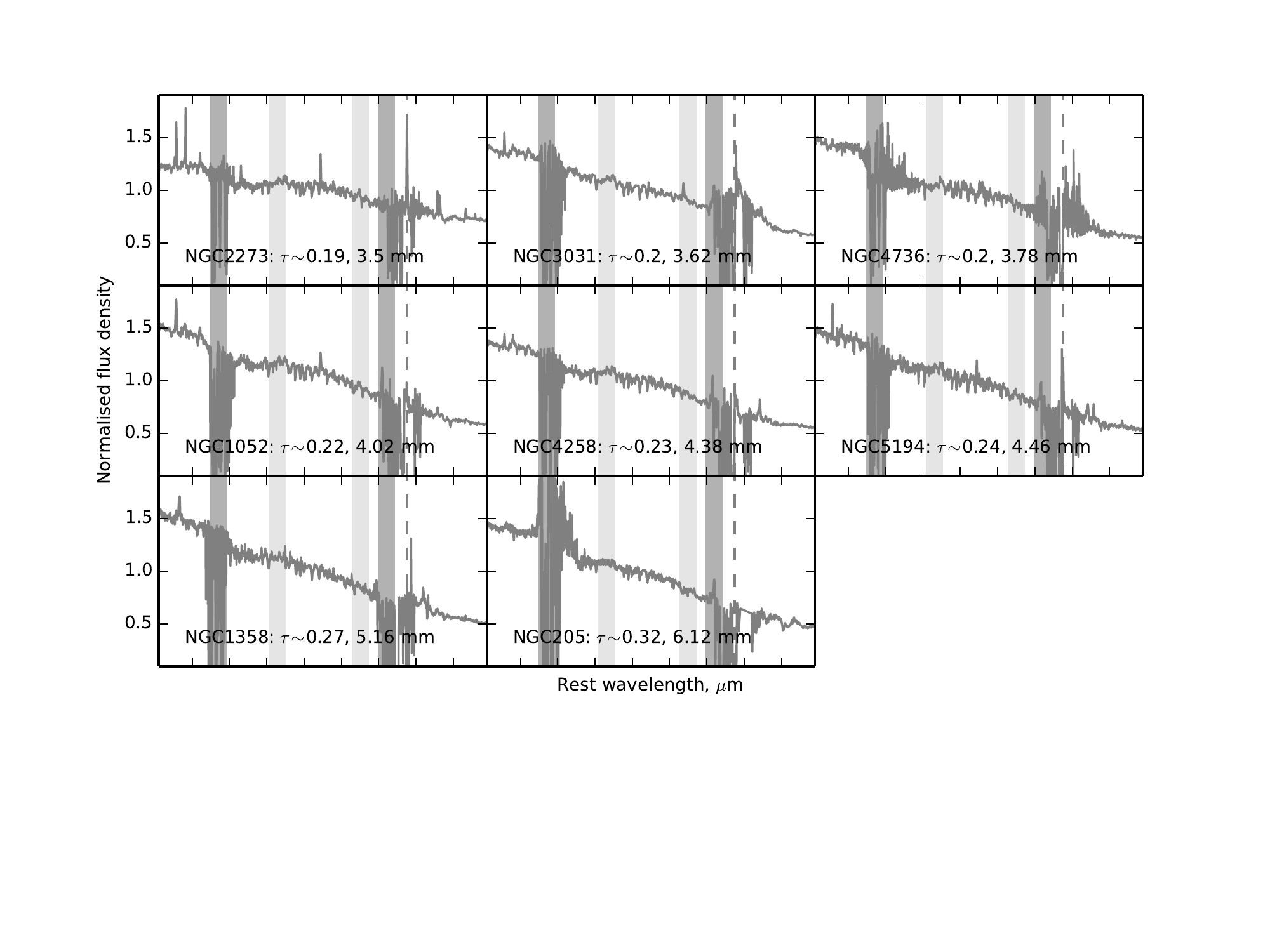}
\caption{ {\small As for Figure \ref{tau1}.}}
\label{tau4}
\end{figure*}

First, Figures \ref{tau1} - \ref{tau4} present the reduced spectra arranged in order of increasing 225 GHz zenith optical depth ($\tau_{225 \; GHz}$), or atmospheric water vapor column (WV), at the start of the observation\footnote{excluding NGC~2639, NGC~2655, NGC~2832, and NGC~3079, for which no $\tau$ data are available, and NGC~315 and NGC~4395, whose poor sky subtraction makes them unsuitable for this analysis.}. The 225 GHz optical depth is measured every 10 minutes by a tipping radiometer at the Caltech Submillimeter Observatory \citep{Radford11}, and is related to the water vapor column above Mauna Kea as WV = 20($\tau_{225 \; GHz}$ - 0.016) mm \citep[][]{Davis97}\footnote{See \citet{Thomas-Osip07} for other observing sites.}. The atmospheric water vapor column varies with elevation and azimuth angle \citep{Naylor02}, so $\tau_{225 \; GHz}$ is not necessarily an accurate indication of the water vapor column affecting the GNIRS observations. Nonetheless, the trend to lower S/N in the inter-band regions with increasing $\tau_{225 \; GHz}$ is clear. NGC~2768, for instance, observed at $\tau_{225 \; GHz} \sim 0.05$ (WV$\sim$0.7 mm; Figure \ref{tau1}), displays a very clean spectrum, whereas the inter-band regions of NGC~1358 ($\tau_{225 \; GHz} \sim 0.27$; WV$\sim$5 mm; Figure \ref{tau4}) are unusably noisy. Roughly speaking, the inter-band regions tend to be fairly free of residuals from dividing by the standard star up to $\tau_{225 \; GHz} \sim 0.07$ (WV$\sim$1.1 mm). Regardless of the water vapor, however, almost all spectra display at least a small residual just shortwards of the Pa$\alpha$ wavelength of 1.875 $\mu$m, where the atmosphere is virtually opaque even in the best conditions.

\begin{figure*}
\hspace*{-15mm}
\includegraphics[scale=0.95, clip, trim=0 200 0 0]{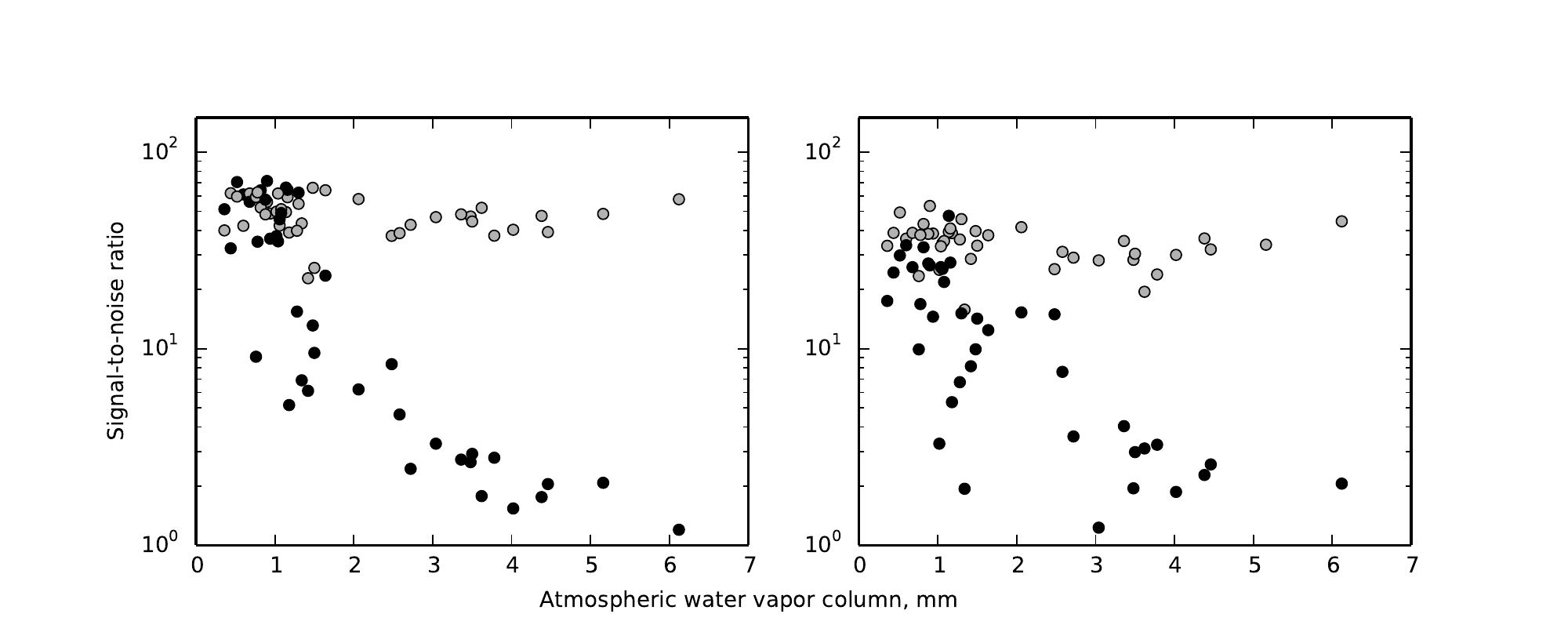}
\caption{ {\small  S/N vs telluric water column, in the spectral regions shown in Figures \ref{trans} - \ref{tau4}. Left: 1.35 - 1.39 $\mu$m (black) and 1.51 -- 1.55 $\mu$m (gray). Right: 1.80 -- 1.84 $\mu$m (black) and 1.73 -- 1.77 $\mu$m (gray).}}
\label{snvtau}
\end{figure*}

In Fig. \ref{snvtau} we show the S/N achieved in the galaxy spectra in two areas of strong H$_2$O absorption. On the whole, the S/N in these inter-band regions increases markedly as the WV decreases. At 1.35 -- 1.39 $\mu$m (between the J and H bands), S/N$\sim$30-70 is often achieved below WV$\sim$1.3 mm. Improvement is also seen at 1.81 -- 1.85 $\mu$m (between H and K), with S/N$\sim$20 quite frequently being reached in dry conditions. There is considerable scatter in the plots, though, particularly at 1.81 -- 1.85 $\mu$m. The point at 1.2 mm, S/N=5.4 in the right-hand plot is NGC~3169. The airmass match between this galaxy and its standard star was unusually poor, which may account for its unusually low S/N given the water vapor. There is no obvious reason for the other points in the lower left corner of the plot, which may instead be related to random spatial and temporal variations in the atmosphere.

While the trend in S/N vs $\tau$ is clear, the absolute values of the S/N obtained in very dry conditions are less straightforward to interpret. They may either indicate the limit on the achievable S/N imposed by systematic errors from the division by the standard star, or they may reflect the S/N that could be achieved for each object in the exposure time used for the observations, or other effects. For comparison, we also show the S/N in neighboring, relatively ``clean'' spectral regions. As expected, the S/N in those regions is not related to the water vapor column. In this particular data set, the S/N achieved at 1.35 -- 1.39 $\mu$m in dry conditions is comparable to that obtained in a relatively absorption-free region of the H band at 1.51 -- 1.55 $\mu$m. At 1.81 -- 1.85 $\mu$m, the S/N is systematically lower than that reached at 1.73 -- 1.77 $\mu$m, regardless of the conditions.

 The fact that we obtain similar S/N at 1.35 -- 1.39 and 1.51 -- 1.55 $\mu$m is likely an artifact of our simple definition of the S/N as the mean/rms of the points in that region. All else being equal, the S/N at $\sim$1.35 $\mu$m will be lower than that at $\sim$1.55 $\mu$m, simply because the strong atmospheric absorption means that the telescope receives fewer photons from the astronomical object at those wavelengths. The fact that the S/N is similar in regions of high and low transmission implies that systematic effects are limiting the measured S/N in the regions of high transmission. In this case, this likely results from the fact that the galaxy spectra contain real spectral structure that contributes to the measurements of the rms in the spectrum.  When similar S/N is reached in regions of low transmission, we are probably also measuring similar spectral structure.

Observations of pairs of featureless standard stars, which were not obtained as part of this observing program, would enable a more in-depth exploration of the limiting S/N that can be achieved in dry conditions. Nonetheless, we can draw some general conclusions. On one hand, it is possible to achieve S/N of several tens even in regions of very poor atmospheric transmission when the atmospheric water vapor column $\lesssim$1 mm. On the other hand, observing when the local $\tau$ meter indicates low water vapor is not a foolproof means of attaining high S/N: the relation between S/N and $\tau$ in Figure \ref{snvtau} contains considerable scatter. Also, it is important to realize that the atmospheric transmission is zero under many of the narrow lines that are blended together in Fig. \ref{trans} \citep[see e.g. the high-resolution Mauna Kea transmission spectrum shown by][]{Cotton14}. This means that if a narrow spectral feature lies at one of these wavelengths, it will not be detected in the final spectrum even if the telluric line cancellation is very good.

A number of groups have been investigating correcting for telluric absorption using atmospheric models \citep[e.g.][]{Bailey07,Seifahrt10,Cotton14,Gullikson14}. Until these methods are in widespread use, we hope that the analysis presented here will aid the planning of observations at the edges of the NIR atmospheric windows. 

\acknowledgments
REM is grateful to NRC-Herzberg Institute of Astrophysics for their hospitality during a portion of this work, and would like to acknowledge Else Starkenburg and Tom Geballe for helpful discussions. We also thank the anonymous referee for a timely report that helped improve the paper. Supported by the Gemini Observatory, operated by the Association of Universities for Research in Astronomy, Inc., on behalf of the international Gemini partnership of Argentina, Australia, Brazil, Canada, Chile, and the USA. CRA is supported by a Marie Curie Intra European Fellowship within the 7th European Community Framework Programme (PIEF-GA-2012-327934). LCH acknowledges support from the Kavli Foundation, Peking University, and the Chinese Academy of Science through grant No. XDB09030100 (Emergence of Cosmological Structures) from the Strategic Priority Research Program.  A.A.-H. acknowledges support from the Spanish Plan Nacional de Astronom\'{\i}a y Astrof\'{\i}sica under grant AYA2012-31447. ARA thanks the Conselho Nacional de Desenvolvimento Cient\'{i}fico e
Tecnol\'{o}gico (CNPq) for partial support of this work through grant
307403/2012-2.
This research used the facilities of the Canadian Astronomy Data Centre operated by the National Research Council of Canada with the support of the Canadian Space Agency.

\pagebreak


\end{document}